\documentclass[a4paper,11pt]{article}
\pdfoutput=1 
\usepackage{jheppub}	
\usepackage[utf8]{inputenc}
\usepackage[english]{babel}
\usepackage{makeidx}
\usepackage{amsfonts}
\usepackage{enumerate}
\usepackage{mathrsfs}
\usepackage{tensor}
\usepackage{xcolor,tikz,pgfplots}
\usepackage[autostyle]{csquotes}

\usepackage{youngtab}
\usepackage{caption}
\usepackage{subcaption}
\newcommand\subcap[1]{\phantomcaption%
       \caption*{\textbf{\figurename~\thefigure\thesubfigure:} #1}} 

\usepackage{braket}

\usetikzlibrary{matrix,calc,positioning,decorations.markings,decorations.pathmorphing,decorations.pathreplacing}
\usetikzlibrary{arrows,cd}
\usetikzlibrary{shapes.misc}
\usepackage{bbding}
\tikzset{
    >=stealth',
    punkt/.style={
           rectangle,
           rounded corners,
           draw=black, very thick,
           text width=7.2em,
           minimum height=2em,
           text centered},
    punkt2/.style={
           rectangle,
           rounded corners,
           draw=black!20!red, very thick,
           text width=7em,
           minimum height=2em,
           text centered},
    punktL/.style={
           rectangle,
           rounded corners,
           draw=black!20!red, very thick,
           text width=8.8em,
           minimum height=2em,
           text centered},
    pil/.style={
           ->,
           thick,
           shorten <=2pt,
           shorten >=2pt,},
    pil2/.style={
           <->,
           thick,
           shorten <=2pt,
           shorten >=2pt,}}

\usepackage{float}
\usepackage{booktabs}

\usepackage{bm}
\usepackage{anyfontsize}

\numberwithin{equation}{section}

\title{{\fontsize{16.75pt}{19pt}\selectfont $\mathcal{N}=1$ conformal dualities from unoriented chiral quivers}}

\author[a]{\fontsize{12.5pt}{13pt}\selectfont Antonio Amariti,}
\author[b]{Massimo Bianchi,}
\author[c]{Marco Fazzi,} 
\author[d,e]{Salvo Mancani,}
\author[e]{Fabio Riccioni,}
\author[a,f]{Simone Rota}

\affiliation[a]{INFN, Sezione di Milano, Via Celoria 16, I-20133 Milano, Italy}
\affiliation[b]{Dipartimento  di  Fisica,  Universit\`a  di  Roma  ``Tor  Vergata'' \& INFN, Sezione di Roma ``Tor Vergata'', Via della Ricerca Scientifica 1, I-00133, Roma, Italy}
\affiliation[c]{Dipartimento di Fisica, Universit\`a di Milano-Bicocca \& INFN, Sezione di Milano-Bicocca, Piazza della Scienza 3, I-20126 Milano, Italy}
\affiliation[d]{Dipartimento di Fisica, Università di Roma ``La Sapienza'', Piazzale Aldo Moro 2, I-00185 Roma, Italy}
\affiliation[e]{INFN, Sezione di  Roma, Piazzale Aldo Moro 2, I-00185 Roma, Italy}
\affiliation[f]{Dipartimento di Fisica, Universit\`a degli Studi di Milano, Via Celoria 16, I-20133 Milano, Italy}

\emailAdd{antonio.amariti@mi.infn.it,  marco.fazzi@mib.infn.it,  simone.rota@mi.infn.it, massimo.bianchi@roma2.infn.it, salvo.mancani@uniroma1.it, Fabio.Riccioni@roma1.infn.it}

\abstract{
We study various orientifold projections of 4d $\mathcal{N}=1$ toric gauge theories, associated with CY singularities known as $L^{a,b,a}/\mathbb{Z}_2$, with $a+b$ even. We obtain superconformal chiral theories that have the same central charge, anomalies and superconformal index, whereas they were different before the orientifold. Some of these projections are implemented by a novel type of orientifold without fixed loci, known as glide orientifold. We claim that these theories flow to the same conformal manifold, and they are connected by quadratic exactly marginal deformations. The latter can be written in terms of conjugate pairs of bifundamental fields of $R$-charge one, generalizing previous results for unoriented non-chiral theories.
}


\preprint{PREPRINT}
 \clearpage

\begin{document}

\maketitle

\section{Introduction}\label{sec:Intro}

D3-branes probing local singularities in Calabi--Yau (CY) spaces provide a very large, in fact infinite, class of superconformal theories~\cite{Maldacena_1999, Gubser_1998, Witten:1998qj, Beasley_2000}. The near-horizon geometry is a product of five-dimensional AdS spacetime and a five-dimensional Sasaki--Einstein manifold~\cite{Morrison:1998cs, Klebanov:1998hh}, the `base of the CY cone'. Many chiral and non-chiral superconformal quiver theories with several unitary gauge groups and matter in the bifundamental or adjoint representations can be realized this way~\cite{Morrison:1998cs}, which proved to be an unprecedented laboratory to study the behavior of strongly coupled gauge theories using holography \cite{Witten:1998qj, Gubser_1998, Aharony:1999ti}. If one adds orientifold planes  to the brane configuration  new interesting physics arises. Orthogonal and symplectic gauge groups enter the game together with  (anti)symmetric tensors or bi(anti)fundamentals of the unitary gauge group factors, giving rise to `unoriented quiver theories' \cite{Bianchi:2013gka, GarciaEtxebarria:2012qx, Garcia-Etxebarria:2013tba, Garcia-Etxebarria:2015hua, Garcia-Etxebarria:2016bpb, Bianchi:2020fuk}. This results from the orientifold projection in string theory that reverses the orientation of the strings and induces a $\mathbb{Z}_2$ involution in the gauge theory~\cite{Sagnotti:1987tw, Pradisi:1988xd, Bianchi:1990yu, Bianchi:1990tb, Polchinski:1995mt, Angelantonj:2002ct}. Unoriented quiver theories get close to providing a local embedding of the Standard Model \cite{Ibanez:2001nd, Wijnholt:2007vn,Cicoli:2021dhg} as well as other interesting scenarios \cite{Addazi:2014ila,Addazi:2015rwa,Addazi:2015hka, Addazi:2015yna}. The construction may be further specialized to toric CY singularities, that allow for a brane tiling or dimer description \cite{Franco:2007ii} (see also \cite{Argurio:2020dko} for recent developments based only on geometric data), or enriched by including non-compact `flavor' branes \cite{Bianchi:2013gka} and even taking into account non-perturbative stringy effects associated to Euclidean D-branes and their bound-states \cite{Bianchi:2009ij, Bianchi:2012ud, Bianchi:2012kt}.
Moreover orientifolds on brane tilings have been recently shown to lead to stable supersymmetry breaking gauge theory vacua in \cite{Argurio:2019eqb, Argurio:2020dkg, Argurio:2020npm, Argurio:2022vfq}.
In spite of these important results, realizing new superconformal theories with orientifolds remains  challenging, since O-planes tend to spoil conformal symmetry \cite{Bianchi:2013gka, Argurio:2017upa, Bianchi:2020fuk}, at least to subleading order in $1/N$ in case of a large number $N$ of D3-branes. 

Quite remarkably, it was recently shown that specific orientifold models not only give rise to conformal fixed points at strong coupling in the infrared (IR), but also that these models belong to the same conformal manifold. The mechanism was first identified and discussed in \cite{Antinucci:2020yki,Antinucci:2021edv} and then extended in \cite{Amariti:2021lhk} to non-chiral $L^{a,b,a}$ quiver gauge theories, a subfamily of the $L^{p,q,r}$ models \cite{Benvenuti:2005ja,Butti:2005sw,Franco:2005sm} characterized by the presence of orbifold singularities. The models admit both a brane tiling description and a Hanany--Witten one \cite{Hanany:1996ie} in type IIA string theory in terms of $N$ D4-branes, extended along the directions $x^{0123}$ and compactified along one direction, say $x^6$, and $n_G$ NS5-branes extended along $x^{012345}$ and separated along $x^6$. Such non-chiral models, known as elliptic models~\cite{Witten:1997sc, Uranga:1998vf}, have $n_G$ $SU(N)$ gauge groups\footnote{The $U(1)_\text{center-of-mass}$ is free while the other $U(1)$'s decouple in the IR.} and $\mathcal{N}=2$ supersymmetry if $a=0$, otherwise $\mathcal{N}=1$. Pairs of six-dimensional orientifold planes with opposite charge (O6$^{\pm}$) extended along $x^{0123457}$ can be placed symmetrically on the circle without breaking further supersymmetry. For the case with extended supersymmetry this description has been extensively studied in~\cite{Uranga:1998uj}, where four families of models have been identified. The classification depends on the presence of an odd or even number $n_G$ of gauge groups and on the possibility of placing, or not placing, an O6$^{+}$ and/or an O6$^{-}$ on top of an NS5-brane.
Breaking supersymmetry down to $\mathcal{N}=1$ in presence of orientifold planes and suitable choices of fractional branes has been shown to lead to models with conjugate pairs of chiral multiplets with $R$-charge $R=1$ in tensor representations of the gauge group~\cite{Bianchi:2020fuk, Antinucci:2021edv, Amariti:2021lhk}.
Supersymmetry is broken in general by tilting some of the NS5-branes and/or O6-planes, such that the orientifold projection can be still applied consistently.\footnote{In the IIA elliptic engineering of $L^{a,b,a}$, $a+b$ corresponds to the total number of untilted NS5-branes, whereas $a$ to that of the tilted NS5's. When $a=0$ we are left with $b=n_G$ NS5's and $\mathcal{N}=2$ supersymmetry.}

From the perspective of brane tilings associated to $L^{a,b,a}$, we can visualize the process with orientifolds acting with fixed loci on the tiling, as discussed in~\cite{Franco:2007ii}. Proper choices of fractional branes, often dictated by the constraints on the $\beta$-functions, lead to models with the same central charges and superconformal index after integrating out the chiral fields with $R$-charge $R=1$ \cite{Antinucci:2021edv, Amariti:2021lhk}. The mass terms for these fields have indeed $R$-charge $R=2$ and they are exactly marginal deformations. 
From a purely field theory perspective, a similar situation occurs when breaking $\mathcal{N}=2$ by a mass term for the adjoint field, given an $\mathcal{N}=1$ description. The resulting theory develops a quartic term in the superpotential and the Seiberg dual theory has mesons with a marginal mass term, so that these mesons can be integrated out. In this context, Seiberg duality relates strongly-coupled and weakly-coupled regimes of theories whose matter content differs only for mesons with marginal mass~\cite{Leigh:1995ep, Strassler:1995xm}, inheriting the action of $S$-duality from the mother $\mathcal{N}=2$ theory.\footnote{Or better, being $\eta$ the coupling of the quartic term, there is a line of conformal theories described by the equation $\gamma(g,\eta)=-1/2$ and Seiberg duality relates opposite regimes on this line, see~\cite{Strassler:2005qs}.} 

Observe that parent theories $L^{a_1,b_1,a_1}$, described either by elliptic models or brane tilings, admit relevant mass terms~\cite{Bianchi:2014qma} that deform the model into an $L^{a_2,b_2,a_2}$ with $a_1+b_1=a_2+b_2$ and $b_1>b_2$. Clearly, they are not Seiberg dual to each other, as can be seen from the fact that their toric geometry is different. The novel aspect here is the presence of the orientifold, because once we reach $\mathcal{N}=1$, all projected models with constant $a+b$ have the same number of gauge groups and of non-anomalous $U(1)$'s but differ only by the presence of fields, tensors or adjoints, that admit marginal mass terms. Therefore, the consequence of the orientifold projection is that the two theories flow to 
the same conformal manifold. 
For this reason, we borrow nomenclature from the literature \cite{Razamat:2019vfd} and say that certain orientifolds of $L^{a,b,a}$ with constant $a+b$ are \emph{conformally dual}.
They are not Seiberg dual, for they cannot be related by Seiberg dualities known in the literature~\cite{Antinucci:2021edv, Amariti:2021lhk}, but they inherit part of the $S$-duality action on the marginal masses from the mother $\mathcal{N}=2$ models through the mechanism of inherited duality introduced in~\cite{Argyres:1999xu,Halmagyi:2004ju}.

It is natural to wonder whether the mechanism discussed so far can be generalized to other $\mathcal{N}=1$ models, extending the notion of conformal dualities to toric quiver gauge theories with a chiral field content. 
The first necessary ingredient in the recipe is the presence of internal points in the toric diagram. Indeed these are associated to anomalous $U(1)$ global (baryonic) symmetries  and they require the presence of a chiral field content, i.e. there are bi-fundamental fields connecting two nodes of the quiver, without the corresponding anti-bifundamental~\cite{Douglas:1996sw}.
Another necessary  ingredient is the presence of points on the perimeter of the toric diagram, because this allows to RG flow from one model to another even before the orientifold projection, through a mass deformation~\cite{Bianchi:2014qma}.

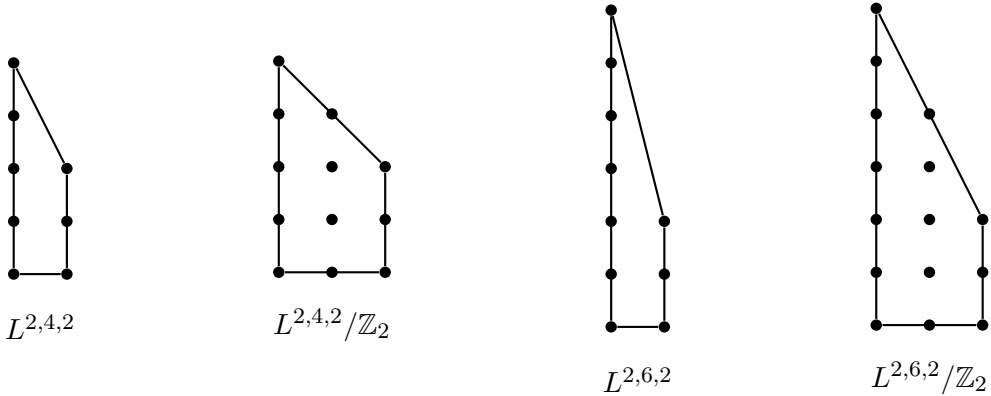
\begin{figure}[ht!]
    \hfill
    \begin{subfigure}{0.2\textwidth}
      \begin{tikzpicture}[auto, scale=0.7]
		\node [circle, fill=black, inner sep=0pt, minimum size=1.5mm] (0) at (0,0) {}; 
		\node [circle, fill=black, inner sep=0pt, minimum size=1.5mm] (1) at (0,1) {};
		\node [circle, fill=black, inner sep=0pt, minimum size=1.5mm] (2) at (0,2) {}; 
		\node [circle, fill=black, inner sep=0pt, minimum size=1.5mm] (3) at (0,3) {}; 
		\node [circle, fill=black, inner sep=0pt, minimum size=1.5mm] (4) at (0,4) {}; 
		\node [circle, fill=black, inner sep=0pt, minimum size=1.5mm] (r0) at (1,0) {};
		\node [circle, fill=black, inner sep=0pt, minimum size=1.5mm] (r1) at (1,1) {};
		\node [circle, fill=black, inner sep=0pt, minimum size=1.5mm] (r2) at (1,2) {};
        \draw (0) to (4) [thick];
        \draw (4) to (r2) [thick];
        \draw (r2) to (r0) [thick];
        \draw (r0) to (0) [thick];
        \node (a) at (0.5,-1) {$L^{2,4,2}$}; 
      \end{tikzpicture}
    \end{subfigure}
    \hspace{7pt} 
    \begin{subfigure}{0.2\textwidth}
      \begin{tikzpicture}[auto, scale=0.7]
		\node [circle, fill=black, inner sep=0pt, minimum size=1.5mm] (0) at (0,0) {}; 
		\node [circle, fill=black, inner sep=0pt, minimum size=1.5mm] (1) at (0,1) {};
		\node [circle, fill=black, inner sep=0pt, minimum size=1.5mm] (2) at (0,2) {}; 
		\node [circle, fill=black, inner sep=0pt, minimum size=1.5mm] (3) at (0,3) {}; 
		\node [circle, fill=black, inner sep=0pt, minimum size=1.5mm] (4) at (0,4) {}; 
		\node [circle, fill=black, inner sep=0pt, minimum size=1.5mm] (c0) at (1,0) {};
		\node [circle, fill=black, inner sep=0pt, minimum size=1.5mm] (c1) at (1,1) {};
		\node [circle, fill=black, inner sep=0pt, minimum size=1.5mm] (c2) at (1,2) {};
		\node [circle, fill=black, inner sep=0pt, minimum size=1.5mm] (c3) at (1,3) {};
		\node [circle, fill=black, inner sep=0pt, minimum size=1.5mm] (r0) at (2,0) {};
		\node [circle, fill=black, inner sep=0pt, minimum size=1.5mm] (r1) at (2,1) {};
		\node [circle, fill=black, inner sep=0pt, minimum size=1.5mm] (r2) at (2,2) {};
        \draw (0) to (4) [thick];
        \draw (4) to (r2) [thick];
        \draw (r2) to (r0) [thick];
        \draw (r0) to (0) [thick];
        \node (a) at (1,-1) {$L^{2,4,2}/\mathbb{Z}_2$};
      \end{tikzpicture}
    \end{subfigure}
    \hspace{30pt} 
    \begin{subfigure}{0.2\textwidth}
      \begin{tikzpicture}[auto, scale=0.7]
		\node [circle, fill=black, inner sep=0pt, minimum size=1.5mm] (0) at (0,0) {}; 
		\node [circle, fill=black, inner sep=0pt, minimum size=1.5mm] (1) at (0,1) {};
		\node [circle, fill=black, inner sep=0pt, minimum size=1.5mm] (2) at (0,2) {}; 
		\node [circle, fill=black, inner sep=0pt, minimum size=1.5mm] (3) at (0,3) {}; 
		\node [circle, fill=black, inner sep=0pt, minimum size=1.5mm] (4) at (0,4) {}; 
		\node [circle, fill=black, inner sep=0pt, minimum size=1.5mm] (5) at (0,5) {};
		\node [circle, fill=black, inner sep=0pt, minimum size=1.5mm] (6) at (0,6) {}; 
		\node [circle, fill=black, inner sep=0pt, minimum size=1.5mm] (r0) at (1,0) {};
		\node [circle, fill=black, inner sep=0pt, minimum size=1.5mm] (r1) at (1,1) {};
		\node [circle, fill=black, inner sep=0pt, minimum size=1.5mm] (r2) at (1,2) {};
        \draw (0) to (6) [thick];
        \draw (6) to (r2) [thick];
        \draw (r2) to (r0) [thick];
        \draw (r0) to (0) [thick];
        \node (a) at (0.5,-1) {$L^{2,6,2}$}; 
      \end{tikzpicture}
    \end{subfigure}
    \hspace{7pt} 
    \begin{subfigure}{0.2\textwidth}
      \begin{tikzpicture}[auto, scale=0.7]
		\node [circle, fill=black, inner sep=0pt, minimum size=1.5mm] (0) at (0,0) {}; 
		\node [circle, fill=black, inner sep=0pt, minimum size=1.5mm] (1) at (0,1) {};
		\node [circle, fill=black, inner sep=0pt, minimum size=1.5mm] (2) at (0,2) {}; 
		\node [circle, fill=black, inner sep=0pt, minimum size=1.5mm] (3) at (0,3) {}; 
		\node [circle, fill=black, inner sep=0pt, minimum size=1.5mm] (4) at (0,4) {}; 
		\node [circle, fill=black, inner sep=0pt, minimum size=1.5mm] (5) at (0,5) {};
		\node [circle, fill=black, inner sep=0pt, minimum size=1.5mm] (6) at (0,6) {}; 
		\node [circle, fill=black, inner sep=0pt, minimum size=1.5mm] (c0) at (1,0) {};
		\node [circle, fill=black, inner sep=0pt, minimum size=1.5mm] (c1) at (1,1) {};
		\node [circle, fill=black, inner sep=0pt, minimum size=1.5mm] (c2) at (1,2) {};
		\node [circle, fill=black, inner sep=0pt, minimum size=1.5mm] (c3) at (1,3) {};
		\node [circle, fill=black, inner sep=0pt, minimum size=1.5mm] (c4) at (1,4) {};
		\node [circle, fill=black, inner sep=0pt, minimum size=1.5mm] (r0) at (2,0) {};
		\node [circle, fill=black, inner sep=0pt, minimum size=1.5mm] (r1) at (2,1) {};
		\node [circle, fill=black, inner sep=0pt, minimum size=1.5mm] (r2) at (2,2) {};
        \draw (0) to (6) [thick];
        \draw (6) to (r2) [thick];
        \draw (r2) to (r0) [thick];
        \draw (r0) to (0) [thick];
        \node (a) at (1,-1) {$L^{2,6,2}/\mathbb{Z}_2$};
      \end{tikzpicture}
    \end{subfigure}
    \hfill
    \caption{Examples of chiral orbifolds for the $L^{a,b,a}$ family.} 
\label{fig:ToricExamples}
\end{figure}

A natural set of models where to look for a generalization of the mechanism of conformal duality in presence of orientifolds consists of $L^{a,b,a}/\mathbb{Z}_2$ orbifolds, leading generically to a chiral field content (with the exception of $L^{0,2,0}$ and $L^{1,1,1}$) -- see Fig. \ref{fig:ToricExamples} for examples of such orbifolds on the toric diagram.
In this paper we will see  that such orbifolds admit a generalization of the mechanism of  conformal duality similar to the one obtained for the four $\mathcal{N}=2$ families of~\cite{Uranga:1998uj} broken to $\mathcal{N}=1$. We will distinguish two out of these four families (corresponding to families \textbf{i)} and \textbf{iv)} of \cite{Uranga:1998uj}) allowing for the presence of marginal mass terms. Another difference with the construction presented in~\cite{Antinucci:2021edv, Amariti:2021lhk} is that here we will observe that in some cases fractional branes will not be required. Observe also that in these $L^{a,b,a}/\mathbb{Z}_2$ models the type IIA description is not readily available and the orientifold projection can be performed on the dimer model with the techniques of~\cite{Franco:2007ii} and the recent extension of~\cite{Garcia-Valdecasas:2021znu} in terms of a Klein bottle. This is the other main novelty of the construction performed here. We will observe that the projections are implemented on the brane tiling either in terms of fixed points and fixed lines or by maps on the dimer without fixed points that lead to Klein bottles. 

The paper is organized as follows. Section \ref{sec:conf} contains a lightning review on how to perform orientifold projections on toric dimer models and summarizes the main findings of \cite{Antinucci:2020yki, Antinucci:2021edv, Amariti:2021lhk} on 4d $\mathcal{N}=1$ conformal dualities. In section \ref{sec:glide} we introduce the glide orientifold and its role in the two families of unoriented orbifold models, family $\mathcal{A}$ and family $\mathcal{B}$, to which we devote section \ref{sec:famA} and \ref{sec:famB} respectively. Finally in section \ref{sec:conc} we present our conclusions and discuss some open research avenues.

\section{Orientifolds and conformal duality}
\label{sec:conf}

The aim of this section is to summarize the results of~\cite{Antinucci:2020yki, Antinucci:2021edv, Amariti:2021lhk}, where specific $\mathcal{N}=1$ supersymmetric gauge theories obtained as orientifold projections of toric quivers were shown to be related to one another by conformal duality. We will first concentrate on the simplest models wherein this occurs, namely the orientifold of $\mathbb{C}^2/\mathbb{Z}_2 \times\mathbb{C}$ and the orientifold of the conifold, obtained by mass deformation of the former~\cite{Klebanov:1998hh}. This example contains all relevant information and, as we will see, it can be naturally generalized to the more elaborate models discussed in~\cite{Antinucci:2020yki, Antinucci:2021edv, Amariti:2021lhk}. 

\subsection{A picture of five-branes}

A toric diagram in a $\mathbb{Z}^d$ lattice encodes the information about a $d$-dimensional complex toric variety, admitting the action of a complex torus $(\mathbb{C}^*)^d$ \cite{Bouchard:2007ik, Leung:1997tw, Closset:2009sv, Book:Cox}. For toric CY threefolds (i.e. $d=3$) it is enough to focus on a two-dimensional diagram, thanks to the vanishing of the first Chern class \cite{Leung:1997tw}. The toric data can be translated into a well defined supersymmetric gauge theory in four dimensions, using a five-brane diagram as an intermediate step, from which the corresponding brane tiling or dimer can be immediately drawn \cite{Hanany:2005ve, Franco:2005rj, Feng:2001bn, Franco:2005zu, Hanany:2005ss, Hanany:2006nm, Kennaway:2007tq, Yamazaki_2008}.
More explicitly, one identifies the vectors with coordinates $(p,q)$ that are dual (outgoing normal) to the sides of the toric diagram and represents them as one-cycles of a two-dimensional torus with $(p,q)$ winding numbers. The resulting five-brane diagram owes its name to the fact that in the IIB picture, whereby D3-branes are T-dualized into D5-branes wrapping the two-torus, the one-cycles are NS5-branes that emerge from T-dualizing the toric singularity. As an example, see the toric diagram of $\mathbb{C}^3/\mathbb{Z}_2$ and the associated five-brane diagram in Figs.~\ref{fig:ToricDiagramC3Z2Secondo}-\ref{fig:five-braneC3Z2}.

\begin{figure}
       \begin{subfigure}{0.4\textwidth}
    \centering{
    \begin{tikzpicture}[auto, scale=0.7]
		\node [circle, fill=black, inner sep=0pt, minimum size=1.5mm] (0) at (0,0) {}; 
		\node [circle, fill=black, inner sep=0pt, minimum size=1.5mm] (1) at (2,0) {};
		\node [circle, fill=black, inner sep=0pt, minimum size=1.5mm] (2) at (0,2) {}; 
		\node [circle, fill=black, inner sep=0pt, minimum size=1.5mm] (3) at (0,4) {}; 
        \draw (0) to (1) [thick];
        \draw (1) to (3) [thick];
        \draw (3) to (2) [thick];
        \draw (2) to (0) [thick];
        \draw (1,0) to (1,-2) [->, thick, blue!50];
        \draw (0,1) to (-2,1) [->, thick, red!70];
        \draw (0,3) to (-2,3) [->, thick, red!70];
        \draw (1,2) to (3,3) [->, thick, green!70!black];
\end{tikzpicture}}
\vspace{20pt}
\subcap{The toric diagram of $\mathbb{C}^2/\mathbb{Z}_2 \times \mathbb{C}$ and the vectors orthogonal to the edges.}\label{fig:ToricDiagramC3Z2Secondo}
    \end{subfigure}
    \hfill
       \begin{subfigure}{0.4\textwidth}
    \centering{
    \begin{tikzpicture}[auto, scale=0.7]
		\node (0) at (0,0) {}; 
		\node (1) at (6,0) {};
		\node (2) at (6,6) {}; 
		\node (3) at (0,6) {}; 
		\node (c) at (3,3) {};
		\node (c1) at (3,4.5) {};
		\node (c2) at (3,1.5) {};
		\node[cross out, minimum size=2mm, draw=red, inner sep=1mm, thick] (0) at (0,0) {}; 
		\node[cross out, minimum size=2mm, draw=red, inner sep=1mm, thick] (O2) at (0,3) {};
		\node[cross out, minimum size=2mm, draw=red, inner sep=1mm, thick] (O3) at (3,3) {};
		\node[cross out, minimum size=2mm, draw=red, inner sep=1mm, thick] (O4) at (3,0) {};
        \draw (0) to (1) [shorten >=-0.15cm, shorten <=-0.15cm, thick];
        \draw (1) to (2) [shorten >=-0.15cm, shorten <=-0.15cm, thick];
        \draw (2) to (3) [shorten >=-0.15cm, shorten <=-0.15cm, thick];
        \draw (3) to (0) [shorten >=-0.15cm, shorten <=-0.15cm, thick];
        \draw (3,6) to (3,0) [->, thick, blue!50];
        \draw (6,1.5) to (0,1.5) [->, thick, red!70];
        \draw (6,4.5) to (0,4.5) [->, thick, red!70];
        \draw (3,0) to (6,1.5) [->, thick, green!70!black];
        \draw (0,1.5) to (6,4.5) [->, thick, green!70!black];
        \draw (0,4.5) to (3,6) [->, thick, green!70!black];
        \draw[fill=gray!80!white, nearly transparent]  (3,0) -- (6,1.5) -- (3,1.5) -- cycle;
        \draw[fill=gray!30!white, nearly transparent]  (0,1.5) -- (3,3) -- (3,1.5) -- cycle;
        \draw[fill=gray!80!white, nearly transparent]  (3,4.5) -- (3,3) -- (6,4.5) -- cycle;
        \draw[fill=gray!30!white, nearly transparent]  (0,4.5) -- (3,4.5) -- (3,6) -- cycle;
        \node [above right=0.1 cm of 0] {1};
        \node [above left=0.05 pt of 1] {1};
        \node [below right=0.1 cm of c] {0};
        \node [above left=0.1cm of c] {0};
        \node [below left=0.1 cm of 2] {1};
        \node [below right=0.05 pt of 3] {1};
        \node [below right=0.05 pt of c1] {\tiny{$(-1)$}};
        \node [above left=0.05 pt of c2] {\tiny{$(+1)$}};
        \node [above left=0.05 pt of c1] {\tiny{$(+1)$}};
        \node [below right=0.05 pt of c2] {\tiny{$(-1)$}};
        \node [left=0.05pt of O2, red] {$\tau_0$};
        \node [right=0.05pt of O3, red] {$\tau_{00}$};
        \node [below left=0.05pt of 0, red] {$\tau_1$};
        \node [below left=0.05pt of O4, red] {$\tau_{11}$};
\end{tikzpicture}}
\subcap{The five-brane diagram of $\mathbb{C}^2/\mathbb{Z}_2 \times \mathbb{C}$, with the four fixed points of the orientifold projection.}\label{fig:five-braneC3Z2}
    \end{subfigure}
\end{figure}

The one-cycles divide the planar graph in different regions, highlighted in white, gray or black in Fig.~\ref{fig:five-braneC3Z2}. White regions are $SU(N)$ gauge factors, whereas bifundamental fields $X_{ab}$, where $a,b$ are labels for the gauge factors, correspond to arrows  crossing the white regions $a$ and $b$ at the  intersection points. The direction of the arrows determines if a field transforms as a fundamental (out)  or antifundamental (in) of a gauge factor. Gray and black regions are encircled by the arrows in clockwise ($+1$) or counterclockwise ($-1$) direction, respectively. They correspond to gauge-invariant interaction terms, that involve all the fields surrounding the region and give rise to a trace-like `mesonic' operator. In our example, the five-brane diagram in Fig.~\ref{fig:five-braneC3Z2} gives two gauge factors, which have been labelled by 0 and 1. The bifundamental fields are
\begin{align}
\begin{array}{lcl}
    X_{01}^1 = \left( \tiny{\yng(1)}_0, \, \tiny{\overline{\yng(1)}}_1 \right)^1 & , &  X_{10}^1 = \left( \tiny{\yng(1)}_1, \, \tiny{\overline{\yng(1)}}_0 \right)^1 \;  \\[5pt]
    X_{01}^2 = \left( \tiny{\yng(1)}_0, \, \tiny{\overline{\yng(1)}}_1 \right)^2 & , &  X_{10}^2 = \left( \tiny{\yng(1)}_1, \, \tiny{\overline{\yng(1)}}_0 \right)^2 \;  \\[5pt]
    X_{00} = \phi_0 = \left( \tiny{\yng(1)}_0, \, \tiny{\overline{\yng(1)}}_0 \right) & , &  X_{11} = \phi_1 = \left( \tiny{\yng(1)}_1, \, \tiny{\overline{\yng(1)}}_1 \right) \;  
\end{array}
\end{align}
where the latter two fields are  adjoints and the upper index labels  different fields with the same transformation rules. The superpotential reads 
\begin{align}
    W_{_{\mathbb{C}^2/\mathbb{Z}_2\times \mathbb{C}}} = \epsilon_{ij} \left( \phi_1 X_{10}^i X_{01}^j + \phi_0 X_{01}^i X_{10}^j \right) \; .
\end{align}
As a final step one turns the five-brane diagram into the brane tiling or dimer by shrinking the gray and black regions into points, white points for $(+1)$ and black for $(-1)$, and connects white points to black ones by edges, so as to obtain a bipartite graph. This is in a sense dual to the five-brane diagram, as the edges represent the bifundamental fields. The brane tiling encodes the information about the toric CY geometry and completely defines the dual gauge theory in the sense of the AdS/CFT correspondence. The brane tiling of  $\mathbb{C}^2/\mathbb{Z}_2\times \mathbb{C}$ is drawn in Fig.~\ref{fig:BraneTilingC3Z2}.

\begin{figure}
\centering{
    \begin{tikzpicture}[auto, scale=0.7]
		\node (0) at (0,0) {}; 
		\node (1) at (6,0) {};
		\node (2) at (6,6) {}; 
		\node (3) at (0,6) {}; 
		\node (c) at (3,3) {};
		\node (c1) at (3,4.5) {};
		\node (c2) at (3,1.5) {};
		\node[cross out, minimum size=2mm, draw=red, inner sep=1mm, thick] (0) at (0,0) {}; 
		\node[cross out, minimum size=2mm, draw=red, inner sep=1mm, thick] (O2) at (0,3) {};
		\node[cross out, minimum size=2mm, draw=red, inner sep=1mm, thick] (O3) at (3,3) {};
		\node[cross out, minimum size=2mm, draw=red, inner sep=1mm, thick] (O4) at (3,0) {};
        \node [above left=7 pt of c1, circle, draw=black!50, fill=white, inner sep=0pt, minimum size=2mm] (I1) {};
        \node [below right=7 pt of c1, circle, draw=black, fill=black, inner sep=0pt, minimum size=2mm] (I2) {};
        \node [above left=7 pt of c2, circle, draw=black!50, fill=white, inner sep=0pt, minimum size=2mm] (I3) {};
        \node [below right=7 pt of c2, circle, draw=black, fill=black, inner sep=0pt, minimum size=2mm] (I4) {};
        \draw (0) to (1) [shorten >=-0.15cm, shorten <=-0.15cm, thick, black!80];
        \draw (1) to (2) [shorten >=-0.15cm, shorten <=-0.15cm, thick, black!80];
        \draw (2) to (3) [shorten >=-0.15cm, shorten <=-0.15cm, thick, black!80];
        \draw (3) to (0) [shorten >=-0.15cm, shorten <=-0.15cm, thick, black!80];
        \draw (3,6) to (3,0) [->, thick, blue!50, nearly transparent];
        \draw (6,1.5) to (0,1.5) [->, thick, red!70, nearly transparent];
        \draw (6,4.5) to (0,4.5) [->, thick, red!70, nearly transparent];
        \draw (3,0) to (6,1.5) [->, thick, green!70!black, nearly transparent];
        \draw (0,1.5) to (6,4.5) [->, thick, green!70!black, nearly transparent];
        \draw (0,4.5) to (3,6) [->, thick, green!70!black, nearly transparent];
        \node [above right=0.1 cm of 0] {1};
        \node [above left=0.05 pt of 1] {1};
        \node [below right=0.1 cm of c] {0};
        \node [above left=0.1cm of c] {0};
        \node [below left=0.1 cm of 2] {1};
        \node [below right=0.05 pt of 3] {1};
        \draw (I1) to (I2) [-, thick];
        \draw (I2) to (I3) [-, thick];
        \draw (I3) to (I4) [-, thick];
        \draw (I4) to (3,0) [-, thick];
        \draw (I1) to (3,6) [-, thick];
        \draw (I4) to (6,1.5) [-, thick];
        \draw (I3) to (0,1.5) [-, thick];
        \draw (I2) to (6,4.5) [-, thick];
        \draw (I1) to (0,4.5) [-, thick];
        \node [below left=0.05pt of O2, red] {$\tau_0$};
        \node [below left=0.05pt of O3, red] {$\tau_{00}$};
        \node [below left=0.05pt of 0, red] {$\tau_1$};
        \node [below left=0.05pt of O4, red] {$\tau_{11}$};
\end{tikzpicture}}
\subcap{The brane tiling of $\mathbb{C}^2/\mathbb{Z}_2 \times \mathbb{C}$ with the four fixed points of the orientifold projection.}\label{fig:BraneTilingC3Z2}
\end{figure}

\begin{figure}
\begin{center}
\begin{subfigure}{0.4\textwidth}
\centering{
\begin{tikzpicture}[auto, scale=0.5]
		\node [circle, draw=blue!50, fill=blue!20, inner sep=0pt, minimum size=5mm] (a) at (0,0) {$a$};
		\node [circle, draw=blue!50, fill=blue!20, inner sep=0pt, minimum size=5mm] (b) at (6,0) {$b$};
		\node (0) at (3,0) {};
        \draw (a) to node [swap] {$X_{ab}$} (b) [->, thick];
        \node [below=0.6cm of 0] {$\left( \tiny{\yng(1)}_a , \, \overline{\tiny{\yng(1)}}_b  \right)$};
\end{tikzpicture}}
\end{subfigure}\\[18pt]
\end{center}
\begin{subfigure}{0.4\textwidth}
\centering{
\begin{tikzpicture}[auto, scale=0.5]
		\node [circle, draw=blue!50, fill=blue!20, inner sep=0pt, minimum size=5mm] (a) at (0,0) {$a$};
		\node [circle, draw=blue!50, fill=blue!20, inner sep=0pt, minimum size=5mm] (b) at (6,0) {$b$};
		\node (0) at (3,0) {$|$};
		\draw (a) to node {} (0) [->, shorten >=-0.15cm, thick];
		\draw (b) to node {} (0) [->, shorten >=-0.15cm, thick];
        \node [below=0.02cm of 0] {$Y_{ab}$};
        \node [below=0.6cm of 0] {$\left( \tiny{\yng(1)}_a , \, \tiny{\yng(1)}_b  \right)$};
\end{tikzpicture}
}
\end{subfigure}
\hfill
\begin{subfigure}{0.4\textwidth}
\centering{
\begin{tikzpicture}[auto, scale=0.5]
		\node [circle, draw=blue!50, fill=blue!20, inner sep=0pt, minimum size=5mm] (a) at (0,0) {$a$};
		\node [circle, draw=blue!50, fill=blue!20, inner sep=0pt, minimum size=5mm] (b) at (6,0) {$b$};
		\node (0) at (3,0) {$|$};
		\draw (a) to node {} (0) [<-, shorten >=-0.15cm, thick];
		\draw (b) to node {} (0) [<-, shorten >=-0.15cm, thick];
        \node [below=0.02cm of 0] {$\widetilde{Y}_{ab}$};
        \node [below=0.6cm of 0] {$\left( \overline{\tiny{\yng(1)}}_a , \, \overline{\tiny{\yng(1)}}_b \right) $};
\end{tikzpicture}
}
\end{subfigure}\\[18pt]
\hfill
\begin{subfigure}{0.35\textwidth}
\centering{
\begin{tikzpicture}[auto, scale=0.5]
		\node [circle, draw=blue!50, fill=blue!20, inner sep=0pt, minimum size=5mm] (a) at (0,0) {$a$};
		\node (A) at (-3,0) {$\tiny{\yng(1,1)}$};
		\node (Ac) at (3,0) {$\overline{\tiny{\yng(1,1)}}$};
        \draw (a) to node {$A_{a}$} (A) [->, thick, shorten >=-3.5pt];
        \draw (a) to node [swap] {$\widetilde{A}_{a}$} (Ac) [<-, thick, shorten >=-3.5pt];
\end{tikzpicture}
}
\end{subfigure}
\begin{subfigure}{0.25\textwidth}
\centering{
\begin{tikzpicture}[auto, scale=0.5]
		\node [circle, draw=blue!50, fill=blue!20, inner sep=0pt, minimum size=5mm] (a) at (0,0) {$a$};
		\node (0) at (3,0) {};
		\draw (a) to [out=50, in=135, looseness=10] (a) [->, thick];
        \node [above=0.6cm of a] {$\; \phi_a =\mathrm{Adj}_a$};
\end{tikzpicture}
}
\end{subfigure}
\begin{subfigure}{0.35\textwidth}
\centering{
\begin{tikzpicture}[auto, scale=0.5]
		\node [circle, draw=blue!50, fill=blue!20, inner sep=0pt, minimum size=5mm] (a) at (0,0) {$a$};
		\node (S) at (-3,0) {$\tiny{\yng(2)}$};
		\node (Sc) at (3,0) {$\overline{\tiny{\yng(2)}}$};
        \draw (a) to node {$S_{a}$} (S) [->, thick, shorten >=-3.5pt];
        \draw (a) to node [swap] {$\widetilde{S}_{a}$} (Sc) [<-, thick, shorten >=-3.5pt];
\end{tikzpicture}
}
\end{subfigure}
\hspace{20pt}
\caption{The various matter fields and their representations that we will use in quiver diagrams. We draw multiple arrows for multiple fields connecting the same pair of nodes. When nodes are both $SO$ and/or $USp$ groups, we drop the arrow and connect them with an edge, signaling the fact that representations are real. For tensor representations, when not specified if they are symmetric or antisymmetric, we simply denote them by $T_a$.}\label{fig:Legenda}
\end{figure}
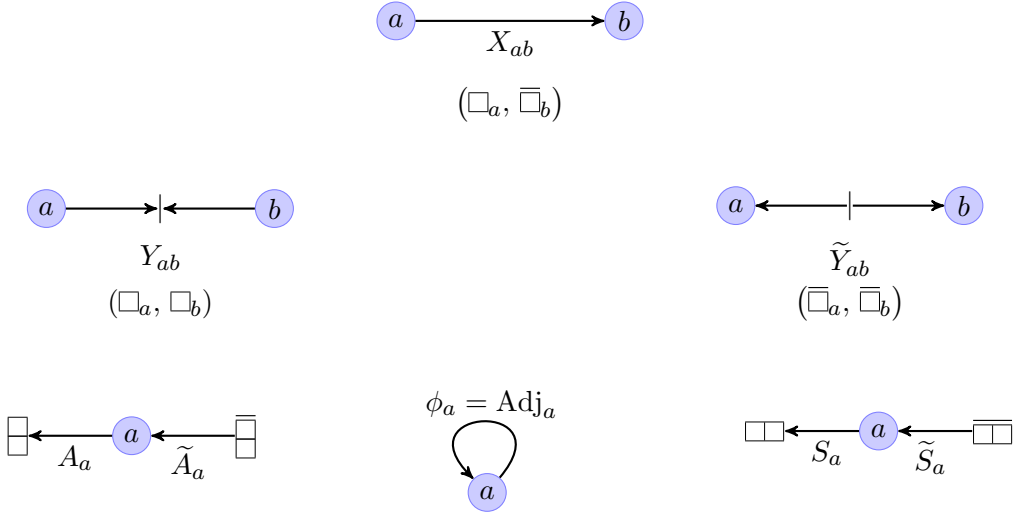

To sum up, in a toric variety the geometry is encoded in a toric diagram whose discrete data allows to construct the brane tiling, which translates geometric information into a 4d gauge theory. The dictionary of this bipartite graph is as follows. Each face represents a gauge group factor $SU(N)_a$, each edge represents a bifundamental field $X_{ab}$ transforming under the adjacent faces, with an orientation given by the direction black to white, each node represents a gauge-invariant term in the superpotential. The gauge factors and matter fields can be translated from the five-branes to a quiver representation in form of nodes and arrows, respectively. The legend in Fig.~\ref{fig:Legenda} shows the various matter fields that appear in subsequent sections.

\subsection{Inherited $S$-duality}

\begin{figure}
\begin{subfigure}{0.4\textwidth}
\centering{
\begin{tikzpicture}[auto, scale=0.5]
		\node [circle, draw=blue!50, fill=blue!20, inner sep=0pt, minimum size=5mm] (0) at (0,0) {$0$}; 
		\node [circle, draw=blue!50, fill=blue!20, inner sep=0pt, minimum size=5mm] (1) at (6,0) {$1$};
        \draw (0) to [out=40, in=140, looseness=1] (1) [->>, thick];
        \draw (1) to [out=220, in=320, looseness=1] (0) [->>, thick];
		\draw (0) to [out=140, in=220, looseness=10] (0) [->, thick];
		\draw (1) to [out=40, in=320, looseness=10] (1) [->, thick];
		\draw (-2.5,0) to node [pos=0.01, red, swap] {$\Omega$} (8.5,0) [-, thick, dashed, red];
\end{tikzpicture}}
\vspace{17pt}
\end{subfigure}
\hfill
\begin{subfigure}{0.4\textwidth}
\centering{
\begin{tikzpicture}[auto, scale=0.5]
		\node [circle, draw=blue!50, fill=blue!20, inner sep=0pt, minimum size=5mm] (0) at (0,0) {$0$}; 
		\node [circle, draw=blue!50, fill=blue!20, inner sep=0pt, minimum size=5mm] (1) at (6,0) {$1$};
		\node [left=10pt of 0] (A) {$\tiny{\yng(1,1)}$};
		\node [right=10pt of 1] (S) {$\tiny{\yng(2)}$};
		\node [below=0.2cm of 0] {\small{$SO(N)$}};
		\node [below=0.2cm of 1] {\small{$\; \; USp(N-2)$}};
        \draw (0) to [out=40, in=140, looseness=1] (1) [-, thick];
        \draw (1) to [out=220, in=320, looseness=1] (0) [-, thick];
        \draw (0) to (A) [-, thick, shorten >=-3.5pt];
        \draw (1) to (S) [-, thick, shorten >=-3.5pt];
\end{tikzpicture}}
\end{subfigure}\\[10pt]
\begin{center}
\begin{subfigure}{0.4\textwidth}
\centering{
\begin{tikzpicture}[auto, scale=0.5]
		\node [circle, draw=blue!50, fill=blue!20, inner sep=0pt, minimum size=5mm] (0) at (0,0) {$0$}; 
		\node [circle, draw=blue!50, fill=blue!20, inner sep=0pt, minimum size=5mm] (1) at (6,0) {$1$};
		\node [left=10pt of 0] (A) {$\tiny{\yng(2)}$};
		\node [right=10pt of 1] (S) {$\tiny{\yng(1,1)}$};
		\node [below=0.2cm of 0] {\small{$SO(N)$}};
		\node [below=0.2cm of 1] {\small{$\; \; USp(N-2)$}};
        \draw (0) to [out=40, in=140, looseness=1] (1) [-, thick];
        \draw (1) to [out=220, in=320, looseness=1] (0) [-, thick];
        \draw (0) to (A) [-, thick, shorten >=-3.5pt];
        \draw (1) to (S) [-, thick, shorten >=-3.5pt];
\end{tikzpicture}}
\end{subfigure}
\end{center}
\caption{The quiver diagram for $\mathbb{C}^2/\mathbb{Z}_2\times \mathbb{C}$ with the orientifold projection $\Omega$ on the top left, while on the top right the associated (linear) `unoriented' quiver after the $\mathcal{N}=2$ choice for the orientifold and the $\mathcal{N}=1$ choice at the bottom.}\label{fig:LinearQuiver}
\end{figure}
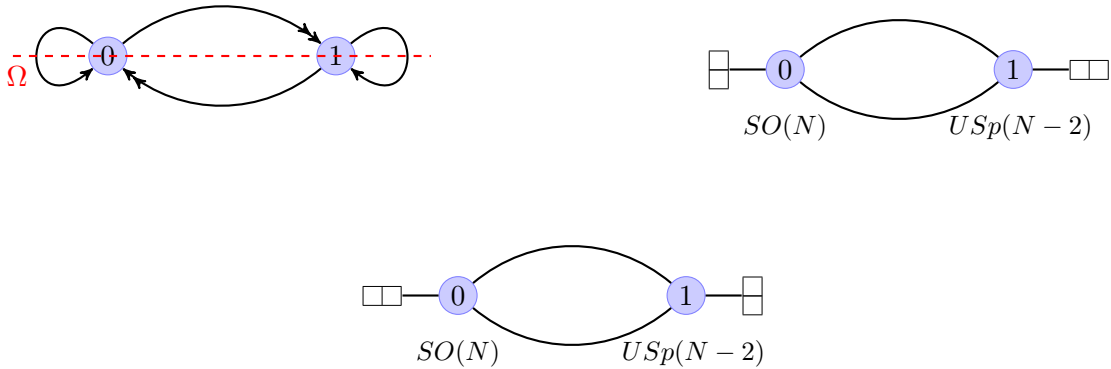

The quiver of $\mathbb{C}^2/\mathbb{Z}_2\times \mathbb{C}$ is drawn on the top left of  Fig.~\ref{fig:LinearQuiver}, where the red dashed line highlights the orientifold projection. (Let us call the projection $\Omega$.) On the brane tiling, the projection is realized by four fixed points, whose charges are encoded in a vector $\vec{\tau} = \left( \tau_{0} \, , \tau_{00} \, , \tau_{1} \, , \tau_{11} \right)$ and project in turn the gauge group $SU(N_0)$, the field $\phi_0$, the gauge group $SU(N_1)$ and the field $\phi_1$. Positive $\tau$'s give either orthogonal groups or symmetric representations, while negative $\tau$'s give either symplectic groups or antisymmetric representations. The product of the $\tau$'s is constrained by the condition \cite{Franco:2007ii}
\begin{align}
    \prod \tau = (-1)^{\frac{N_W}{2}} \; ,
\end{align}
where $N_W$ is the number of terms in the superpotential of the parent theory. In the present $\mathbb{C}^2/\mathbb{Z}_2\times \mathbb{C}$ case $N_W =4$, thus the product of the $\tau$'s is $+1$. The superpotential reads
\begin{align}
    W_{_{\mathbb{C}^2/\mathbb{Z}_2\times \mathbb{C}}}^{\Omega} = - T_0 X_{01} X_{10} + T_1 X_{10} X_{01} \; .
\end{align}
We observe that in the five-brane diagram in Fig.~\ref{fig:five-braneC3Z2}, the fixed points that project the adjoint fields are located at the intersections between the `vertical' (blue) brane and the `oblique' (green) brane, and the two `horizontal' (red) branes are identified by the orientifold.

Given the above orientifold, the field content of the model can easily be read from the linear quiver either on the top right or bottom of Fig. \ref{fig:LinearQuiver}, the two representing two different choices for $\vec{\tau}$. We can find out whether the model has a conformal fixed point by analyzing the beta functions. Let us identify the $R$-charges $R_{01}$ and $R_{10}$ of the two bifundamental fields, while imposing $R(W)=2$ that identifies the $R$-charges $R_0$ and $R_1$ of the two projected fields. This gives the condition $r_{0} + 2 r_{01} = -1$, where 
\begin{equation}
r=R-1
\end{equation}
is the $R$-charge of the fermion in the chiral multiplet.\footnote{$R=2/3$ is the `canonical' $R$-charge of a free field.} Together with the condition that the $\beta$-functions of the two gauge group vanish, we have
\begin{align}
    & r_{0} (m + 2 \tau_{00}) = - (m - 2 \tau_0) \; , \nonumber \\[5pt]
    & r_{0} (m - 2 \tau_{11}) = - (m + 2 \tau_1) \; ,\label{eq:r0c3z2}
\end{align}
where $m= N_0 - N_1$. We now study the solutions to these equations for the different possible choices of the $\tau$'s.

It is straightforward to see that there is a choice that preserves $\mathcal{N}=2$ supersymmetry, choosing $\tau_{00}= -\tau_0$ and $\tau_{11} = -\tau_1$, and we denote it as $\vec{\tau}_{A}$. (Let us call $\Omega_A$ the associated orientifold projection.) In this case the solution demands $m=2\tau_0$ and $\tau_1= -\tau_0$, so that $\vec{\tau}_A = (\pm,\, \mp,\, \mp,\, \pm)$. 
These conditions leave $r_0$ undetermined, and we find that at large $N$ the value of $r_0$ that maximizes the central charge $a$ is $r_0 =-\frac{1}{3}$. Hence, at large $N$ the $R$-charges of all the fields are $\frac{2}{3}$, and the central charge 
\begin{align}
    a^{\Omega_A}_{_{\mathbb{C}^2/\mathbb{Z}_2\times \mathbb{C}}} = \frac{1}{4 } N^2 \; 
\end{align}
is half the value of the central charge of the  parent theory. To summarize, and choosing   without loss of generality $\tau_0=1$, one gets the gauge groups $SO(N)$ and $USp(N -2)$,\footnote{Here $N_0=N$, which is assumed to be even. The ranks of the two groups are $N/2$ and $N/2 -1$.}  with projected fields $A_0$ and $S_1$ in the adjoint of each group.

There are however other possible solutions to Eq.~\eqref{eq:r0c3z2}. These are $\mathcal{N}=1$ solutions with $\tau_{00}=\tau_0$, $\tau_{11}= \tau_1$ and $\tau_1 =-\tau_0$, which we denote as $\vec{\tau}_B = (\pm,\, \pm,\, \mp,\, \mp)$ giving  
\begin{align}
    & r_{00} = - \frac{m - 2\tau_0}{m + 2\tau_0} \; , \nonumber \\[5pt]
    & r_{01} = - \frac{2 \tau_0}{m + 2\tau_0} \; . 
\end{align}
In general for these solutions one should worry about the presence of gauge-invariant composite operators that hit the unitarity bound $R=2/3$ and decouple. This analysis was performed in \cite{Antinucci:2021edv} and it was shown that in the range  $1<m<10$ this cannot occur. There is a particular value of $m$ in this range, namely $m =2 \tau_0$, that gives $R_{0}=1$ and $R_{01}=1/2$. We observe that the value of $m$ is the same as the one of the $\mathcal{N}=2$ preserving orientifold, i.e. $\tau_{0}=-\tau_{00}=-\tau_{1}=\tau_{11}$, and the value of the central charge $a= {3\over 32}(3\text{Tr}R^3-\text{Tr} R)\sim c$ at large $N$ is  
\begin{align}\label{eq:C3Z2OmegaB}
    a^{\Omega_B}_{_{\mathbb{C}^2/\mathbb{Z}_2\times \mathbb{C}}} = \frac{27}{128} N^2 \ ,
\end{align}
which is $\frac{27}{32}$ times the central charge of the $\mathcal{N}=2 $ orientifold. Again, choosing $\tau_0 =1$ gives the gauge groups  $SO(N)$ and $USp(N -2)$, but now the  projected fields are $S_0$ and $A_1$ and both have $R=1$, while the bifundamental fields have $R=1/2$. We observe that the fields with $R=1$ do not contribute to the central charge.

The other value for $m$ in Eq. \eqref{eq:r0c3z2} we are interested in is $m=4\tau_0$, which gives $R_{00}=R_{01}=2/3$ as in the $\mathcal{N}=2$ model, and as a consequence gives also the same central charge at large $N$, although the difference between the ranks at finite $N$ is not the same, and the projected fields are not in the adjoint. As we will see, the presence of this kind of solution will be systematic in the models discussed in this paper. 

\begin{figure}
\begin{subfigure}{0.4\textwidth}
\centering{
\begin{tikzpicture}[auto, scale=0.5]
		\node [circle, draw=blue!50, fill=blue!20, inner sep=0pt, minimum size=5mm] (0) at (0,0) {$0$}; 
		\node [circle, draw=blue!50, fill=blue!20, inner sep=0pt, minimum size=5mm] (1) at (6,0) {$1$};
        \draw (0) to [out=40, in=140, looseness=1] (1) [->>, thick];
        \draw (1) to [out=220, in=320, looseness=1] (0) [->>, thick];
		\draw (-1.5,0) to node [pos=0.01, red, swap] {$\Omega$} (7.5,0) [-, thick, dashed, red];
\end{tikzpicture}}
\end{subfigure}
\hfill
\begin{subfigure}{0.4\textwidth}
\vspace{15pt}
    \centering{
\begin{tikzpicture}[auto, scale=0.5]
		\node [circle, draw=blue!50, fill=blue!20, inner sep=0pt, minimum size=5mm] (0) at (0,0) {$0$}; 
		\node [circle, draw=blue!50, fill=blue!20, inner sep=0pt, minimum size=5mm] (1) at (6,0) {$1$};
		\node [below=0.2cm of 0] {\small{$SO(N)$}};
		\node [below=0.2cm of 1] {\small{$\; \; USp(N-2)$}};
        \draw (0) to [out=40, in=140, looseness=1] (1) [-, thick];
        \draw (1) to [out=220, in=320, looseness=1] (0) [-, thick];
\end{tikzpicture}}
\end{subfigure}
\caption{On the left, the quiver diagram for $\mathcal{C}$, where the red line represents the orientifold projection. On the right, the linear `unoriented' quiver for the theory after the orientifold \cite{Bianchi:2013gka}.}\label{fig:QuiverC}
\end{figure}
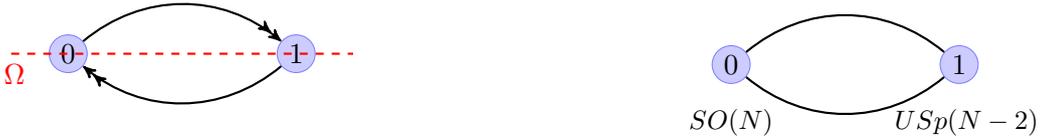

The central charge of the $\mathcal{N}=1 $ model with $m =2\tau_0$ coincides (exactly, that is even at finite $N$) with the central charge of the orientifold of the conifold $\mathcal{C}$. This can be quickly shown by determining the central charges and the difference of the ranks of the two gauge groups at the conformal fixed point.
For completeness, we draw in Figs. \ref{fig:ToricDiagramConifold} and \ref{fig:five-braneConifold} the toric diagram and the five-brane diagram of the conifold. We also draw in Fig. \ref{fig:QuiverC} the corresponding quiver. In this case the orientifold projection cannot be realized on the five-brane diagram by fixed points, but instead by fixed lines. Identifying the $R$-charges of all the fields, the superpotential of the parent theory 
\begin{align}
    W_{_{\mathcal{C}}} = \epsilon_{ab}\epsilon_{cd} X_{01}^{a} X_{10}^c X_{01}^b X_{10}^d \; 
\end{align}
implies that they are all equal to $1/2$, and they do not change after the orientifold projection. Denoting as before $\tau_0$ and $\tau_1$ the orientifold charges that project the two gauge groups, the condition that the $\beta$-functions vanish gives $m = 2\tau_0$ and $\tau_1 =-\tau_0$, where again we denote with $m$ the difference $N_0-N_1$. Choosing $\tau_0=+1$, we end up again with the gauge groups $SO(N)$ and $USp(N -2)$, while the bifundamental fields have $R=1/2$.\footnote{The same solution for the orientifold of the conifold has been constructed in \cite{Naculich:2001xu, Ahn:2001hy}, where the former reference observes that seven-branes should not be present in the type IIB configuration due to the lack of $1/N$ corrections to the $a$-anomaly, whereas the latter constructs a configuration with one O3$^+$ and one O3$^-$ that wrap two homologically different two-cycles, both coming from a fractional O5.} Hence, apart from the absence of the fields  $S_0$ and $A_1$, we get exactly the same ranks and the same $R$-charges as the $\mathcal{N}=1$ $\mathbb{C}^2/\mathbb{Z}_2\times \mathbb{C}$ orientifold with $m=2\tau_0$, implying that we get exactly the same central charge in Eq. \eqref{eq:C3Z2OmegaB}.

\begin{figure}
       \begin{subfigure}{0.4\textwidth}
    \centering{
    \begin{tikzpicture}[auto, scale=0.7]
		\node [circle, fill=black, inner sep=0pt, minimum size=1.5mm] (0) at (0,0) {}; 
		\node [circle, fill=black, inner sep=0pt, minimum size=1.5mm] (1) at (2,0) {};
		\node [circle, fill=black, inner sep=0pt, minimum size=1.5mm] (2) at (2,2) {}; 
		\node [circle, fill=black, inner sep=0pt, minimum size=1.5mm] (3) at (0,2) {}; 
        \draw (0) to (1) [thick];
        \draw (1) to (2) [thick];
        \draw (2) to (3) [thick];
        \draw (3) to (0) [thick];
        \draw (1,0) to (1,-2) [->, thick, blue!50];
        \draw (0,1) to (-2,1) [->, thick, red!50];
        \draw (1,2) to (1,4) [->, thick, brown!70!black];
        \draw (2,1) to (4,1) [->, thick, green!70!black];
\end{tikzpicture}}
\vspace{16pt}
\subcap{The toric diagram of the conifold $\mathcal{C}$ and the vectors orthogonal to the edges.}\label{fig:ToricDiagramConifold}
    \end{subfigure}
    \hfill
       \begin{subfigure}{0.4\textwidth}
    \centering{
    \begin{tikzpicture}[auto, scale=0.7]
		\node (0) at (0,0) {}; 
		\node (1) at (6,0) {};
		\node (2) at (6,6) {}; 
		\node (3) at (0,6) {}; 
		\node (c) at (3,3) {};
		\node (c1) at (2,2) {};
		\node (c2) at (4,2) {};
		\node (c3) at (4,4) {};
		\node (c4) at (2,4) {};
        \draw (0) to (1) [shorten >=-0.15cm, shorten <=-0.15cm, thick];
        \draw (1) to (2) [shorten >=-0.15cm, shorten <=-0.15cm, thick];
        \draw (2) to (3) [shorten >=-0.15cm, shorten <=-0.15cm, thick];
        \draw (3) to (0) [shorten >=-0.15cm, shorten <=-0.15cm, thick];
        \draw (0,3) to (6,3) [thick, red!90!black, dashed];
        \draw (0,0.01) to (6,0.01) [thick, red, dashed];
        \draw (2,6) to (2,0) [->, thick, blue!50];
        \draw (6,4) to (0,4) [->, thick, red!50];
        \draw (0,2) to (6,2) [->, thick, green!70!black];
        \draw (4,0) to (4,6) [->, thick, brown!70!black];
        \draw[fill=gray!80!white, nearly transparent]  (2,2) -- (4,2) -- (4,4) -- (2,4) -- cycle;
        \draw[fill=gray!30!white, nearly transparent]  (2,2) -- (0,2) -- (0,0) -- (2,0) -- cycle;
        \draw[fill=gray!30!white, nearly transparent]  (4,2) -- (6,2) -- (6,0) -- (4,0) -- cycle;
        \draw[fill=gray!30!white, nearly transparent]  (4,4) -- (6,4) -- (6,6) -- (4,6) -- cycle;
        \draw[fill=gray!30!white, nearly transparent]  (2,6) -- (0,6) -- (0,4) -- (2,4) -- cycle;
        \node () at (1,3) {0};
        \node () at (5,3) {0};
        \node () at (3,1) {1};
        \node () at (3,5) {1};
        \node [below=0.05 pt of c] {\tiny{$(-1)$}};
        \node () at (5,1) {\tiny{$(+1)$}};
        \node () at (5,5) {\tiny{$(+1)$}};
        \node () at (1,1) {\tiny{$(+1)$}};
        \node () at (1,5) {\tiny{$(+1)$}};
        \node[red] () at (-0.5,3) {$\tau_0$};
        \node[red] () at (-0.5,0) {$\tau_1$};
\end{tikzpicture}}
\subcap{The five-brane diagram of the conifold $\mathcal{C}$ with the fixed lines of the $\Omega$ projection.}\label{fig:five-braneConifold}
    \end{subfigure}
\end{figure}

As is well-known \cite{Klebanov:1998hh}, if one deforms the  $\mathbb{C}^2/\mathbb{Z}_2\times \mathbb{C}$ parent theory by adding a mass term for the two adjoint fields, this generates a flow that in the IR reaches the conifold model. This explains why the ratio of the central charges of the two parent theories is 27/32 \cite{Tachikawa:2009tt}. This picture is preserved by the orientifold: starting with the $\mathcal{N}=2$  $\mathbb{C}^2/\mathbb{Z}_2\times \mathbb{C}$ orientifold and mass deforming, one ends up with the orientifold of the conifold, and again this explains the value of the ratio of the central charges. The fact that the orientifold of the conifold and the $\mathcal{N}=1$ orientifold $\vec{\tau}_B$ of $\mathbb{C}^2/\mathbb{Z}_2\times \mathbb{C}$ with $m= 2\tau_0$ have the same central charge suggests that the two theories are conformally dual, meaning that they flow to the same conformal manifold, and indeed they are connected by turning on a mass term for the projected fields, which is an exactly marginal deformation because these fields have $R=1$ and the mass operator is not charged under any other global factors. 

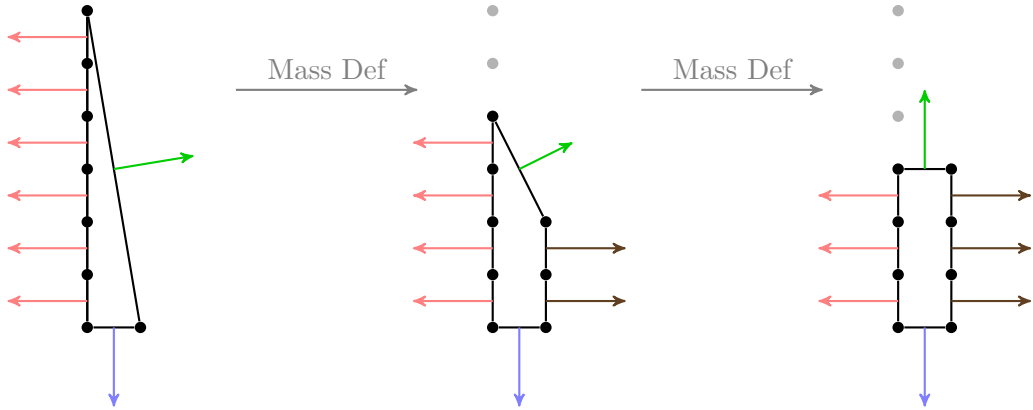
\begin{figure}
\hfill
    \centering{
       \begin{subfigure}{0.3\textwidth}
    \begin{tikzpicture}[auto, scale=0.7]
		\node [circle, fill=black, inner sep=0pt, minimum size=1.5mm] (0) at (0,0) {}; 
		\node [circle, fill=black, inner sep=0pt, minimum size=1.5mm] (1) at (0,1) {};
		\node [circle, fill=black, inner sep=0pt, minimum size=1.5mm] (2) at (0,2) {}; 
		\node [circle, fill=black, inner sep=0pt, minimum size=1.5mm] (3) at (0,3) {}; 
		\node [circle, fill=black, inner sep=0pt, minimum size=1.5mm] (4) at (0,4) {}; 
		\node [circle, fill=black, inner sep=0pt, minimum size=1.5mm] (5) at (0,5) {};
		\node [circle, fill=black, inner sep=0pt, minimum size=1.5mm] (6) at (0,6) {}; 
		\node [circle, fill=black, inner sep=0pt, minimum size=1.5mm] (h1) at (1,0) {}; 
        \draw (0) to (1) [thick];
        \draw (1) to (2) [thick];
        \draw (2) to (3) [thick];
        \draw (3) to (4) [thick];
        \draw (4) to (5) [thick];
        \draw (5) to (6) [thick];
        \draw (6) to (0) [thick];
        \draw (0) to (h1) [thick];
        \draw (h1) to (6) [thick];
        \draw (0.5,0) to (0.5,-1.5) [->, thick, blue!50];
        \draw (0,0.5) to (-1.5,0.5) [->, thick, red!50];
        \draw (0,1.5) to (-1.5,1.5) [->, thick, red!50];
        \draw (0,2.5) to (-1.5,2.5) [->, thick, red!50];
        \draw (0,3.5) to (-1.5,3.5) [->, thick, red!50];
        \draw (0,4.5) to (-1.5,4.5) [->, thick, red!50];
        \draw (0,5.5) to (-1.5,5.5) [->, thick, red!50];
        \draw (0.5,3) to (2,3.25) [->, thick, green!80!black];
        \node (a) at (2.5,4.5) {};
        \node (b) at (6.5,4.5) {};
        \draw (a) [pil, gray] to node {Mass Def} (b); 
\end{tikzpicture}
    \end{subfigure}}
    \hfill
        \centering{
       \begin{subfigure}{0.3\textwidth}
    \begin{tikzpicture}[auto, scale=0.7]
	\node [circle, fill=black, inner sep=0pt, minimum size=1.5mm] (0) at (0,0) {}; 
		\node [circle, fill=black, inner sep=0pt, minimum size=1.5mm] (1) at (0,1) {};
		\node [circle, fill=black, inner sep=0pt, minimum size=1.5mm] (2) at (0,2) {}; 
		\node [circle, fill=black, inner sep=0pt, minimum size=1.5mm] (3) at (0,3) {}; 
		\node [circle, fill=black, inner sep=0pt, minimum size=1.5mm] (4) at (0,4) {}; 
		\node [circle, fill=gray!60, inner sep=0pt, minimum size=1.5mm] (5) at (0,5) {};
		\node [circle, fill=gray!60, inner sep=0pt, minimum size=1.5mm] (6) at (0,6) {}; 
		\node [circle, fill=black, inner sep=0pt, minimum size=1.5mm] (h1) at (1,0) {}; 
		\node [circle, fill=black, inner sep=0pt, minimum size=1.5mm] (h2) at (1,1) {};
		\node [circle, fill=black, inner sep=0pt, minimum size=1.5mm] (h3) at (1,2) {}; 
        \draw (0) to (1) [thick];
        \draw (1) to (2) [thick];
        \draw (2) to (3) [thick];
        \draw (3) to (4) [thick];
        \draw (4) to (h3) [thick];
        \draw (h3) to (h2) [thick];
        \draw (h2) to (h1) [thick];
        \draw (h1) to (0) [thick];
        \draw (0.5,0) to (0.5,-1.5) [->, thick, blue!50];
        \draw (0,0.5) to (-1.5,0.5) [->, thick, red!50];
        \draw (0,1.5) to (-1.5,1.5) [->, thick, red!50];
        \draw (0,2.5) to (-1.5,2.5) [->, thick, red!50];
        \draw (0,3.5) to (-1.5,3.5) [->, thick, red!50];
        \draw (1,0.5) to (2.5,0.5) [->, thick, brown!50!black];
        \draw (1,1.5) to (2.5,1.5) [->, thick, brown!50!black];
        \draw (0.5,3) to (1.5,3.5) [->, thick, green!80!black];
        \node (a) at (2.5,4.5) {};
        \node (b) at (6.5,4.5) {};
        \draw (a) [pil, gray] to node {Mass Def} (b); 
\end{tikzpicture}
    \end{subfigure}}
    \hfill
        \centering{
       \begin{subfigure}{0.25\textwidth}
    \begin{tikzpicture}[auto, scale=0.7]
	\node [circle, fill=black, inner sep=0pt, minimum size=1.5mm] (0) at (0,0) {}; 
		\node [circle, fill=black, inner sep=0pt, minimum size=1.5mm] (1) at (0,1) {};
		\node [circle, fill=black, inner sep=0pt, minimum size=1.5mm] (2) at (0,2) {}; 
		\node [circle, fill=black, inner sep=0pt, minimum size=1.5mm] (3) at (0,3) {}; 
		\node [circle, fill=gray!60, inner sep=0pt, minimum size=1.5mm] (4) at (0,4) {}; 
		\node [circle, fill=gray!60, inner sep=0pt, minimum size=1.5mm] (5) at (0,5) {};
		\node [circle, fill=gray!60, inner sep=0pt, minimum size=1.5mm] (6) at (0,6) {}; 
		\node [circle, fill=black, inner sep=0pt, minimum size=1.5mm] (h1) at (1,0) {}; 
		\node [circle, fill=black, inner sep=0pt, minimum size=1.5mm] (h2) at (1,1) {};
		\node [circle, fill=black, inner sep=0pt, minimum size=1.5mm] (h3) at (1,2) {}; 
		\node [circle, fill=black, inner sep=0pt, minimum size=1.5mm] (h4) at (1,3) {}; 
        \draw (0) to (1) [thick];
        \draw (1) to (2) [thick];
        \draw (2) to (3) [thick];
        \draw (3) to (h4) [thick];
        \draw (h4) to (h3) [thick];
        \draw (h3) to (h2) [thick];
        \draw (h2) to (h1) [thick];
        \draw (h1) to (0) [thick];
        \draw (0.5,0) to (0.5,-1.5) [->, thick, blue!50];
        \draw (0,0.5) to (-1.5,0.5) [->, thick, red!50];
        \draw (0,1.5) to (-1.5,1.5) [->, thick, red!50];
        \draw (0,2.5) to (-1.5,2.5) [->, thick, red!50];
        \draw (1,0.5) to (2.5,0.5) [->, thick, brown!50!black];
        \draw (1,1.5) to (2.5,1.5) [->, thick, brown!50!black];
        \draw (1,2.5) to (2.5,2.5) [->, thick, brown!50!black];
        \draw (0.5,3) to (0.5,4.5) [->, thick, green!80!black];
\end{tikzpicture}
    \end{subfigure}}
    \caption{The chain of toric diagrams connected by mass deformation for $n=6$: $\mathbb{C}^2/\mathbb{Z}_6 \times \mathbb{C} \to L^{2,4,2} \to L^{3,3,3}$, the number of points remains $n+2=8$ all along. It holds also with the orientifold projection, mutatis mutandis.}\label{fig:ToricChain}
\end{figure}

\begin{figure}
    \centering
    \includegraphics[width=1\textwidth]{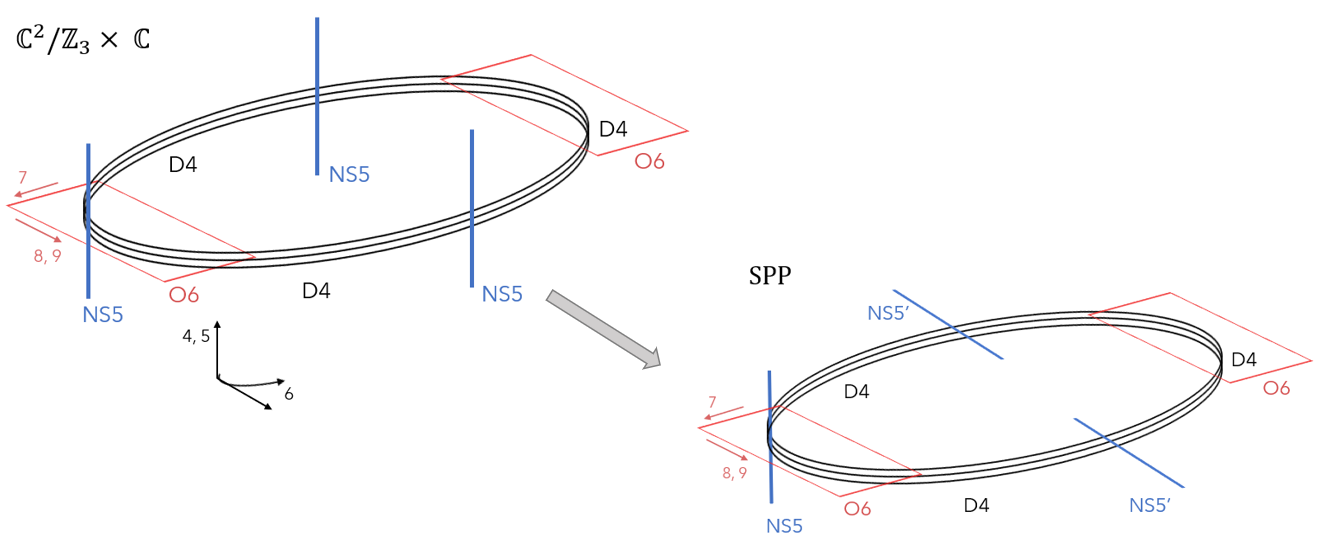}
    \caption{The theories $\mathbb{C}^2/\mathbb{Z}_3 \times \mathbb{C}$ and suspended pinch point (SPP) represented via IIA elliptic models. Rotating NS5 branes results in mass deforming the theory.}\label{fig:Elliptics}
\end{figure}

The results of \cite{Antinucci:2021edv, Amariti:2021lhk} are a natural generalization of the mechanism described above to the infinite class of $L^{a,b,a}$ theories $(a\leq b)$. The parent $L^{a,n-a,a}$ toric models can all be obtained by mass deformations of $L^{0,n,0}$, which is the $\mathcal{N}=2$ orbifold  $\mathbb{C}^2/\mathbb{Z}_{n}\times \mathbb{C}$. Specifically, starting from $L^{0,n,0}$, one can add a mass to a pair of adjoints\footnote{Giving mass to pairs is necessary but not sufficient to preserve toricity \cite{Bianchi:2014qma}.} of gauge groups that are connected in the quiver, and integrating out this mass term gives the $L^{1,n-1,1}$ theory \cite{Bianchi:2014qma}. This can be iterated to produce a chain that ends with $L^{\frac{n}{2},\frac{n}{2},\frac{n}{2}}$ for $n$ even and  $L^{\frac{n-1}{2},\frac{n+1}{2},\frac{n-1}{2}}$ for $n$ odd. We represent some steps of this chain for $n=6$ in Fig. \ref{fig:ToricChain}. In fact, these theories can be embedded in IIA elliptic models, where stacks of D4-branes are wrapped around a circle and their worldvolume is divided by orthogonal NS5's. The rotation of the five-branes describes the mass deformation, see for example Fig.~\ref{fig:Elliptics}. This construction holds also in presence of orientifold planes. 

For the orientifold of $L^{0,n,0}$ theory one can choose the ranks in such a way that the choice of $\tau$'s preserving $\mathcal{N}=2$ supersymmetry has a fixed point with `canonical' $R$-charges 2/3 and half the central charge of the parent theory at large $N$. On the other hand, for all such theories it is also possible to perform an $\mathcal{N}=1$ orientifold projection, with the same ranks of the gauge groups, and such that all the adjoints and projected fields have $R=1$ and all the bifundamental fields have $R=1/2$.

\begin{figure}
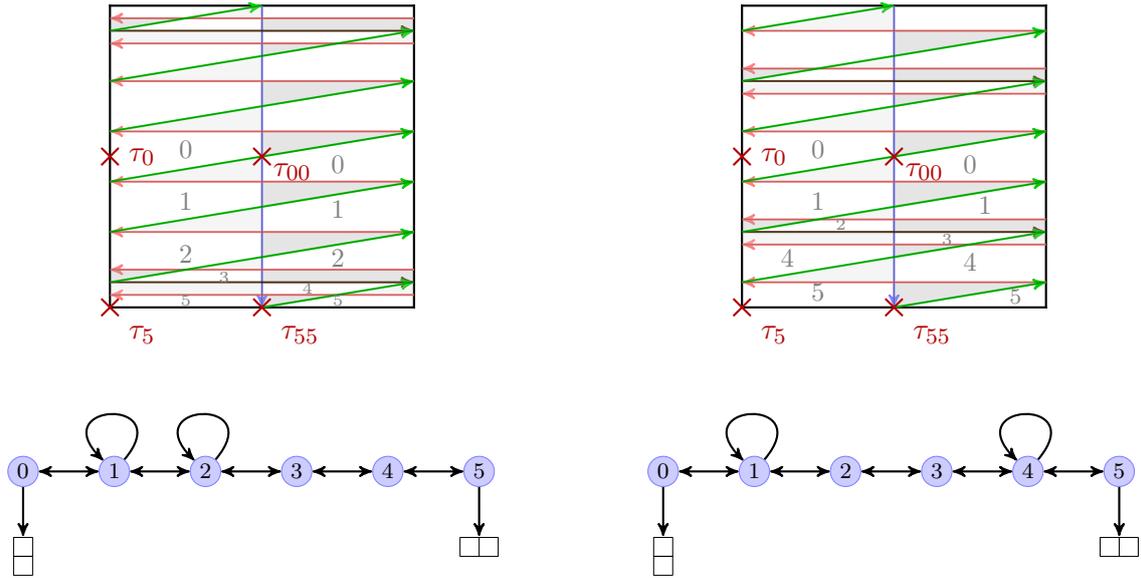

\begin{subfigure}{0.45\textwidth}
\centering{
    %
}
\end{subfigure}\hspace{40pt}
\caption{The two inequivalent ways of accommodating the horizontal branes for $L^{2,8,2}$ and the associated linear quivers.}\label{fig:five-braneChoices}
\end{figure}

Note also that, for higher values of $n$,  the $L^{a,n-a,a}$ theory admits different toric phases, which means different ways of integrating out pairs of adjoints, compatible with the orientifold. These correspond to different ways of accommodating the horizontal branes in the five-brane diagram, see Fig.~\ref{fig:five-braneChoices} for an example. 

\begin{figure}
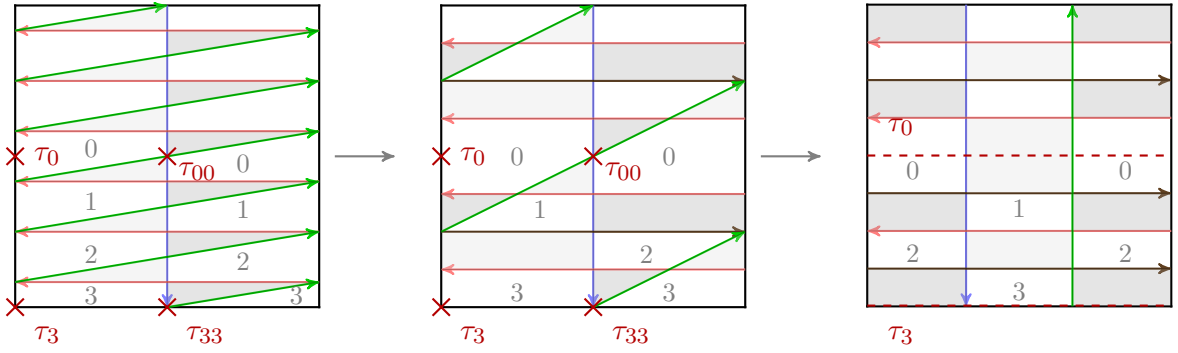

\begin{subfigure}{0.25\textwidth}
\centering{
}
\end{subfigure}
\hfill
\caption{The chain of five-brane diagrams connected by mass deformation for $n=6$: $\mathbb{C}^2/\mathbb{Z}_6 \times \mathbb{C} \to L^{2,4,2} \to L^{3,3,3}$, the number of vectors remains $n+2=8$ all along. It holds also with the orientifold projection, provided a $\mathbb{Z}_2$ symmetry is preserved.}\label{fig:five-braneChain}
\end{figure}

For a family with $n$ even, there are  orientifold projections that generalize the ones of $\mathbb{C}^2/\mathbb{Z}_2 \times \mathbb{C}$ and the conifold, corresponding to the fixed points in Fig. \ref{fig:five-braneC3Z2} and the fixed lines in Fig. \ref{fig:five-braneConifold}. As an example, we represent in Fig. \ref{fig:five-braneChain} the $n=6$ case. The five-brane diagrams are constructed from the toric diagrams in Fig. \ref{fig:ToricChain}. The parent theories describe six gauge groups, with vector-like bifundamentals connecting group $i$ and $i+1$. 
In the first and second diagram, the fixed points with charge $\tau_0$ and $\tau_3$ project the groups $0$ and $3$, while $\tau_{00}$ and $\tau_{33}$ project the adjoints of these groups. The $L^{1,5,1}$ theory is missing in the chain because it does not allow this orientifold. Indeed, in $L^{1,5,1}$ we would have five horizontal red branes (pointing towards the left) and one horizontal brown brane (pointing towards the right), and there is no way to accommodate them compatibly with the desired orientifold. Finally, the last diagram represents the orientifold of the non-chiral $\mathbb{Z}_3$ orbifold of the conifold, which is realized by fixed lines that project groups 0 and 3. In general, given a generic even $n$, all of the $\mathcal{N}=1$ orientifold theories in the family have the same central charge $a$, 't Hooft anomalies and superconformal index of the orientifold of the orbifold of the conifold, which is the theory at the end of the chain. The value of the central charge is always 27/32 the value of the $\mathcal{N}=2$ theory. This was shown originally in \cite{Antinucci:2021edv} for $n=3p$ studying the orientifold of the non-chiral orbifold SPP/$\mathbb{Z}_p$, i.e. the $L^{p,2p,p}$ theory and, then generalized in \cite{Amariti:2021lhk} to any even $n$.

\begin{figure}
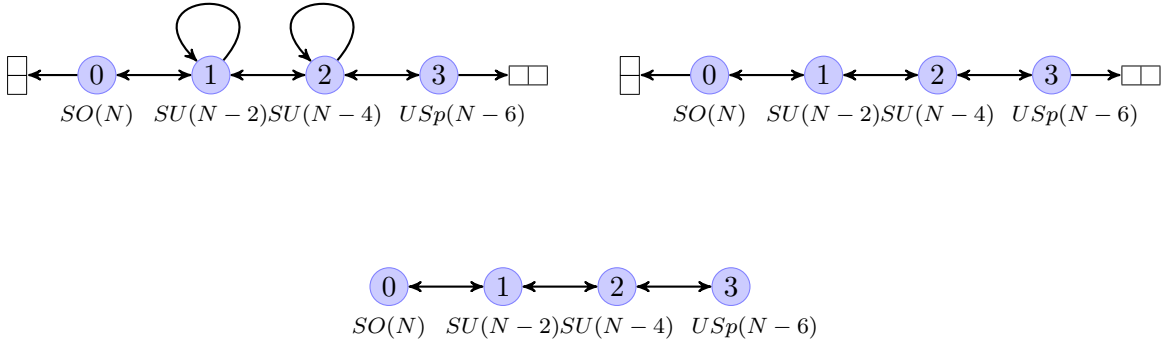

\begin{subfigure}{0.4\textwidth}
\centering{
%
}
\end{subfigure} 
\end{center}
\caption{The chain of linear quiver diagrams after the orientifold connected by mass deformation for $n=6$: $\mathbb{C}^2/\mathbb{Z}_6 \times \mathbb{C} \to L^{2,4,2} \to L^{3,3,3}$.}\label{fig:QuiverC3Z6}
\end{figure}

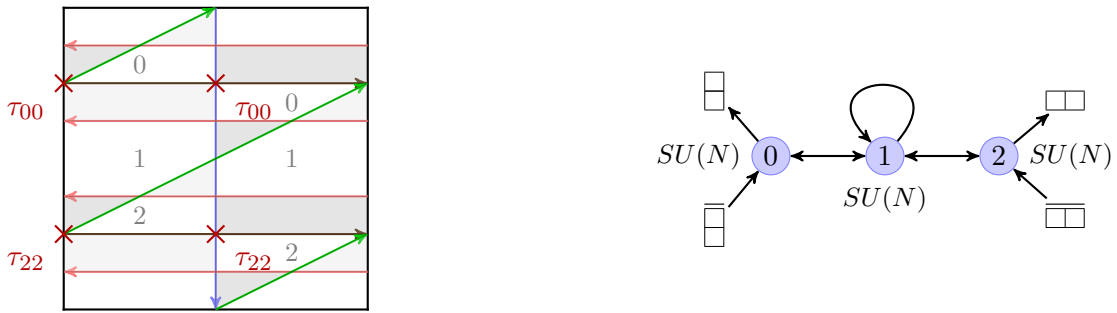
\begin{figure}
\begin{subfigure}{0.4\textwidth}
\centering{
    \begin{tikzpicture}
		\node (0) at (0,0) {}; 
		\node (1) at (4,0) {};
		\node (2) at (4,4) {}; 
		\node (3) at (0,4) {}; 
		\node (c) at (2,2) {};
        \draw (0) to (1) [shorten >=-0.15cm, shorten <=-0.15cm, thick];
        \draw (1) to (2) [shorten >=-0.15cm, shorten <=-0.15cm, thick];
        \draw (2) to (3) [shorten >=-0.15cm, shorten <=-0.15cm, thick];
        \draw (3) to (0) [shorten >=-0.15cm, shorten <=-0.15cm, thick];
        \draw (2,4) to (2,0) [->, thick, blue!50];
        \draw (4,0.5) to (0,0.5) [->, thick, red!50];
        \draw (4,1) to (0,1) [<-, thick, brown!50!black];
        \draw (4,1.5) to (0,1.5) [->, thick, red!50];
        \draw (4,2.5) to (0,2.5) [->, thick, red!50];
        \draw (4,3) to (0,3) [<-, thick, brown!50!black];
        \draw (4,3.5) to (0,3.5) [->, thick, red!50];
        \draw (2,0) to (4,1) [->, thick, green!80!black];
        \draw (0,1) to (4,3) [->, thick, green!80!black];
        \draw (0,3) to (2,4) [->, thick, green!80!black];
        \draw[fill=gray!80!white, nearly transparent]  (2,0) -- (3,0.5) -- (2,0.5) -- cycle;
        \draw[fill=gray!80!white, nearly transparent]  (2,1) -- (2,1.5) -- (4,1.5) -- (4,1) -- cycle;
        \draw[fill=gray!80!white, nearly transparent]  (0,1.5) -- (1,1.5) -- (0,1) -- cycle;
        \draw[fill=gray!80!white, nearly transparent]  (2,2) -- (2,2.5) -- (3,2.5) -- cycle;
        \draw[fill=gray!80!white, nearly transparent]  (2,3) -- (2,3.5) -- (4,3.5) -- (4,3) -- cycle;
        \draw[fill=gray!80!white, nearly transparent]  (0,3.5) -- (1,3.5) -- (0,3) -- cycle;
        \draw[fill=gray!30!white, nearly transparent]  (0,0.5) -- (0,1) -- (2,1) -- (2,0.5) -- cycle;
        \draw[fill=gray!30!white, nearly transparent]  (3,0.5) -- (4,0.5) -- (4,1) -- cycle;
        \draw[fill=gray!30!white, nearly transparent]  (1,1.5) -- (2,1.5) -- (2,2) -- cycle;
        \draw[fill=gray!30!white, nearly transparent]  (0,2.5) -- (0,3) -- (2,3) -- (2,2.5) -- cycle;
        \draw[fill=gray!30!white, nearly transparent]  (3,2.5) -- (4,2.5) -- (4,3) -- cycle;
        \draw[fill=gray!30!white, nearly transparent]  (1,3.5) -- (2,3.5) -- (2,4) -- cycle;
		\node[cross out, minimum size=2mm, draw=red!70!black, inner sep=1mm, thick] (0) at (0,1) {}; 
		\node[cross out, minimum size=2mm, draw=red!70!black, inner sep=1mm, thick] (O2) at (0,3) {};
		\node[cross out, minimum size=2mm, draw=red!70!black, inner sep=1mm, thick] (O3) at (2,3) {};
		\node[cross out, minimum size=2mm, draw=red!70!black, inner sep=1mm, thick] (O4) at (2,1) {};
        \node [below left=0.05pt of O2, red!70!black] {$\tau_{00}$};
        \node [below right=0.05pt of O3, red!70!black] {$\tau_{00}$};
        \node [below left=0.05pt of 0, red!70!black] {$\tau_{22}$};
        \node [below right=0.05pt of O4, red!70!black] {$\tau_{22}$};
        \node[gray] (00) at (1,2) {\small{$1$}};
        \node[gray] (000) at (3,2) {\small{$1$}};
        \node[gray] (11) at (1,3.25) {\small{$0$}};
        \node[gray] (22) at (3,2.75) {\small{$0$}};
        \node[gray] (33) at (1,1.25) {\small{$2$}};
        \node[gray] (44) at (3,0.75) {\small{$2$}};
    \end{tikzpicture}}
\end{subfigure}
\hfill
\begin{subfigure}{0.4\textwidth}
\centering{
\begin{tikzpicture}[auto, scale=0.5]
		\node [circle, draw=blue!50, fill=blue!20, inner sep=0pt, minimum size=5mm] (0) at (0,0) {$0$};
		\node [circle, draw=blue!50, fill=blue!20, inner sep=0pt, minimum size=5mm] (1) at (3,0) {$1$};
		\node [circle, draw=blue!50, fill=blue!20, inner sep=0pt, minimum size=5mm] (2) at (6,0) {$2$};
		\node [above left=12pt of 0] (A) {$\tiny{\yng(1,1)}$};
		\node [below left=12pt of 0] (Ac) {$\overline{\tiny{\yng(1,1)}}$};
		\node [above right=12pt of 2] (S) {${\tiny{\yng(2)}}$};
		\node [below right=12pt of 2] (Sc) {$\overline{\tiny{\yng(2)}}$};
		\node [left=0.1pt of 0] {\small{$SU(N)$}};
		\node [below=0.1pt of 1] {\small{$SU(N)$}};
		\node [right=0.1pt of 2] {\small{$SU(N)$}};
        \draw (0) to (1) [<->, thick];
        \draw (1) to (2) [<->, thick];
        \draw (0) to (A) [->, thick, shorten >=-3.5pt];
        \draw (Ac) to (0) [->, thick, shorten <=-3.5pt];
        \draw (2) to (S) [->, thick, shorten >=-3.5pt];
        \draw (Sc) to (2) [->, thick, shorten <=-3.5pt];
        \draw (1) to [out=50, in=135, looseness=10] (1) [->, thick];
\end{tikzpicture}}
\end{subfigure}    
\caption{The orientifold projection of $L^{2,4,2}$ with four fixed points that yields unitary groups and pairs of conjugate tensor fields and the choice $\tau_{00} = -1$, $\tau_{22} = +1$.}\label{fig:five-braneOmega2}
\end{figure}

For $n$ even another orientifold projection is allowed, with all of the four fixed points lying on the intersections between NS5-branes, i.e. generating two conjugate pairs of tensor matter fields. In this way, all gauge factors remain unitary. We can construct such an orientifold by shifting the fixed points by a quarter of a period in the five-brane diagram (see Fig.~\ref{fig:five-braneOmega2} for the $L^{2,4,2}$ example). These orientifolds were also analyzed in \cite{Amariti:2021lhk}, and shown to realize the same mechanism as the models above. Note that when the toric diagram is a rectangle, such a projection cannot be obtained either with fixed points or with fixed lines, which implies that in order to include the last step of the chain the orientifold projection must be realized differently on the five-brane diagram. As we will see in the next section, this occurs for a chain of $L^{a,b,a}/\mathbb{Z}_2$ models with a particular orientifold projection introduced in \cite{Garcia-Valdecasas:2021znu} known as \emph{glide orientifold}. This will play an important role in the remainder.

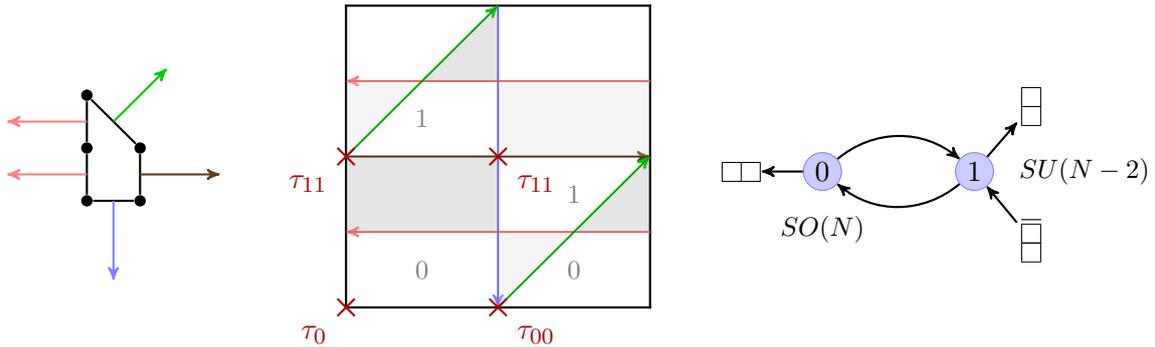
\begin{figure}
\begin{subfigure}{0.2\textwidth}
\centering{
\begin{tikzpicture}[auto, scale=0.7]
	\node [circle, fill=black, inner sep=0pt, minimum size=1.5mm] (0) at (0,0) {}; 
		\node [circle, fill=black, inner sep=0pt, minimum size=1.5mm] (1) at (0,1) {};
		\node [circle, fill=black, inner sep=0pt, minimum size=1.5mm] (2) at (0,2) {}; 
		\node [circle, fill=black, inner sep=0pt, minimum size=1.5mm] (h1) at (1,0) {}; 
		\node [circle, fill=black, inner sep=0pt, minimum size=1.5mm] (h2) at (1,1) {};
        \draw (0) to (1) [thick];
        \draw (1) to (2) [thick];
        \draw (2) to (h2) [thick];
        \draw (h2) to (h1) [thick];
        \draw (h1) to (0) [thick];
        \draw (0.5,0) to (0.5,-1.5) [->, thick, blue!50];
        \draw (0,0.5) to (-1.5,0.5) [->, thick, red!50];
        \draw (0,1.5) to (-1.5,1.5) [->, thick, red!50];
        \draw (1,0.5) to (2.5,0.5) [->, thick, brown!50!black];
        \draw (0.5,1.5) to (1.5,2.5) [->, thick, green!80!black];
\end{tikzpicture}
}    
\end{subfigure}
\begin{subfigure}{0.4\textwidth}
\centering{
    \begin{tikzpicture}
		\node (0) at (0,0) {}; 
		\node (1) at (4,0) {};
		\node (2) at (4,4) {}; 
		\node (3) at (0,4) {}; 
		\node (c) at (2,2) {};
        \draw (0) to (1) [shorten >=-0.15cm, shorten <=-0.15cm, thick];
        \draw (1) to (2) [shorten >=-0.15cm, shorten <=-0.15cm, thick];
        \draw (2) to (3) [shorten >=-0.15cm, shorten <=-0.15cm, thick];
        \draw (3) to (0) [shorten >=-0.15cm, shorten <=-0.15cm, thick];
        \draw (2,4) to (2,0) [->, thick, blue!50];
        \draw (4,1) to (0,1) [->, thick, red!50];
        \draw (4,2) to (0,2) [<-, thick, brown!50!black];
        \draw (4,3) to (0,3) [->, thick, red!50];
        \draw (2,0) to (4,2) [->, thick, green!80!black];
        \draw (0,2) to (2,4) [->, thick, green!80!black];
        \draw[fill=gray!80!white, nearly transparent]  (0,1) -- (0,2) -- (2,2) -- (2,1) -- cycle;
        \draw[fill=gray!80!white, nearly transparent]  (4,2) -- (4,1) -- (3,1) -- cycle;
        \draw[fill=gray!80!white, nearly transparent]  (1,3) -- (2,3) -- (2,4) -- cycle;
        \draw[fill=gray!30!white, nearly transparent]  (2,2) -- (4,2) -- (4,3) -- (2,3) -- cycle;
        \draw[fill=gray!30!white, nearly transparent]  (0,2) -- (0,3) -- (1,3) -- cycle;
        \draw[fill=gray!30!white, nearly transparent]  (2,0) -- (2,1) -- (3,1) -- cycle;
		\node[cross out, minimum size=2mm, draw=red!70!black, inner sep=1mm, thick] (0) at (0,0) {}; 
		\node[cross out, minimum size=2mm, draw=red!70!black, inner sep=1mm, thick] (O2) at (0,2) {};
		\node[cross out, minimum size=2mm, draw=red!70!black, inner sep=1mm, thick] (O3) at (2,2) {};
		\node[cross out, minimum size=2mm, draw=red!70!black, inner sep=1mm, thick] (O4) at (2,0) {};
        \node [below left=0.05pt of 0, red!70!black] {$\tau_0$};
        \node [below right=0.05pt of O4, red!70!black] {$\tau_{00}$};
        \node [below right=0.05pt of O3, red!70!black] {$\tau_{11}$};
        \node [below left=0.05pt of O2, red!70!black] {$\tau_{11}$};
        \node[gray] (00) at (1,0.5) {\small{$0$}};
        \node[gray] (000) at (3,0.5) {\small{$0$}};
        \node[gray] (11) at (1,2.5) {\small{$1$}};
        \node[gray] (22) at (3,1.5) {\small{$1$}};
\end{tikzpicture}}
\end{subfigure}
\hfill
\begin{subfigure}{0.4\textwidth}
\centering{
\begin{tikzpicture}[auto, scale=0.5]
		\node [circle, draw=blue!50, fill=blue!20, inner sep=0pt, minimum size=5mm] (0) at (0,0) {$0$};
		\node [circle, draw=blue!50, fill=blue!20, inner sep=0pt, minimum size=5mm] (1) at (4,0) {$1$};
		\node [above right=12pt of 1] (A) {$\tiny{\yng(1,1)}$};
		\node [below right=12pt of 1] (Ac) {$\overline{\tiny{\yng(1,1)}}$};
		\node [left=12pt of 0] (S) {$\tiny{\yng(2)}$};
		\node [below=0.2cm of 0] {\small{$SO(N)$}};
		\node [right=0.2cm of 1] {\small{$SU(N-2)$}};
        \draw (0) to [out=40, in=140, looseness=1] (1) [->, thick];
        \draw (1) to [out=220, in=320, looseness=1] (0) [->, thick];
        \draw (1) to (A) [->, thick, shorten >=-3.5pt];
        \draw (Ac) to (1) [->, thick, shorten <=-3.5pt];
        \draw (0) to (S) [->, thick, shorten >=-3.5pt];
\end{tikzpicture}}
\end{subfigure}    
    \caption{An example with $n$ odd obtained from mass deformation of $\mathbb{C}^2/\mathbb{Z}_3 \times \mathbb{C}$ and the orientifold projection with fixed points. On the left, the toric diagram of $L^{1,2,1}$ or SPP, center its five-brane and on the right the quiver after the orientifold with $\tau_{0}=+1$, $\tau_{00}=+1$, $\tau_{11}=-1$.}\label{fig:five-braneOdd}
\end{figure}

In the case of $n$ odd the process is similar, but at the end of  the chain a single adjoint field remains, corresponding to the $L^{\frac{n-1}{2},\frac{n+1}{2},\frac{n-1}{2}}$ toric model. We show in Fig. \ref{fig:five-braneOdd} the example of SPP, that is $L^{1,2,1}$, obtained by a mass deformation of  $\mathbb{C}^2/\mathbb{Z}_3 \times \mathbb{C}$. The figure reveals the general feature of these orientifolds, in which one group and its adjoint are projected, while the other two $\tau$'s give two conjugate fields which are symmetric or antisymmetric under a unitary group. In \cite{Antinucci:2021edv} these models where studied for $n=3p$ and generalized in \cite{Amariti:2021lhk} to any $n$, showing that again the same mechanism occurs.  

Finally, \cite{Amariti:2021lhk} shows that the conformal duality discussed above is an `inherited $S$-duality' from the mother $\mathcal{N}=2$ theory that is subsequently mass-deformed.

In the rest of this paper we will show how the same construction works for another infinite class of toric models, which are $\mathbb{Z}_2$ chiral orbifolds of the models of this section. In particular, the next section is devoted to the description of the models in terms of five-brane diagrams. 

\section{Glide orientifold and $L^{a,b,a}/\mathbb{Z}_2$ models}
\label{sec:glide}

In this section we are going to introduce the class of orientifold models $L^{a,b,a}/\mathbb{Z}_2$, which we will focus on hereafter. In the first subsection we  discuss the glide orientifold projection introduced in \cite{Garcia-Valdecasas:2021znu}, which we will perform in the case $a=b$. The second subsection is devoted to a description of the models involved and the results. The following two sections will then contain a detailed analysis of such models. 

\subsection{Orientifolds and Klein bottles}

Among the $\mathbb{Z}_2$ involutions of a torus, there are some choices that do not leave fixed points. This is interesting, as the five-brane diagram and the brane tiling are embedded in a torus and the orientifold projection is realized as a $\mathbb{Z}_2$ involution. While the past literature mostly focuses on projections with fixed loci, the case of a glide orientifold was recently analyzed in \cite{Garcia-Valdecasas:2021znu}. This involution maps a point to another by combining a shift by half a period of the fundamental cell with a reflection about one of its axes, see Fig.~\ref{fig:five-braneGlideC3Z2}. The topology obtained after this involution is that of a Klein bottle. This operation does not leave any fixed points and this can be understood by the fact that the net orientifold charge in the system is zero. As a consequence, the glide orientifold yields only $SU(N)$ gauge factors and tensor representations, if any, in conjugate pairs. Hence, the resulting model is automatically free of any gauge anomaly. We denote this projection by $\Omega_{\mathrm{gl}}$. 

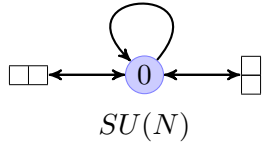
\begin{figure}
    \centering{
\begin{tikzpicture}[auto, scale=0.6]
		\node [circle, draw=blue!50, fill=blue!20, inner sep=0pt, minimum size=5mm] (0) at (0,0) {$0$};
		\node [left=25pt of 0] (S) {$\tiny{\yng(2)}$};
		\node [right=25pt of 0] (A) {$\tiny{\yng(1,1)}$};
		\node [below=3pt of 0] {$SU(N)$};
        \draw (0) to (S) [->, thick, shorten >=-3.5pt];
        \draw (0) to (S) [<-, thick, shorten >=-3.5pt];
        \draw (0) to (A) [->, thick, shorten >=-3.5pt];
        \draw (0) to (A) [<-, thick, shorten >=-3.5pt];
        \draw (0) to [out=50, in=135, looseness=10] (0) [->, thick];
\end{tikzpicture}}
\caption{The linear quiver resulting from the glide orientifold of $\mathbb{C}^3/\mathbb{Z}_2$.} \label{fig:QuiverGlideC3Z2}
\end{figure}

As an example, consider again the original orbifold $\mathbb{C}^2/\mathbb{Z}_2 \times \mathbb{C}$. First, we need to move to a different toric phase by means of an $SL(2, \mathbb{Z})$ transformation, since the toric diagram needs to be symmetric about an axis that crosses at least two points of the toric diagram. In other words, if one consider the vectors dual to the sides of the toric diagram, an even number of them must be parallel to the symmetry axis of the glide. For our example, we choose the phase displayed in Fig.~\ref{fig:ToricGlideC3Z2}, whose associated five-brane and its symmetry axis is drawn in Fig.~\ref{fig:five-braneGlideC3Z2}. Finally, the Klein bottle is explicitly displayed in Fig.~\ref{fig:KleinC3Z2}. The parent theory has two gauge factors, labelled by 0 and 1. The projection maps one factor into the other, while representation are conjugate. Moreover, the glide projection maps fields $\phi_0 \to \phi_1$, $X_{01}^1 \to X_{01}^2$ and $X_{10}^1 \to X_{10}^2$, so the upper index can be dropped. Since $\tiny{\yng(1)}_1 \to \overline{\tiny{\yng(1)}}_0 $, we can split $X_{01} = \left( \tiny{\yng(1)}_0, \, \overline{\tiny{\yng(1)}}_1 \right)$ in $\tiny{\yng(1,1)}_0 = A$, $\tiny{\yng(2)}_0 = S$, and $X_{10} = \left( \tiny{\yng(1)}_1, \, \overline{\tiny{\yng(1)}}_0 \right)$ in $\overline{\tiny{\yng(1,1)}}_0 = \widetilde{A}$, $\overline{\tiny{\yng(2)}}_0 = \widetilde{S}$. The linear quiver that summarizes this matter content is drawn in Fig.~\ref{fig:QuiverGlideC3Z2}. Finally, the superpotential reads 
\begin{align}
    W_{_{\mathbb{C}^2/\mathbb{Z}_2 \times \mathbb{C}}}^{\Omega_{\mathrm{gl}}} = \phi_0 \widetilde{A} S - \phi_0 \widetilde{S} A \; . 
\end{align}

\begin{figure}
\begin{subfigure}{0.25\textwidth}
\centering{
   \begin{tikzpicture}[auto, scale=0.7]
		\node [circle, fill=black, inner sep=0pt, minimum size=1.5mm] (0) at (0,0) {}; 
		\node [circle, fill=black, inner sep=0pt, minimum size=1.5mm] (1) at (2,0) {};
		\node [circle, fill=black, inner sep=0pt, minimum size=1.5mm] (2) at (4,0) {}; 
		\node [circle, fill=black, inner sep=0pt, minimum size=1.5mm] (3) at (2,2) {}; 
        \draw (0) to (1) [thick];
        \draw (1) to (2) [thick];
        \draw (2) to (3) [thick];
        \draw (3) to (0) [thick];
        \draw (1,0) to (1,-1.5) [->, thick, blue!50];
        \draw (3,0) to (3,-1.5) [->, thick, blue!50];
        \draw (1,1) to (0,2) [->, thick, red!50];
        \draw (3,1) to (4,2) [->, thick, green!80!black];
\end{tikzpicture}}
\subcap{The toric diagram of $\mathbb{C}^2/\mathbb{Z}_2 \times \mathbb{C}$.}\label{fig:ToricGlideC3Z2}
\end{subfigure}
\hfill
\begin{subfigure}{0.25\textwidth}
\centering{
    \begin{tikzpicture}
		\node (0) at (0,0) {}; 
		\node (1) at (4,0) {};
		\node (2) at (4,4) {}; 
		\node (3) at (0,4) {}; 
		\node (c) at (2,2) {};
		\node (c1) at (0,2) {};
        \draw (0) to (1) [shorten >=-0.15cm, shorten <=-0.15cm, thick];
        \draw (1) to (2) [shorten >=-0.15cm, shorten <=-0.15cm, thick];
        \draw (2) to (3) [shorten >=-0.15cm, shorten <=-0.15cm, thick];
        \draw (3) to (0) [shorten >=-0.15cm, shorten <=-0.15cm, thick];
        \draw (1,4) to (1,0) [->, thick, blue!50];
        \draw (3,4) to (3,0) [->, thick, blue!50];
        \draw (3,0) to (0,3) [->, thick, red!50];
        \draw (4,3) to (3,4) [->, thick, red!50];
        \draw (1,0) to (4,3) [->, thick, green!80!black];
        \draw (0,3) to (1,4) [->, thick, green!80!black];
        \draw[fill=gray!80!white, nearly transparent]  (1,0) -- (2,1) -- (1,2) -- cycle;
        \draw[fill=gray!80!white, nearly transparent]  (3,2) -- (4,3) -- (3,4) -- cycle;
        \draw[fill=gray!30!white, nearly transparent]  (3,0) -- (2,1) -- (3,2) -- cycle;
        \draw[fill=gray!30!white, nearly transparent]  (1,2) -- (1,4) -- (0,3) -- cycle;
		\draw (0,2) to (4,2) [thick, dashed, purple];
        \node [below right=0.05pt of c1, purple] {$\Omega_{\mathrm{gl}}$};
        \node[gray] (00) at (0.5,0.8) {\small{$0$}};
        \node[gray] (000) at (1.7,0.4) {\small{$0$}};
        \node[gray] (11) at (1.7,1.7) {\small{$0$}};
    \end{tikzpicture}}
\subcap{The five-brane diagram of $\mathbb{C}^2/\mathbb{Z}_2 \times \mathbb{C}$ and the horizontal axis that provides the glide orientifold.}\label{fig:five-braneGlideC3Z2}
\end{subfigure}
\hfill
\begin{subfigure}{0.25\textwidth}
\centering{
    \begin{tikzpicture}
		\node (0) at (0,0) {}; 
		\node (1) at (4,0) {};
		\node (2) at (4,4) {}; 
		\node (3) at (0,4) {}; 
		\node (c) at (2,2) {};
		\node (c1) at (0,2) {};
        \draw (0) to (1) [shorten >=-0.15cm, shorten <=-0.15cm, thick];
        \draw (1) to (2) [shorten >=-0.15cm, shorten <=-0.15cm, thick];
        \draw (2) to (3) [shorten >=-0.15cm, shorten <=-0.15cm, thick];
        \draw (3) to (0) [shorten >=-0.15cm, shorten <=-0.15cm, thick];
        \draw (1,4) to (1,0) [->, thick, blue!50];
        \draw (3,4) to (3,0) [->, thick, blue!50];
        \draw (3,0) to (0,3) [->, thick, red!50];
        \draw (4,3) to (3,4) [->, thick, red!50];
        \draw (1,0) to (4,3) [->, thick, green!80!black];
        \draw (0,3) to (1,4) [->, thick, green!80!black];
        \draw[fill=gray!80!white, nearly transparent]  (1,0) -- (2,1) -- (1,2) -- cycle;
        \draw[fill=gray!80!white, nearly transparent]  (3,2) -- (4,3) -- (3,4) -- cycle;
        \draw[fill=gray!30!white, nearly transparent]  (3,0) -- (2,1) -- (3,2) -- cycle;
        \draw[fill=gray!30!white, nearly transparent]  (1,2) -- (1,4) -- (0,3) -- cycle;
		\draw (0,0) to (2,0) [->>, thick, purple];
		\draw (2,4) to (2,0) [->, thick, purple];
		\draw (0,4) to (2,4) [->>, thick, purple];
		\draw (0,0) to (0,4) [->, thick, purple];
        \node[gray] (00) at (0.5,0.8) {\small{$0$}};
        \node[gray] (000) at (1.7,0.4) {\small{$0$}};
        \node[gray] (11) at (1.7,1.7) {\small{$0$}};
    \end{tikzpicture}}
    \subcap{The five-brane diagram of $\mathbb{C}^2/\mathbb{Z}_2 \times \mathbb{C}$ and the Klein bottle resulting from the glide orientifold.}\label{fig:KleinC3Z2}
\end{subfigure}
\hfill
\end{figure}

\subsection{Families of orientifolds of $L^{a,b,a}/\mathbb{Z}_2$}

In the previous section we argued about a conformal duality between projected non-chiral toric theories, connected by a deformation that changes the shape of the toric diagram while preserving the number of points \cite{Bianchi:2014qma}. We constructed a chain of toric polygons that belong to the $L^{a,b,a}$ family, where $a$ and $b$ can take generic values with $a \leq b$.

In order to generalize the duality to chiral theories, we consider the $\mathbb{Z}_2$ orbifold $L^{a,b,a}/\mathbb{Z}_2$ of that family, and infer some general behaviors that may be useful for a complete classification of the phenomenon of conformal duality among toric quivers after an orientifold projection. Indeed we observe that all such models are chiral, except for $L^{0,2,0}$ and $L^{1,1,1}$. Note that in order to have chiral theories, we lose the elliptic model description of these model. Perhaps, one could still construct a similar one along the lines of \cite{Garcia-Valdecasas:2021znu}.

Moreover, for constructing a chain of toric diagrams as before, the number of external points of the toric diagram, i.e. on the perimeter, and the number of internal points must separately coincide. This implies that the models before the projection have the same number of gauge nodes and of non-anomalous $U(1)$ symmetries. This imposes constraints on the possible orbifold that we can use to generate an infinite family of dual models. Interestingly the conformally dual models discussed in \cite{Antinucci:2020yki}, that are not related by an $R=2$ mass deformation, respect these constraints as well. It may be a hint to a more general phenomenon whereby the conformally dual models obtained after orientifold projections are related by non-quadratic superpotential deformations. 

In the following we identify three families in terms of the parity of $a$ and $b$ and in terms of the type of projection that we will realize on the five-brane web, though we will focus only on the first two.

\subsection*{Family $\mathcal{A}$}
The first family that we study is the orientifold projection of $L^{a,b,a}/\mathbb{Z}_2$ with $a+b$ even that leads to quiver gauge theories with only unitary gauge groups and two pairs of conjugate tensor matter fields at the extremal gauge nodes, and we call it \emph{family} $\mathcal{A}$. This is the generalization of family \emph{i)} of~\cite{Amariti:2021lhk} and we analyze it in Section~\ref{sec:famA}. Imposing $a+b=2k$, two extreme possibilities are $L^{0,2k,0}/\mathbb{Z}_2$ and $L^{k,k,k}/\mathbb{Z}_2$. As anticipated above, the case with $k=1$ has a non-chiral field content before the orientifold projection, while for $k>1$ internal points on the toric diagram necessarily arise. A generic model in the family is $L^{2p,2k-2p,2p}/\mathbb{Z}_2$ with $4k-4p$ hexagons and $4p$ squares. We can construct a chain of toric diagrams that describe, upon orientifold projection, theories connected by conformal duality, i.e. a quadratic exactly marginal deformation:
\begin{align}\label{eq:ChainZ2}
L^{0,2k,0}/\mathbb{Z}_2 \; \to \; L^{2,2(k-1),2}/\mathbb{Z}_2 \ldots \to \; L^{2p,2k-2p,2p}/\mathbb{Z}_2 \; \to \ldots \; \to \;  L^{k,k,k}/\mathbb{Z}_2 \; , 
\end{align}
where $p=1,\, \ldots ,\,  \lfloor \frac{k}{2} \rfloor$. Along the chain, pairs of vector-like fields with $R=1$ are integrated out thanks to quadratic marginal deformation, until the last step where tensor fields are deformed. Note that the number of vector-like fields is $k-1$, so when $k$ is even the last step requires that one remaining vector-like field is integrated out together with the tensors.

\begin{figure}[h]
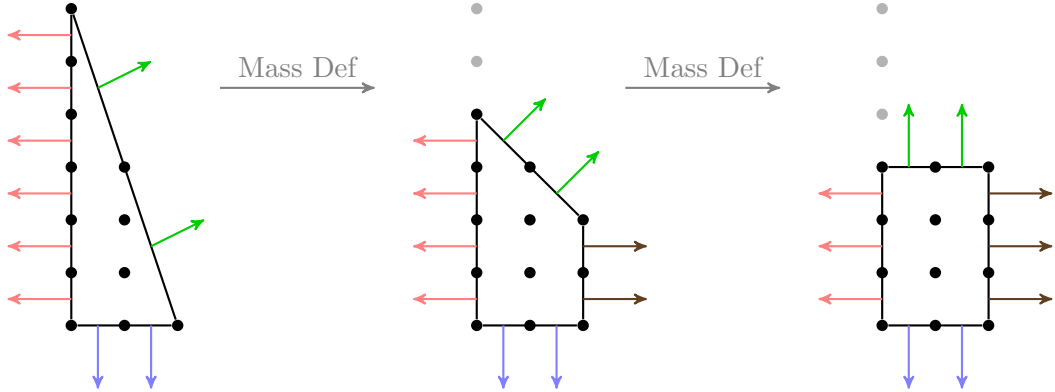

\hfill
    \centering{
       \begin{subfigure}{0.3\textwidth}

    \end{subfigure}}
    \caption{An example of a chain of toric diagrams of family $\mathcal{A}$ connected by mass deformation for $k=3$: $L^{0,6,0}/\mathbb{Z}_2 \to L^{2,4,2}/\mathbb{Z}_2 \to L^{3,3,3}/\mathbb{Z}_2$. It holds also with the orientifold projection, mutatis mutandis.}\label{fig:ToricFamA}
\end{figure}

In all steps but the last the orientifold projection is given by fixed points. On the five-brane, these lie at the intersection between a vertical brane and a skew brane. On the other hand, the last step requires a glide orientifold. An example of a chain of models, upon orientifold, is showed in Fig.~\ref{fig:ToricFamA}-\ref{fig:five-braneFamA}, and the quiver in Fig.~\ref{fig:ExampleQuiverFamA}.

Note that the orientifold projection allows also for another configuration with four tensor fields at four different nodes, associated to five-branes where the four fixed points lie at the intersection of the green vectors and horizontal red ones. In this way, orientifolds of $L^{p,2k-p,p}/\mathbb{Z}_2$ are also allowed. However, such a configuration does not feature the $R$-charges and dualities we want to discuss here, at least for the first examples we worked out, hence we will not consider it in the following. 

\begin{figure}[h]
\begin{subfigure}{0.25\textwidth}
\centering{
    %
}
\end{subfigure}
\hfill
\caption{The chain of five-brane diagrams of family $\mathcal{A}$ connected by mass deformation for $k=3$: $L^{0,6,0}/\mathbb{Z}_2 \to L^{2,4,2}/\mathbb{Z}_2 \to L^{3,3,3}/\mathbb{Z}_2$. It holds also with the orientifold projection, provided a $\mathbb{Z}_2$ symmetry is preserved.}\label{fig:five-braneFamA}
\end{figure}

\begin{figure}[h]
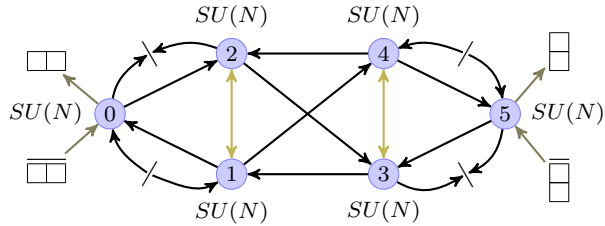

\centering{
%
}
\caption{The quiver for the orientifold theory of $L^{0,6,0}/\mathbb{Z}_2 \to L^{2,4,2}/\mathbb{Z}_2 \to L^{3,3,3}/\mathbb{Z}_2$ in Family $\mathcal{A}$ with choice $(\tau_{00} = +, \, \tau_{55} = -)$. Colored fields represent the pairs that are mass deformed in the chain, the color match the chain of five-branes in Fig.~\ref{fig:five-braneFamA}.}\label{fig:ExampleQuiverFamA}
\end{figure}

\break

\subsection*{Family $\mathcal{B}$}

We will refer to the second family as {\it family} $\mathcal{B}$, which generalizes family \emph{iv)} of~\cite{Amariti:2021lhk}. Also in this case $a+b=2k$, but the orientifold projection acts either with fixed points or with fixed lines, lying on top of the faces of the five-brane, either hexagons or squares depending on the fixed loci. Contrary to the previous family, four gauge factors are now real groups,\footnote{We loosely refer to $SO(N)$ or $USp(N)$ as real gauge groups since, contrary to $SU(N)$, they do not admit complex representations. Complex spinors do not appear in perturbative open string constructions.} $SO$ or $USp$ depending on the signs of the fixed loci. The chain of toric diagrams is the same as in the previous family, Eq.~\eqref{eq:ChainZ2}, but the orientifold projection is realized differently on the five-branes, as we need to move the fixed points by a quarter period, see for example Fig.~\ref{fig:five-braneFamB} and the associated quiver in Fig.~\ref{fig:ExampleQuiverFamB}. The extremal cases are $L^{0,2k,0}/\mathbb{Z}_2$ and  $L^{k,k,k}/\mathbb{Z}_2$. Moreover, the last model in the chain is projected by fixed lines. As we will observe in Section~\ref{sec:famB}, in this family marginal quadratic superpotential deformations with $R=2$ can be generated only with a specific shift between the ranks of the gauge factors.

\begin{figure}
\begin{subfigure}{0.25\textwidth}
\centering{
    \begin{tikzpicture}
        \node (a) at (4,2) {};
        \node (b) at (5.2,2) {};
        \draw (a) [pil, ultra thick, yellow!70!black] to node {} (b); 
		\node (0) at (0,0) {}; 
		\node (1) at (4,0) {};
		\node (2) at (4,4) {}; 
		\node (3) at (0,4) {}; 
		\node (c) at (2,2) {};
        \draw (0) to (1) [shorten >=-0.15cm, shorten <=-0.15cm, thick];
        \draw (1) to (2) [shorten >=-0.15cm, shorten <=-0.15cm, thick];
        \draw (2) to (3) [shorten >=-0.15cm, shorten <=-0.15cm, thick];
        \draw (3) to (0) [shorten >=-0.15cm, shorten <=-0.15cm, thick];
        \draw (0,4) to (0,0) [->, thick, blue!50];
        \draw (4,4) to (4,0) [->, thick, blue!50];
        \draw (2,4) to (2,0) [->, thick, blue!50];
        \draw (4,0.333) to (0,0.333) [->, thick, red!50];
        \draw (4,1) to (0,1) [->, thick, red!50];
        \draw (4,1.667) to (0,1.667) [->, thick, red!50];
        \draw (4,2.333) to (0,2.333) [->, thick, red!50];
        \draw (4,3) to (0,3) [->, thick, red!50];
        \draw (4,3.667) to (0,3.667) [->, thick, red!50];
        \draw (0,0) to (4,1.333) [->, thick, green!80!black];
        \draw (0,1.333) to (4,2.667) [->, thick, green!80!black];
        \draw (0,2.667) to (4,4) [->, thick, green!80!black];
        \draw (2,0) to (4,0.667) [->, thick, green!80!black];
        \draw (0,0.667) to (4,2) [->, thick, green!80!black];
        \draw (0,2) to (4,3.333) [->, thick, green!80!black];
        \draw (0,3.333) to (2,4) [->, thick, green!80!black];        
        \draw[fill=gray!80!white, nearly transparent]  (2,0) -- (3,0.333) -- (2,0.333) -- cycle;
        \draw[fill=gray!80!white, nearly transparent]  (2,0.667) -- (2,1) -- (3,1) -- cycle;
        \draw[fill=gray!80!white, nearly transparent]  (2,1.333) -- (2,1.667) -- (3,1.667) -- cycle;
        \draw[fill=gray!80!white, nearly transparent]  (2,2) -- (2,2.333) -- (3,2.333) -- cycle;
        \draw[fill=gray!80!white, nearly transparent]  (2,2.667) -- (2,3) -- (3,3) -- cycle;
        \draw[fill=gray!80!white, nearly transparent]  (2,3.333) -- (2,3.667) -- (3,3.667) -- cycle;
        \draw[fill=gray!80!white, nearly transparent]  (0,0) -- (1,0.333) -- (0,0.333) -- cycle;
        \draw[fill=gray!80!white, nearly transparent]  (0,0.667) -- (0,1) -- (1,1) -- cycle;
        \draw[fill=gray!80!white, nearly transparent]  (0,1.333) -- (0,1.667) -- (1,1.667) -- cycle;
        \draw[fill=gray!80!white, nearly transparent]  (0,2) -- (0,2.333) -- (1,2.333) -- cycle;
        \draw[fill=gray!80!white, nearly transparent]  (0,2.667) -- (0,3) -- (1,3) -- cycle;
        \draw[fill=gray!80!white, nearly transparent]  (0,3.333) -- (0,3.667) -- (1,3.667) -- cycle;
        \draw[fill=gray!30!white, nearly transparent]  (1,0.333) -- (2,0.667) -- (2,0.333) -- cycle;
        \draw[fill=gray!30!white, nearly transparent]  (1,1) -- (2,1.333) -- (2,1) -- cycle;
        \draw[fill=gray!30!white, nearly transparent]  (1,1.667) -- (2,2) -- (2,1.667) -- cycle;
        \draw[fill=gray!30!white, nearly transparent]  (1,2.333) -- (2,2.667) -- (2,2.333) -- cycle;
        \draw[fill=gray!30!white, nearly transparent]  (1,3) -- (2,3.333) -- (2,3) -- cycle;
        \draw[fill=gray!30!white, nearly transparent]  (1,3.667) -- (2,4) -- (2,3.667) -- cycle;
        \draw[fill=gray!30!white, nearly transparent]  (3,0.333) -- (4,0.667) -- (4,0.333) -- cycle;
        \draw[fill=gray!30!white, nearly transparent]  (3,1) -- (4,1.333) -- (4,1) -- cycle;
        \draw[fill=gray!30!white, nearly transparent]  (3,1.667) -- (4,2) -- (4,1.667) -- cycle;
        \draw[fill=gray!30!white, nearly transparent]  (3,2.333) -- (4,2.667) -- (4,2.333) -- cycle;
        \draw[fill=gray!30!white, nearly transparent]  (3,3) -- (4,3.333) -- (4,3) -- cycle;
        \draw[fill=gray!30!white, nearly transparent]  (3,3.667) -- (4,4) -- (4,3.667) -- cycle;
		\node[cross out, minimum size=2mm, draw=red!70!black, inner sep=1mm, thick] (0) at (1,0) {}; 
		\node[cross out, minimum size=2mm, draw=red!70!black, inner sep=1mm, thick] (O2) at (1,2) {};
		\node[cross out, minimum size=2mm, draw=red!70!black, inner sep=1mm, thick] (O3) at (3,2) {};
		\node[cross out, minimum size=2mm, draw=red!70!black, inner sep=1mm, thick] (O4) at (3,0) {};
        \node [right=0.05pt of O2, red!70!black] {$\tau_{0}$};
        \node [right=0.05pt of O3, red!70!black] {$\tau_{1}$};
        \node [below right=0.05pt of 0, red!70!black] {$\tau_{7}$};
        \node [below right=0.05pt of O4, red!70!black] {$\tau_{6}$};
        \node[gray] (00) at (0.6,2) {\small{$0$}};
        \node[gray] (000) at (2.6,2) {\small{$1$}};
        \node[gray] (11) at (3,1.3) {\small{$2$}};
        \node[gray] (22) at (1,1.3) {\small{$3$}};
        \node[gray] (33) at (1,0.65) {\small{$4$}};
        \node[gray] (44) at (3,0.65) {\small{$5$}};
        \node[gray] (55) at (3.4,0.15) {\small{$6$}};
        \node[gray] (555) at (1.4,0.15) {\small{$7$}};
    \end{tikzpicture}}
\end{subfigure}
\hfill
\begin{subfigure}{0.25\textwidth}
\centering{
    \begin{tikzpicture}
        \node (a) at (4,2) {};
        \node (b) at (5.2,2) {};
        \draw (a) [pil, ultra thick, yellow!40!black] to node {} (b); 
		\node (0) at (0,0) {}; 
		\node (1) at (4,0) {};
		\node (2) at (4,4) {}; 
		\node (3) at (0,4) {}; 
		\node (c) at (2,2) {};
        \draw (0) to (1) [shorten >=-0.15cm, shorten <=-0.15cm, thick];
        \draw (1) to (2) [shorten >=-0.15cm, shorten <=-0.15cm, thick];
        \draw (2) to (3) [shorten >=-0.15cm, shorten <=-0.15cm, thick];
        \draw (3) to (0) [shorten >=-0.15cm, shorten <=-0.15cm, thick];
        \draw (0,4) to (0,0) [->, thick, blue!50];
        \draw (4,4) to (4,0) [->, thick, blue!50];
        \draw (2,4) to (2,0) [->, thick, blue!50];
        \draw (4,0.5) to (0,0.5) [->, thick, red!50];
        \draw (4,1) to (0,1) [<-, thick, brown!50!black];
        \draw (4,1.5) to (0,1.5) [->, thick, red!50];
        \draw (4,2.5) to (0,2.5) [->, thick, red!50];
        \draw (4,3) to (0,3) [<-, thick, brown!50!black];
        \draw (4,3.5) to (0,3.5) [->, thick, red!50];
        \draw (2,0) to (4,2) [->, thick, green!80!black];
        \draw (0,2) to (2,4) [->, thick, green!80!black];
        \draw (0,0) to (4,4) [->, thick, green!80!black];
        \draw[fill=gray!80!white, nearly transparent]  (2,0) -- (2.5,0.5) -- (2,0.5) -- cycle;
        \draw[fill=gray!80!white, nearly transparent]  (2,1) -- (2,1.5) -- (3.5,1.5) -- (3,1) -- cycle;
        \draw[fill=gray!80!white, nearly transparent]  (2,2) -- (2,2.5) -- (2.5,2.5) -- cycle;
        \draw[fill=gray!80!white, nearly transparent]  (2,3) -- (2,3.5) -- (3.5,3.5) -- (3,3) -- cycle;
        \draw[fill=gray!80!white, nearly transparent]  (0,0) -- (0.5,0.5) -- (0,0.5) -- cycle;
        \draw[fill=gray!80!white, nearly transparent]  (0,1) -- (0,1.5) -- (1.5,1.5) -- (1,1) -- cycle;
        \draw[fill=gray!80!white, nearly transparent]  (0,2) -- (0,2.5) -- (0.5,2.5) -- cycle;
        \draw[fill=gray!80!white, nearly transparent]  (0,3) -- (0,3.5) -- (1.5,3.5) -- (1,3) -- cycle;
        \draw[fill=gray!30!white, nearly transparent]  (0.5,0.5) -- (1,1) -- (2,1) -- (2,0.5) -- cycle;
        \draw[fill=gray!30!white, nearly transparent]  (2.5,0.5) -- (3,1) -- (4,1) -- (4,0.5) -- cycle;
        \draw[fill=gray!30!white, nearly transparent]  (1.5,1.5) -- (2,1.5) -- (2,2) -- cycle;
        \draw[fill=gray!30!white, nearly transparent]  (0.5,2.5) -- (1,3) -- (2,3) -- (2,2.5) -- cycle;
        \draw[fill=gray!30!white, nearly transparent]  (2.5,2.5) -- (3,3) -- (4,3) -- (4,2.5) -- cycle;
        \draw[fill=gray!30!white, nearly transparent]  (1.5,3.5) -- (2,3.5) -- (2,4) -- cycle;
        \draw[fill=gray!30!white, nearly transparent]  (3.5,1.5) -- (4,1.5) -- (4,2) -- cycle;
        \draw[fill=gray!30!white, nearly transparent]  (3.5,3.5) -- (4,3.5) -- (4,4) -- cycle;
		\node[cross out, minimum size=2mm, draw=red!70!black, inner sep=1mm, thick] (0) at (1,0) {}; 
		\node[cross out, minimum size=2mm, draw=red!70!black, inner sep=1mm, thick] (O2) at (1,2) {};
		\node[cross out, minimum size=2mm, draw=red!70!black, inner sep=1mm, thick] (O3) at (3,2) {};
		\node[cross out, minimum size=2mm, draw=red!70!black, inner sep=1mm, thick] (O4) at (3,0) {};
        \node [right=0.05pt of O2, red!70!black] {$\tau_{0}$};
        \node [right=0.05pt of O3, red!70!black] {$\tau_{1}$};
        \node [below right=0.05pt of 0, red!70!black] {$\tau_{7}$};
        \node [below right=0.05pt of O4, red!70!black] {$\tau_{6}$};
        \node[gray] (00) at (0.6,2) {\small{$0$}};
        \node[gray] (000) at (2.6,2) {\small{$1$}};
        \node[gray] (11) at (3.6,1.2) {\small{$2$}};
        \node[gray] (22) at (1.6,1.2) {\small{$3$}};
        \node[gray] (33) at (0.4,0.8) {\small{$4$}};
        \node[gray] (44) at (2.4,0.8) {\small{$5$}};
        \node[gray] (55) at (3.4,0.25) {\small{$6$}};
        \node[gray] (555) at (1.4,0.25) {\small{$7$}};
    \end{tikzpicture}}
\end{subfigure}
\hfill
\begin{subfigure}{0.25\textwidth}
\centering{
    \begin{tikzpicture}
		\node (0) at (0,0) {}; 
		\node (1) at (4,0) {};
		\node (2) at (4,4) {}; 
		\node (3) at (0,4) {}; 
		\node (c) at (2,2) {};
		\node (c1) at (0,2) {};
        \draw (0) to (1) [shorten >=-0.15cm, shorten <=-0.15cm, thick];
        \draw (1) to (2) [shorten >=-0.15cm, shorten <=-0.15cm, thick];
        \draw (2) to (3) [shorten >=-0.15cm, shorten <=-0.15cm, thick];
        \draw (3) to (0) [shorten >=-0.15cm, shorten <=-0.15cm, thick];
        \draw (0,4) to (0,0) [->, thick, blue!50];
        \draw (4,4) to (4,0) [->, thick, blue!50];
        \draw (2,4) to (2,0) [->, thick, blue!50];
        \draw (4,0.5) to (0,0.5) [<-, thick, brown!50!black];
        \draw (4,1) to (0,1) [->, thick, red!50];
        \draw (4,1.5) to (0,1.5) [<-, thick, brown!50!black];
        \draw (4,2.5) to (0,2.5) [->, thick, red!50];
        \draw (4,3) to (0,3) [<-, thick, brown!50!black];
        \draw (4,3.5) to (0,3.5) [->, thick, red!50];
        \draw (1,0) to (1,4) [->, thick, green!80!black];
        \draw (3,0) to (3,4) [->, thick, green!80!black];
        \draw[fill=gray!80!white, nearly transparent]  (2,0.5) -- (2,1) -- (3,1) -- (3,0.5) -- cycle;
        \draw[fill=gray!80!white, nearly transparent]  (2,1.5) -- (2,2.5) -- (3,2.5) -- (3,1.5) -- cycle;
        \draw[fill=gray!80!white, nearly transparent]  (2,3) -- (2,3.5) -- (3,3.5) -- (3,3) -- cycle;
        \draw[fill=gray!80!white, nearly transparent]  (0,0.5) -- (0,1) -- (1,1) -- (1,0.5) -- cycle;
        \draw[fill=gray!80!white, nearly transparent]  (0,1.5) -- (0,2.5) -- (1,2.5) -- (1,1.5) -- cycle;
        \draw[fill=gray!80!white, nearly transparent]  (0,3) -- (0,3.5) -- (1,3.5) -- (1,3) -- cycle;
        \draw[fill=gray!30!white, nearly transparent]  (1,0) -- (1,0.5) -- (2,0.5) -- (2,0) -- cycle;
        \draw[fill=gray!30!white, nearly transparent]  (1,1) -- (1,1.5) -- (2,1.5) -- (2,1) -- cycle;
        \draw[fill=gray!30!white, nearly transparent]  (1,2.5) -- (1,3) -- (2,3) -- (2,2.5) -- cycle;
        \draw[fill=gray!30!white, nearly transparent]  (1,3.5) -- (1,4) -- (2,4) -- (2,3.5) -- cycle;
        \draw[fill=gray!30!white, nearly transparent]  (3,0) -- (3,0.5) -- (4,0.5) -- (4,0) -- cycle;
        \draw[fill=gray!30!white, nearly transparent]  (3,1) -- (3,1.5) -- (4,1.5) -- (4,1) -- cycle;
        \draw[fill=gray!30!white, nearly transparent]  (3,2.5) -- (3,3) -- (4,3) -- (4,2.5) -- cycle;
        \draw[fill=gray!30!white, nearly transparent]  (3,3.5) -- (3,4) -- (4,4) -- (4,3.5) -- cycle;
        \draw (0,2) to (4,2) [thick, red!90!black, dashed];
        \draw (0,0.01) to (4,0.01) [thick, red, dashed];
        \node[red!70!black] () at (-0.3,2) {$\tau_0$};
        \node[red!70!black] () at (-0.3,0) {$\tau_6$};
        \node (O4) at (3,0) {};
        \node [below=0.05pt of O4, transparent] {};
        \node[gray] (00) at (1.5,1.7) {\small{$0$}};
        \node[gray] (000) at (3.5,1.7) {\small{$1$}};
        \node[gray] (11) at (2.4,1.2) {\small{$2$}};
        \node[gray] (22) at (0.4,1.2) {\small{$3$}};
        \node[gray] (33) at (1.5,0.8) {\small{$4$}};
        \node[gray] (44) at (3.5,0.8) {\small{$5$}};
        \node[gray] (55) at (2.4,0.25) {\small{$6$}};
        \node[gray] (555) at (0.4,0.25) {\small{$7$}};
    \end{tikzpicture}}
\end{subfigure}
\hfill
\caption{The chain of five-brane diagrams of family $\mathcal{B}$ connected by mass deformation for $k=3$: $L^{0,6,0}/\mathbb{Z}_2 \to L^{2,4,2}/\mathbb{Z}_2 \to L^{3,3,3}/\mathbb{Z}_2$. It holds also with the orientifold projection, provided a $\mathbb{Z}_2$ symmetry is preserved.}\label{fig:five-braneFamB}
\end{figure}
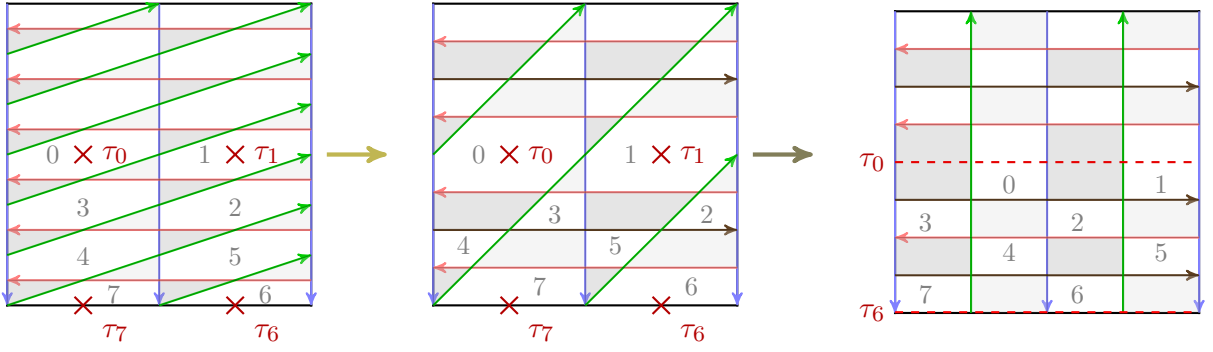

\begin{figure}
\centering{
\begin{tikzpicture}[auto, scale=0.4]
		\node [circle, draw=blue!50, fill=blue!20, inner sep=0pt, minimum size=4mm] (0) at (0,2) {\scriptsize{0}};
		\node [circle, draw=blue!50, fill=blue!20, inner sep=0pt, minimum size=4mm] (1) at (0,-2) {\scriptsize{1}};
		\node [circle, draw=blue!50, fill=blue!20, inner sep=0pt, minimum size=4mm] (2) at (5,2) {\scriptsize{2}};
		\node [circle, draw=blue!50, fill=blue!20, inner sep=0pt, minimum size=4mm] (3) at (5,-2) {\scriptsize{3}};
		\node [circle, draw=blue!50, fill=blue!20, inner sep=0pt, minimum size=4mm] (4) at (10,2) {\scriptsize{4}};
		\node [circle, draw=blue!50, fill=blue!20, inner sep=0pt, minimum size=4mm] (5) at (10,-2) {\scriptsize{5}};
		\node [circle, draw=blue!50, fill=blue!20, inner sep=0pt, minimum size=4mm] (6) at (15,2) {\scriptsize{6}};
		\node [circle, draw=blue!50, fill=blue!20, inner sep=0pt, minimum size=4mm] (7) at (15,-2) {\scriptsize{7}};
		\node [above=0.02pt of 0] {\scriptsize{$SO(N)$}};
		\node [above=0.02pt of 2] {\scriptsize{$SU(N-2)$}};
		\node [above=0.02pt of 4] {\scriptsize{$SU(N-4)$}};
		\node [above=0.02pt of 6] {\scriptsize{$USp(N-6)$}};
		\node [below=0.02pt of 1] {\scriptsize{$SO(N)$}};
		\node [below=0.02pt of 3] {\scriptsize{$SU(N-2)$}};
		\node [below=0.02pt of 5] {\scriptsize{$SU(N-4)$}};
		\node [below=0.02pt of 7] {\scriptsize{$USp(N-6)$}};
        \draw (0) to (2) [<-, thick];
        \draw (2) to (4) [<-, thick];
        \draw (4) to (6) [<-, thick];
        \draw (1) to (3) [<-, thick];
        \draw (3) to (5) [<-, thick];
        \draw (5) to (7) [<-, thick];
        \draw (0) to (1) [-, thick, yellow!40!black];
        \draw (2) to (3) [<->, thick, yellow!70!black];
        \draw (4) to (5) [<->, thick, yellow!70!black];
        \draw (6) to (7) [-, thick, yellow!40!black];
        \draw (3) to (0) [<-, thick];
        \draw (5) to (2) [<-, thick];
        \draw (7) to (4) [<-, thick];
        \draw (2) to (1) [<-, thick];
        \draw (4) to (3) [<-, thick];
        \draw (6) to (5) [<-, thick];
\end{tikzpicture}}
\caption{The quiver for the orientifold theory of $L^{0,6,0}/\mathbb{Z}_2 \to L^{2,4,2}/\mathbb{Z}_2 \to L^{3,3,3}/\mathbb{Z}_2$ in Family $\mathcal{B}$ with choice $(\tau_0 = \tau_1 = + , \, \tau_6 = \tau_7 = -)$. Colored fields represent the pairs that are mass deformed in the chain, the color match the chain of five-branes in Fig.~\ref{fig:five-braneFamB}.}\label{fig:ExampleQuiverFamB}
\end{figure}
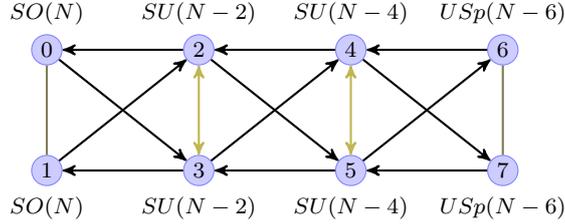

\section{Family $\mathcal{A}$}
\label{sec:famA}

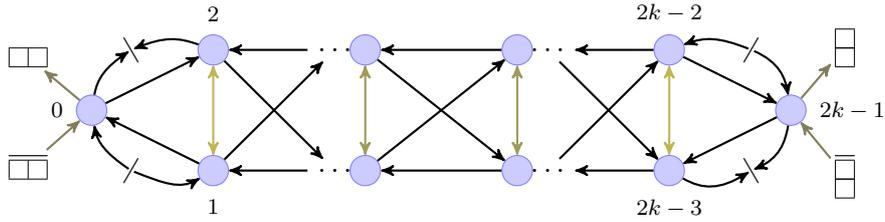
\begin{figure}
\centering{
\begin{tikzpicture}[auto, scale=0.4]
		\node [circle, draw=blue!50, fill=blue!20, inner sep=0pt, minimum size=4mm] (0) at (0,0) {};
		\node [circle, draw=blue!50, fill=blue!20, inner sep=0pt, minimum size=4mm] (1) at (4,-2) {};
		\node [circle, draw=blue!50, fill=blue!20, inner sep=0pt, minimum size=4mm] (2) at (4,2) {};
		\node [circle, draw=blue!50, fill=blue!20, inner sep=0pt, minimum size=4mm] (a) at (9,-2) {};
		\node [circle, draw=blue!50, fill=blue!20, inner sep=0pt, minimum size=4mm] (b) at (9,2) {};
		\node [circle, draw=blue!50, fill=blue!20, inner sep=0pt, minimum size=4mm] (c) at (14,2) {};
		\node [circle, draw=blue!50, fill=blue!20, inner sep=0pt, minimum size=4mm] (d) at (14,-2) {};
		\node [circle, draw=blue!50, fill=blue!20, inner sep=0pt, minimum size=4mm] (3) at (19,-2) {};
		\node [circle, draw=blue!50, fill=blue!20, inner sep=0pt, minimum size=4mm] (4) at (19,2) {};
		\node [circle, draw=blue!50, fill=blue!20, inner sep=0pt, minimum size=4mm] (5) at (23,0) {};
		\node [left=1pt of 0] (k0) {\scriptsize{0}};
		\node [below=1pt of 1] (k1) {\scriptsize{1}};
		\node [below=1pt of 3] (k3) {\scriptsize{$2k-3$}};
		\node [above=1pt of 2] (k2) {\scriptsize{2}};
		\node [above=1pt of 4] (k4) {\scriptsize{$2k-2$}};
		\node [right=1pt of 5] (2km1) {\scriptsize{$2k-1$}};
		\node [above left=12pt of 0] (S) {$\tiny{\yng(2)}$};
		\node [below left=12pt of 0] (Sc) {$\overline{\tiny{\yng(2)}}$};
		\node [above right=12pt of 5] (A) {$\tiny{\yng(1,1)}$};
		\node [below right=12pt of 5] (Ac) {$\overline{\tiny{\yng(1,1)}}$};
        \draw (0) to (2) [->, thick];
        \draw (4) to (5) [->, thick];
        \draw (5) to (3) [->, thick];
        \draw (1) to (0) [->, thick];
        \draw (c) to (b) [->, thick];
        \draw (d) to (a) [->, thick];
        \draw (b) to (d) [->, thick];
        \draw (a) to (c) [->, thick];
        \draw (1) to (2) [<->, thick, yellow!70!black];
        \draw (3) to (4) [<->, thick, yellow!70!black];
        \draw (a) to (b) [<->, thick, yellow!55!black];
        \draw (c) to (d) [<->, thick, yellow!55!black];
        \draw (0) to (S) [->, thick, shorten >=-3.5pt, yellow!40!black];
        \draw (Sc) to (0) [->, thick, shorten <=-3.5pt, yellow!40!black];
        \draw (5) to (A) [->, thick, shorten >=-3.5pt, yellow!40!black];
        \draw (Ac) to (5) [->, thick, shorten <=-3.5pt, yellow!40!black];
        \node [] (aa) at (8,-2) {$\ldots$};
        \node [] (bb) at (8,2) {$\ldots$};
        \draw (bb) to (2) [->, thick];
        \draw (aa) to (1) [->, thick];
        \draw (1) to (bb) [->, thick];
        \draw (2) to (aa) [->, thick];
        \node [] (cc) at (15,2) {$\ldots$};
        \node [] (dd) at (15,-2) {$\ldots$};
        \draw (4) to (cc) [->, thick];
        \draw (3) to (dd) [->, thick];
        \draw (dd) to (4) [->, thick];
        \draw (cc) to (3) [->, thick];
		\node [rotate=30] (02) at (1.3,2) {$|$};
		\node [rotate=-30] (01) at (1.3,-2) {$|$};
		\node [rotate=-30] (45) at (21.7,2) {$|$};
		\node [rotate=30] (35) at (21.7,-2) {$|$};
		\draw (0) to [out=80, in=200, looseness=1] (02) [->, shorten >=-0.15cm, thick];
		\draw (2) to [out=160, in=10, looseness=1] (02) [->, shorten >=-0.15cm, thick];
		\draw (0) to [out=280, in=160, looseness=1] (01) [<-, shorten >=-0.15cm, thick];
		\draw (1) to [out=210, in=350, looseness=1] (01) [<-, shorten >=-0.15cm, thick];
		\draw (4) to [out=20, in=170, looseness=1] (45) [<-, shorten >=-0.15cm, thick];
		\draw (5) to [out=100, in=350, looseness=1] (45) [<-, shorten >=-0.15cm, thick];
		\draw (3) to [out=330, in=200, looseness=1] (35) [->, shorten >=-0.15cm, thick];
		\draw (5) to [out=260, in=10, looseness=1] (35) [->, shorten >=-0.15cm, thick];
\end{tikzpicture}}
\caption{The generic quiver of family $\mathcal{A}$ models. Colored fields are the mass deformed pairs. All gauge nodes are $SU(N)$.}\label{fig:GeneralQuiverFamA}
\end{figure}

In this section we study in detail the conformal duality for the $L^{a,b,a}/\mathbb{Z}_2$ models (with $a+b= 2k$) after an orientifold projection that induces only unitary gauge groups. The generic quiver of this family of models is given in Fig.~\ref{fig:GeneralQuiverFamA}, where colored fields have  $R=1$ and they are progressively integrated out by mass deformations that generate the chain of conformally dual models in Eq.~\eqref{eq:ChainZ2}. We realize the orientifolds as fixed points on the five-brane for $a\neq b$ and both even, while for $a=b$ the gauge theory is realized as a glide orientifold \cite{Garcia-Valdecasas:2021znu}. 

The analysis is based on the computation and comparison of the central charges of the different $L^{a,b,a}/\mathbb{Z}_2$ orientifolds, which we denote as $a^{\Omega}_{a,b,a}$. We first analyze in detail the cases with $k=1$ and $k=2$ and then discuss the generalization to any $k$.

\subsection{Orbifold with $k=1$}

The case $k=1$ is the only non-chiral one among the models we discuss in this section. The $L^{0,2,0}/\mathbb{Z}_2$ model is  $\mathbb{C}^3/(\mathbb{Z}_2 \times \mathbb{Z}_2)$ with charges $(0,1,1)\times(1,0,1)$, while $L^{1,1,1}/\mathbb{Z}_2$ is the non-chiral $\mathbb{Z}_2$ orbifold of the conifold $\mathcal{C}$.   

\subsubsection*{Orientifold projection of $L^{0,2,0}/\mathbb{Z}_2$ with fixed points}

\begin{figure}
\begin{subfigure}{0.25\textwidth}
\centering{
    \begin{tikzpicture}[auto, scale=0.8]
		\node [circle, fill=black, inner sep=0pt, minimum size=1.5mm] (0) at (0,0) {}; 
		\node [circle, fill=black, inner sep=0pt, minimum size=1.5mm] (1h) at (1,0) {};
		\node [circle, fill=black, inner sep=0pt, minimum size=1.5mm] (2h) at (2,0) {};
		\node [circle, fill=black, inner sep=0pt, minimum size=1.5mm] (1l) at (0,1) {}; 
		\node [circle, fill=black, inner sep=0pt, minimum size=1.5mm] (2l) at (0,2) {}; 
		\node [circle, fill=black, inner sep=0pt, minimum size=1.5mm] (1c) at (1,1) {}; 
        \draw (0) to (2h) [thick];
        \draw (0) to (2l) [thick];
        \draw (2l) to (2h) [thick];
        \draw (0.5,0) to (0.5,-1.2) [->, thick, blue!50];
        \draw (1.5,0) to (1.5,-1.2) [->, thick, blue!50];
        \draw (0,0.5) to (-1.2,0.5) [->, thick, red!50];
        \draw (0,1.5) to (-1.2,1.5) [->, thick, red!50];
        \draw (0.5,1.5) to (1.3,2.3) [->, thick, green!80!black];
        \draw (1.5,0.5) to (2.3,1.3) [->, thick, green!80!black];
\end{tikzpicture}}
\end{subfigure}
\hfill
\begin{subfigure}{0.35\textwidth}
\centering{
    \begin{tikzpicture}
		\node (0) at (0,0) {}; 
		\node (1) at (4,0) {};
		\node (2) at (4,4) {}; 
		\node (3) at (0,4) {}; 
		\node (c) at (2,2) {};
        \draw (0) to (1) [shorten >=-0.15cm, shorten <=-0.15cm, thick];
        \draw (1) to (2) [shorten >=-0.15cm, shorten <=-0.15cm, thick];
        \draw (2) to (3) [shorten >=-0.15cm, shorten <=-0.15cm, thick];
        \draw (3) to (0) [shorten >=-0.15cm, shorten <=-0.15cm, thick];
        \draw (0,4) to (0,0) [->, thick, blue!50];
        \draw (4,4) to (4,0) [->, thick, blue!50];
        \draw (2,4) to (2,0) [->, thick, blue!50];
        \draw (4,1) to (0,1) [->, thick, red!50];
        \draw (4,3) to (0,3) [->, thick, red!50];
        \draw (0,0) to (4,4) [->, thick, green!80!black];
        \draw (2,0) to (4,2) [->, thick, green!80!black];
        \draw (0,2) to (2,4) [->, thick, green!80!black];
        \draw[fill=gray!80!white, nearly transparent]  (0,0) -- (1,1) -- (0,1) -- cycle;
        \draw[fill=gray!80!white, nearly transparent]  (0,2) -- (1,3) -- (0,3) -- cycle;
        \draw[fill=gray!80!white, nearly transparent]  (2,0) -- (3,1) -- (2,1) -- cycle;
        \draw[fill=gray!80!white, nearly transparent]  (2,2) -- (3,3) -- (2,3) -- cycle;
        \draw[fill=gray!30!white, nearly transparent]  (1,1) -- (2,1) -- (2,2) -- cycle;
        \draw[fill=gray!30!white, nearly transparent]  (1,3) -- (2,3) -- (2,4) -- cycle;
        \draw[fill=gray!30!white, nearly transparent]  (3,1) -- (4,1) -- (4,2) -- cycle;
        \draw[fill=gray!30!white, nearly transparent]  (3,3) -- (4,3) -- (4,4) -- cycle;
		\node[cross out, minimum size=2mm, draw=red!70!black, inner sep=1mm, thick] (0) at (0,0) {}; 
		\node[cross out, minimum size=2mm, draw=red!70!black, inner sep=1mm, thick] (O2) at (0,2) {};
		\node[cross out, minimum size=2mm, draw=red!70!black, inner sep=1mm, thick] (O3) at (2,2) {};
		\node[cross out, minimum size=2mm, draw=red!70!black, inner sep=1mm, thick] (O4) at (2,0) {};
        \node [right=0.05pt of O2, red!70!black] {$\tau_{00}$};
        \node [right=0.05pt of O3, red!70!black] {$\tau_{00}$};
        \node [below right=0.05pt of 0, red!70!black] {$\tau_{11}$};
        \node [below right=0.05pt of O4, red!70!black] {$\tau_{11}$};
        \node[gray] (00) at (1,2) {\small{$0$}};
        \node[gray] (11) at (1,0.3) {\small{$1$}};
        \node[gray] (22) at (1,3.5) {\small{$1$}};
    \end{tikzpicture}}
\end{subfigure}
\hfill
\begin{subfigure}{0.35\textwidth}
\centering{
    \begin{tikzpicture}[auto]
        \node[circle, draw=blue!50, fill=blue!20, inner sep=0pt, minimum size=5mm] (0) at (0,0) {$0$};
        \node[circle, draw=blue!50, fill=blue!20, inner sep=0pt, minimum size=5mm] (1) at (2.5,0) {$1$};
    	\node [left=1pt of 0] (k0) {\small{$SU(N)$}};
		\node [right=1pt of 1] (k1) {\small{$SU(N)$}};
        \draw (0) to (1) [<->, thick];
		\node [above left=12pt of 0] (S) {$\tiny{\yng(2)}$};
		\node [below left=12pt of 0] (Sc) {$\overline{\tiny{\yng(2)}}$};
		\node [above right=12pt of 1] (A) {$\tiny{\yng(1,1)}$};
		\node [below right=12pt of 1] (Ac) {$\overline{\tiny{\yng(1,1)}}$};
		\draw (0) to (S) [->, shorten >=-0.15cm, thick];
		\draw (Sc) to (0) [->, shorten <=-0.15cm, thick];
		\draw (1) to (A) [->, shorten >=-0.15cm, thick];
		\draw (Ac) to (1) [->, shorten <=-0.15cm, thick];
		\node (01) at (1.25,0.7) {$|$};
		\node (10) at (1.25,-0.7) {$|$};
		\draw (0) to [out=60, in=180, looseness=1] (01) [->, shorten >=-0.15cm, thick];
		\draw (1) to [out=130, in=0, looseness=1] (01) [->, shorten >=-0.15cm, thick];
		\draw (0) to [out=300, in=180, looseness=1] (10) [<-, shorten >=-0.15cm, thick];
		\draw (1) to [out=240, in=0, looseness=1] (10) [<-, shorten >=-0.15cm, thick];
    \end{tikzpicture}}
\end{subfigure} 
\caption{The model $L^{0,2,0}/\mathbb{Z}_2$. On the left the toric diagram is drawn, at the center the five-brane and its orientifold projection with fixed points, on the right the quiver resulting from the orientifold projection.}\label{fig:L020Z2Omega}
\end{figure}
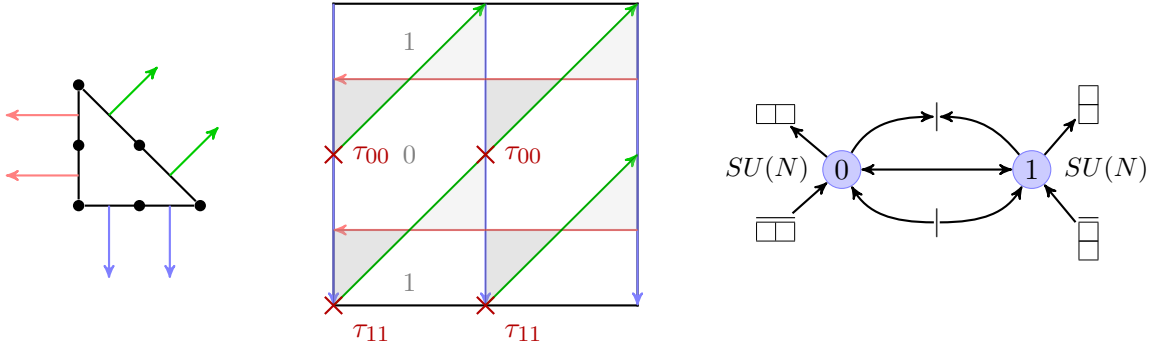

We study the orientifold projections with fixed points of $L^{0,2,0}/\mathbb{Z}_2$. After the projection, the gauge group is $G=SU(N_0)\times SU(N_1)$, whereas the field content is given by 
\begin{align}\label{eq:L020Z2Omegafields}
    &X_{01} = \left( \tiny{\yng(1)_0, \, \overline{\yng(1)}_1} \right) \; , \quad X_{10} = \left( \tiny{\yng(1)_1, \, \overline{\yng(1)}_0} \right) \; \nonumber \\[5pt]
    &Y_{01} = \left( \tiny{\yng(1)_0, \, \yng(1)}_1 \right) \; , \quad \widetilde{Y}_{01} = \left( \overline{ \tiny{\yng(1)} }_0, \, \overline{ \tiny{\yng(1)} }_1 \right) \; \nonumber \\[5pt]
    &T_{00} = \left( \tiny{\yng(1)_0, \, \yng(1)}_0 \right) \; , \quad \widetilde{T}_{00} = \left( \overline{\tiny{\yng(1)} }_0, \, \overline{\tiny{\yng(1)}}_0 \right) \; \nonumber \\[5pt]
    &T_{11} = \left( \tiny{\yng(1)_1, \, \yng(1)}_1 \right) \; , \quad \widetilde{T}_{11} = \left( \overline{\tiny{\yng(1)}}_1, \, \overline{\tiny{\yng(1)}}_1 \right) \; ,
\end{align}
where fields $Y_{01}$ and $\widetilde{Y}_{01}$ arise at the intersection between red and blue vectors on the five-brane diagram in Fig.~\ref{fig:L020Z2Omega}, while the particular representation of the tensor fields $T_{00}$, $\widetilde{T}_{00}$, $T_{11}$ and $\widetilde{T}_{11}$ depends on the signs of the charges $\vec{\tau}$. Since $N_W/2=4$, the product of these charges must be positive. Another constraint for the charges $\vec{\tau}$ comes from gauge anomaly cancellation that requires the presence of conjugate pairs of tensor representations, hence $\tau_{00}=\widetilde{\tau}_{00}$ and $\tau_{11}=\widetilde{\tau}_{11}$. We denote the inequivalent choices of $\vec{\tau}= (\tau_{00},\, \widetilde{\tau}_{00},\, \tau_{11},\, \widetilde{\tau}_{11})$ as $\vec{\tau}_A=(\pm,\, \pm,\, \mp,\, \mp)$ and $\vec{\tau}_B=(\pm,\, \pm,\, \pm,\, \pm)$. The resulting quiver is drawn in Fig.~\ref{fig:L020Z2Omega}, while the superpotential reads 
\begin{align}
    W_{_{0,2,0}}^{\Omega} = X_{01}T_{11}\widetilde{Y}_{01} - Y_{01}\widetilde{T}_{11}X_{10} + X_{10}T_{00}\widetilde{Y}_{01} - Y_{01}\widetilde{T}_{00}X_{01} \; .
\end{align}
Imposing that the $\beta$-functions all vanish one obtains
\begin{align}
     &r_{00} + \widetilde{r}_{00} = r_{11} + \widetilde{r}_{11} \; , \nonumber \\[5pt]
     &r_{01} + \widetilde{r}_{00} + r_{Y} = - 1 \; , \nonumber \\[5pt] 
     &r_{10} + r_{00} + \widetilde{r}_{Y} = - 1 \; , \nonumber \\[5pt] 
     &\left( r_{00} + \widetilde{r}_{00} \right) + \left( r_{01} + r_{10} \right) + \left( r_{Y} + \widetilde{r}_{Y} \right)  = -2 \; , \nonumber \\[5pt]
     &\left( r_{00} + \widetilde{r}_{00} \right) \left( m + 2 \tau_{00} \right) = - 2 m \; , \nonumber \\[5pt]
     &\left( r_{00} + \widetilde{r}_{00} \right) \left( m - 2 \tau_{11} \right) = - 2 m \; ,
\end{align}
where $m=N_0 - N_1$ and $\vec{\tau}_{A}$ is selected. Note that with $m=0$, the choice on $\vec{\tau}$ is no longer constrained.\footnote{The fact that $\vec{\tau}_B$ works here is due to $R=1$. Similarly, in family \emph{ii)} and \emph{iii)} of~\cite{Amariti:2021lhk} one can engineer a configuration of O6-planes that yields pairs of tensors that transform in the same way. The point is that they can still be integrated out.} Further imposing that $R$-charges of conjugate pairs are the same, i.e. $r_{01}=r_{10}$ and $r=\widetilde{r}$ we get
\begin{align}
    &r_{00}=\widetilde{r}_{00}=r_{11}=\widetilde{r}_{11} = - \frac{m}{m + 2 \tau_{00}} \; , \nonumber \\[5pt]
    &r_{01} + r_{00} + r_{Y} = - 1 \; ,
\end{align}
We can select equal ranks by imposing $m=0$, as opposed to the orientifold projection studied in~\cite{Antinucci:2021edv, Amariti:2021lhk}, meaning that this model does not requires the presence of fractional branes. This holds for the whole family $\mathcal{A}$. At large $N$, we have
\begin{align}\label{eq:L020Z2OmegaAnomalies}
    &\mathrm{Tr}R=0 \; , \nonumber \\[5pt]
    &\mathrm{Tr}R^3=2 N^2 \left( r_{01}^3 + (-1 - r_{01})^3 + 1 \right) \; ,
\end{align}
and the central charge is maximized at 
\begin{align}\label{eq:CentralChargeC3Z2Z2Omega}
    &r_{00} = \widetilde{r}_{00} = r_{11} = \widetilde{r}_{11} = 0 \; , \nonumber \\[5pt]
    &r_{01} = r_{10} = r_{Y} = \widetilde{r}_{Y} = - \frac{1}{2} \; , \nonumber \\[5pt] 
    & a_{_{0,2,0}}^{\Omega_{m=0}} = \frac{27}{64} N^2\; ,
\end{align} 
implying superconformal $R=1$ for the tensor fields, and $R=1/2$ for the remaining ones. 

Finally, note that for $m=\tau_{01} $, we have that all the fields have $R=2/3$, and the value of the central charge is
\begin{equation}
 a_{_{0,2,0}}^{\Omega_{m=\tau_{01}}} = \frac{1}{2} N^2 \; .\label{eq:CentralChargeC3Z2Z2Omegatau}
 \end{equation}
 Remarkably, shifting the ranks of the unitary groups, which means that a fractional brane is present in the system, yields the $R$-charge of free fields. This solution is present for all orbifold of flat space $\left(L^{0,2k,0}/\mathbb{Z}_2\right)^\Omega$. Surprisingly, the ratio between the two central charges in Eqs.~\eqref{eq:CentralChargeC3Z2Z2Omega}-\eqref{eq:CentralChargeC3Z2Z2Omegatau} is 27/32. This is supposed to happen when $\mathcal{N}=2$ is broken, via mass deformation, down to $\mathcal{N}=1$~\cite{Tachikawa:2009tt}. We briefly discuss the role of this solution in section \ref{sec:conc}.

\subsubsection*{Glide orientifold of $L^{1,1,1}/\mathbb{Z}_2$}

\begin{figure}
\begin{subfigure}{0.25\textwidth}
\centering{
    \begin{tikzpicture}[auto, scale=0.8]
		\node [circle, fill=black, inner sep=0pt, minimum size=1.5mm] (0) at (0,0) {}; 
		\node [circle, fill=black, inner sep=0pt, minimum size=1.5mm] (1h) at (1,0) {};
		\node [circle, fill=black, inner sep=0pt, minimum size=1.5mm] (2h) at (2,0) {};
		\node [circle, fill=black, inner sep=0pt, minimum size=1.5mm] (0v) at (0,1) {}; 
		\node [circle, fill=black, inner sep=0pt, minimum size=1.5mm] (1v) at (1,1) {}; 
		\node [circle, fill=black, inner sep=0pt, minimum size=1.5mm] (2v) at (2,1) {}; 
        \draw (0) to (2h) [thick];
        \draw (2h) to (2v) [thick];
        \draw (0v) to (2v) [thick];
        \draw (0) to (0v) [thick];
        \draw (0.5,0) to (0.5,-1.2) [->, thick, blue!50];
        \draw (1.5,0) to (1.5,-1.2) [->, thick, blue!50];
        \draw (0,0.5) to (-1.2,0.5) [->, thick, red!50];
        \draw (2,0.5) to (3.2,0.5) [->, thick, brown!50!black];
        \draw (0.5,1) to (0.5,2.2) [->, thick, green!80!black];
        \draw (1.5,1) to (1.5,2.2) [->, thick, green!80!black];
\end{tikzpicture}}
\end{subfigure}
\hfill
\begin{subfigure}{0.35\textwidth}
\centering{
    \begin{tikzpicture}
		\node (0) at (0,0) {}; 
		\node (1) at (4,0) {};
		\node (2) at (4,4) {}; 
		\node (3) at (0,4) {}; 
		\node (c) at (2,2) {};
        \draw (0) to (1) [shorten >=-0.15cm, shorten <=-0.15cm, thick];
        \draw (1) to (2) [shorten >=-0.15cm, shorten <=-0.15cm, thick];
        \draw (2) to (3) [shorten >=-0.15cm, shorten <=-0.15cm, thick];
        \draw (3) to (0) [shorten >=-0.15cm, shorten <=-0.15cm, thick];
        \draw (0,4) to (0,0) [->, thick, blue!50];
        \draw (4,4) to (4,0) [->, thick, blue!50];
        \draw (2,4) to (2,0) [->, thick, blue!50];
        \draw (0,1) to (4,1) [->, thick, brown!50!black];
        \draw (4,3) to (0,3) [->, thick, red!50!];
        \draw (1,0) to (1,4) [->, thick, green!80!black];
        \draw (3,0) to (3,4) [->, thick, green!80!black];
        \draw[fill=gray!80!white, nearly transparent]  (0,1) -- (1,1) -- (1,3) -- (0,3) -- cycle;
        \draw[fill=gray!80!white, nearly transparent]  (2,1) -- (3,1) -- (3,3) -- (2,3) -- cycle;
        \draw[fill=gray!30!white, nearly transparent]  (1,0) -- (2,0) -- (2,1) -- (1,1) -- cycle;
        \draw[fill=gray!30!white, nearly transparent]  (1,3) -- (2,3) -- (2,4) -- (1,4) -- cycle;
        \draw[fill=gray!30!white, nearly transparent]  (3,0) -- (4,0) -- (4,1) -- (3,1) -- cycle;
        \draw[fill=gray!30!white, nearly transparent]  (3,3) -- (4,3) -- (4,4) -- (3,4) -- cycle;
        \draw (-0.2,2) to (4.2,2) [thick, dashed, purple];
        \node [below right=0.05pt of c1, purple] {$\Omega_{\mathrm{gl}}$};
        \node[gray] (00) at (1.5,1.6) {\small{$0$}};
        \node[gray] (11) at (0.5,0.5) {\small{$1$}};
        \node[gray] (22) at (0.5,3.5) {\small{$1$}};
    \end{tikzpicture}}
\end{subfigure}
\hfill
\begin{subfigure}{0.35\textwidth}
\centering{
    \begin{tikzpicture}[auto]
        \node[circle, draw=blue!50, fill=blue!20, inner sep=0pt, minimum size=5mm] (0) at (0,0) {$0$};
        \node[circle, draw=blue!50, fill=blue!20, inner sep=0pt, minimum size=5mm] (1) at (2.5,0) {$1$};
    	\node [left=1pt of 0] (k0) {\small{$SU(N)$}};
		\node [right=1pt of 1] (k1) {\small{$SU(N)$}};
        \draw (0) to (1) [<->, thick];
		\node (01) at (1.25,0.7) {$|$};
		\node (10) at (1.25,-0.7) {$|$};
		\draw (0) to [out=60, in=180, looseness=1] (01) [->, shorten >=-0.15cm, thick];
		\draw (1) to [out=130, in=0, looseness=1] (01) [->, shorten >=-0.15cm, thick];
		\draw (0) to [out=300, in=180, looseness=1] (10) [<-, shorten >=-0.15cm, thick];
		\draw (1) to [out=240, in=0, looseness=1] (10) [<-, shorten >=-0.15cm, thick];
    \end{tikzpicture}}
\end{subfigure} 
\caption{The model $L^{1,1,1}/\mathbb{Z}_2$. On the left the toric diagram is drawn, at the center the five-brane and its glide projection, on the right the quiver resulting from the orientifold projection.}\label{fig:L111Z2Omega}
\end{figure}
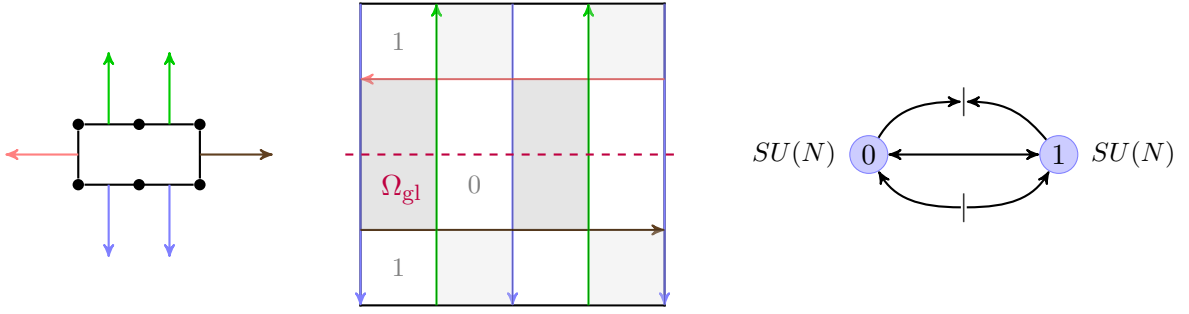
    
We study the glide orientifold of $L^{1,1,1}/\mathbb{Z}_2$, whose field content is 
\begin{align}\label{eq:L111Z2Omegafields}
    &X_{01} = \left( \tiny{\yng(1)_0, \, \overline{\yng(1)}_1} \right) \; , \quad X_{10} = \left( \tiny{\yng(1)_1, \, \overline{\yng(1)}_0} \right) \; \nonumber \\[5pt]
    &Y_{01} = \left( \tiny{\yng(1)_0, \, \yng(1)}_1 \right) \; , \quad \widetilde{Y}_{01} = \left( \overline{ \tiny{\yng(1)} }_0, \, \overline{ \tiny{\yng(1)} }_1 \right) \; ,
\end{align}
which is similar to Eq.~\eqref{eq:L020Z2Omegafields} except for the tensor fields. The projection yields a theory with gauge group $G=SU(N_0)\times SU(N_1)$ and superpotential
\begin{align}
    W_{_{1,1,1}}^{\Omega_{\mathrm{gl}}} = X_{01}X_{10}Y_{01}\widetilde{Y}_{01} - X_{10}X_{01}Y_{01}\widetilde{Y}_{01} \; , 
\end{align}
as can be read from Fig.~\ref{fig:L111Z2Omega}. The set of constraints for the superconformal $R$-charges, together with $r_{01}=r_{10}$ yields
\begin{align}
    & N_0 = N_1 = N \; , \nonumber \\[5pt]
    & r_{01} + r_{Y} = -1 \; ,
\end{align}
and at large $N$ we retrieve Eqs.~\eqref{eq:L020Z2OmegaAnomalies} and at the local maximum
\begin{align}\label{eq:CentralChargeL111Z2Omega}
    &r_{01} = r_{10} = r_{Y} = \widetilde{r}_{Y} = - \frac{1}{2} \; , \nonumber \\[5pt] 
    & a_{_{1,1,1}}^{\Omega_{\mathrm{gl}}} = \frac{27}{64} N^2\; .
\end{align}
The central charge, 't Hooft anomalies and the superconformal index are the same for the orientifold projection with fixed points of $L^{0,2,0}/\mathbb{Z}_2$ and the two orientifold theories are conformally dual.

\subsection{Orbifold with $k=2$}

We now discuss models with $a+b=4$. The parent theories are chiral and their toric diagrams are reflexive polygons, having one internal point~\cite{Hanany:2012hi}. As already mentioned above, we restrict ourselves on orientifolds of models with even $a$ (and $b$), because in the case of $a$, $b$ odd the projection does not give a theory with the the desired features, as we will explicitly see for $L^{1,3,1}/\mathbb{Z}_2$.

\subsubsection*{Orientifold projection of $L^{0,4,0}/\mathbb{Z}_2$ with fixed points}

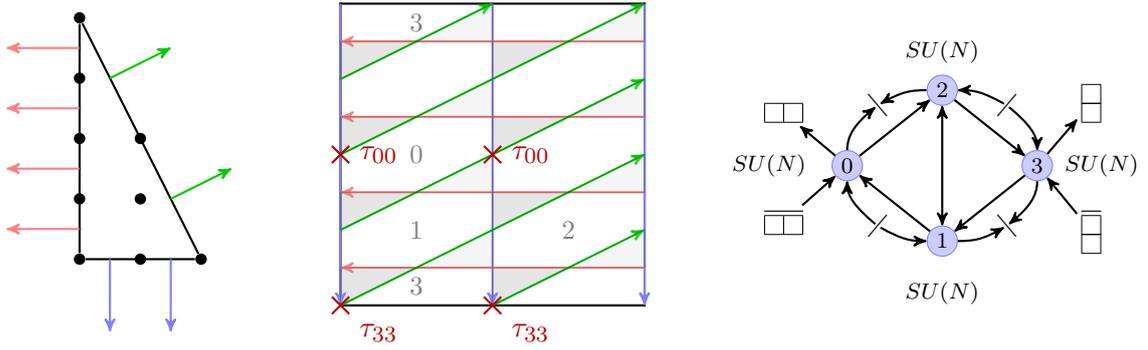
\begin{figure}
\begin{subfigure}{0.25\textwidth}
\centering{
    \begin{tikzpicture}[auto, scale=0.8]
		\node [circle, fill=black, inner sep=0pt, minimum size=1.5mm] (0) at (0,0) {}; 
		\node [circle, fill=black, inner sep=0pt, minimum size=1.5mm] (1h) at (1,0) {};
		\node [circle, fill=black, inner sep=0pt, minimum size=1.5mm] (2h) at (2,0) {};
		\node [circle, fill=black, inner sep=0pt, minimum size=1.5mm] (1l) at (0,1) {}; 
		\node [circle, fill=black, inner sep=0pt, minimum size=1.5mm] (2l) at (0,2) {}; 
		\node [circle, fill=black, inner sep=0pt, minimum size=1.5mm] (3l) at (0,3) {};
		\node [circle, fill=black, inner sep=0pt, minimum size=1.5mm] (4l) at (0,4) {}; 
		\node [circle, fill=black, inner sep=0pt, minimum size=1.5mm] (1c) at (1,1) {};
		\node [circle, fill=black, inner sep=0pt, minimum size=1.5mm] (2c) at (1,2) {}; 
        \draw (0) to (2h) [thick];
        \draw (0) to (4l) [thick];
        \draw (4l) to (2h) [thick];
        \draw (0.5,0) to (0.5,-1.2) [->, thick, blue!50];
        \draw (1.5,0) to (1.5,-1.2) [->, thick, blue!50];
        \draw (0,0.5) to (-1.2,0.5) [->, thick, red!50];
        \draw (0,1.5) to (-1.2,1.5) [->, thick, red!50];
        \draw (0,2.5) to (-1.2,2.5) [->, thick, red!50];
        \draw (0,3.5) to (-1.2,3.5) [->, thick, red!50];
        \draw (0.5,3) to (1.5,3.5) [->, thick, green!80!black];
        \draw (1.5,1) to (2.5,1.5) [->, thick, green!80!black];
\end{tikzpicture}}
\end{subfigure}
\hfill
\begin{subfigure}{0.35\textwidth}
\centering{
    \begin{tikzpicture}
		\node (0) at (0,0) {}; 
		\node (1) at (4,0) {};
		\node (2) at (4,4) {}; 
		\node (3) at (0,4) {}; 
		\node (c) at (2,2) {};
        \draw (0) to (1) [shorten >=-0.15cm, shorten <=-0.15cm, thick];
        \draw (1) to (2) [shorten >=-0.15cm, shorten <=-0.15cm, thick];
        \draw (2) to (3) [shorten >=-0.15cm, shorten <=-0.15cm, thick];
        \draw (3) to (0) [shorten >=-0.15cm, shorten <=-0.15cm, thick];
        \draw (0,4) to (0,0) [->, thick, blue!50];
        \draw (4,4) to (4,0) [->, thick, blue!50];
        \draw (2,4) to (2,0) [->, thick, blue!50];
        \draw (4,0.5) to (0,0.5) [->, thick, red!50];
        \draw (4,1.5) to (0,1.5) [->, thick, red!50];
        \draw (4,2.5) to (0,2.5) [->, thick, red!50];
        \draw (4,3.5) to (0,3.5) [->, thick, red!50];
        \draw (0,0) to (4,2) [->, thick, green!80!black];
        \draw (0,2) to (4,4) [->, thick, green!80!black];
        \draw (2,0) to (4,1) [->, thick, green!80!black];
        \draw (0,1) to (4,3) [->, thick, green!80!black];
        \draw (0,3) to (2,4) [->, thick, green!80!black];
        \draw[fill=gray!80!white, nearly transparent]  (0,0) -- (1,0.5) -- (0,0.5) -- cycle;
        \draw[fill=gray!80!white, nearly transparent]  (0,1) -- (1,1.5) -- (0,1.5) -- cycle;
        \draw[fill=gray!80!white, nearly transparent]  (0,2) -- (1,2.5) -- (0,2.5) -- cycle;
        \draw[fill=gray!80!white, nearly transparent]  (0,3) -- (1,3.5) -- (0,3.5) -- cycle;
        \draw[fill=gray!80!white, nearly transparent]  (2,0) -- (3,0.5) -- (2,0.5) -- cycle;
        \draw[fill=gray!80!white, nearly transparent]  (2,1) -- (3,1.5) -- (2,1.5) -- cycle;
        \draw[fill=gray!80!white, nearly transparent]  (2,2) -- (3,2.5) -- (2,2.5) -- cycle;
        \draw[fill=gray!80!white, nearly transparent]  (2,3) -- (3,3.5) -- (2,3.5) -- cycle;
        \draw[fill=gray!30!white, nearly transparent]  (1,0.5) -- (2,0.5) -- (2,1) -- cycle;
        \draw[fill=gray!30!white, nearly transparent]  (1,1.5) -- (2,1.5) -- (2,2) -- cycle;
        \draw[fill=gray!30!white, nearly transparent]  (1,2.5) -- (2,2.5) -- (2,3) -- cycle;
        \draw[fill=gray!30!white, nearly transparent]  (1,3.5) -- (2,3.5) -- (2,4) -- cycle;
        \draw[fill=gray!30!white, nearly transparent]  (3,0.5) -- (4,0.5) -- (4,1) -- cycle;
        \draw[fill=gray!30!white, nearly transparent]  (3,1.5) -- (4,1.5) -- (4,2) -- cycle;
        \draw[fill=gray!30!white, nearly transparent]  (3,2.5) -- (4,2.5) -- (4,3) -- cycle;
        \draw[fill=gray!30!white, nearly transparent]  (3,3.5) -- (4,3.5) -- (4,4) -- cycle;
        \node[cross out, minimum size=2mm, draw=red!70!black, inner sep=1mm, thick] (0) at (0,0) {}; 
		\node[cross out, minimum size=2mm, draw=red!70!black, inner sep=1mm, thick] (O2) at (0,2) {};
		\node[cross out, minimum size=2mm, draw=red!70!black, inner sep=1mm, thick] (O3) at (2,2) {};
		\node[cross out, minimum size=2mm, draw=red!70!black, inner sep=1mm, thick] (O4) at (2,0) {};
        \node [right=0.05pt of O2, red!70!black] {$\tau_{00}$};
        \node [right=0.05pt of O3, red!70!black] {$\tau_{00}$};
        \node [below right=0.05pt of 0, red!70!black] {$\tau_{33}$};
        \node [below right=0.05pt of O4, red!70!black] {$\tau_{33}$};
        \node[gray] (00) at (1,2) {\small{$0$}};
        \node[gray] (11) at (1,1) {\small{$1$}};
        \node[gray] (22) at (3,1) {\small{$2$}};
        \node[gray] (33) at (1,0.25) {\small{$3$}};
        \node[gray] (333) at (1,3.75) {\small{$3$}};
    \end{tikzpicture}}
\end{subfigure}
\hfill
\begin{subfigure}{0.35\textwidth}
\centering{
    \begin{tikzpicture}[auto]
        \node [circle, draw=blue!50, fill=blue!20, inner sep=0pt, minimum size=4mm] (0) at (0,0) {\scriptsize{$0$}};
		\node [circle, draw=blue!50, fill=blue!20, inner sep=0pt, minimum size=4mm] (1) at (1.25,-1) {\scriptsize{$1$}};
		\node [circle, draw=blue!50, fill=blue!20, inner sep=0pt, minimum size=4mm] (2) at (1.25,1) {\scriptsize{$2$}};
		\node [circle, draw=blue!50, fill=blue!20, inner sep=0pt, minimum size=4mm] (3) at (2.5,0) {\scriptsize{$3$}};
		\node [above left=12pt of 0] (S) {$\tiny{\yng(2)}$};
		\node [below left=12pt of 0] (Sc) {$\overline{\tiny{\yng(2)}}$};
		\node [above right=12pt of 3] (A) {$\tiny{\yng(1,1)}$};
		\node [below right=12pt of 3] (Ac) {$\overline{\tiny{\yng(1,1)}}$};
		\node [left=0.2cm of 0] {\scriptsize{$SU(N)$}};
		\node [below=0.2cm of 1] {\scriptsize{$SU(N)$}};
		\node [above=0.02cm of 2] {\scriptsize{$SU(N)$}};
		\node [right=0.02cm of 3] {\scriptsize{$SU(N)$}};
        \draw (0) to (2) [->, thick];
        \draw (2) to (3) [->, thick];
        \draw (1) to (0) [->, thick];
        \draw (3) to (1) [->, thick];
        \draw (1) to (2) [<->, thick];
        \draw (0) to (S) [->, thick, shorten >=-3.5pt];
        \draw (Sc) to (0) [->, thick, shorten <=-3.5pt];
        \draw (3) to (A) [->, thick, shorten >=-3.5pt];
        \draw (Ac) to (3) [->, thick, shorten <=-3.5pt];
		\node [rotate=40] (02) at (0.4,0.8) {$|$};
		\node [rotate=-40] (01) at (0.4,-0.8) {$|$};
		\node [rotate=-40] (23) at (2.1,0.8) {$|$};
		\node [rotate=40] (13) at (2.1,-0.8) {$|$};
		\draw (0) to [out=90, in=200, looseness=1] (02) [->, shorten >=-0.15cm, thick];
		\draw (2) to [out=180, in=10, looseness=1] (02) [->, shorten >=-0.15cm, thick];
		\draw (0) to [out=270, in=160, looseness=1] (01) [<-, shorten >=-0.15cm, thick];
		\draw (1) to [out=180, in=350, looseness=1] (01) [<-, shorten >=-0.15cm, thick];
		\draw (2) to [out=0, in=170, looseness=1] (23) [<-, shorten >=-0.15cm, thick];
		\draw (3) to [out=90, in=340, looseness=1] (23) [<-, shorten >=-0.15cm, thick];
		\draw (1) to [out=0, in=190, looseness=1] (13) [->, shorten >=-0.15cm, thick];
		\draw (3) to [out=270, in=20, looseness=1] (13) [->, shorten >=-0.15cm, thick];
    \end{tikzpicture}}
\end{subfigure} 
\caption{The model $L^{0,4,0}/\mathbb{Z}_2$. On the left the toric diagram is drawn, at the center the five-brane and its orientifold projection with fixed points, on the right the quiver resulting from the orientifold projection.}\label{fig:L040Z2Omega}
\end{figure}

We study the orientifold of $L^{0,4,0}/\mathbb{Z}_2$ with fixed points such that $\prod \tau = +1$ the projection yields gauge groups $G=SU(N_0) \times SU(N_1) \times SU(N_2) \times SU(N_3)$. The toric diagram of the parent theory, the five-brane and the quiver of the resulting projected theory are shown in Fig~\ref{fig:L040Z2Omega}. The field content is
\begin{align}\label{eq:L040Z2Omegafields}
    &X_{02} = \left( \tiny{\yng(1)_0, \, \overline{\yng(1)}_2} \right) \; , \quad X_{10} = \left( \tiny{\yng(1)_1, \, \overline{\yng(1)}_0} \right) \; \nonumber \\[5pt]
    &X_{12} = \left( \tiny{\yng(1)_1, \, \overline{\yng(1)}_2} \right) \; , \quad X_{21} = \left( \tiny{\yng(1)_2, \, \overline{\yng(1)}_1} \right) \; \nonumber \\[5pt]
    &X_{23} = \left( \tiny{\yng(1)_2, \, \overline{\yng(1)}_3} \right) \; , \quad X_{31} = \left( \tiny{\yng(1)_3, \, \overline{\yng(1)}_1} \right) \; \nonumber \\[5pt]
    &Y_{02} = \left( \tiny{\yng(1)_0, \, \yng(1)}_2 \right) \; , \quad \widetilde{Y}_{01} = \left( \overline{ \tiny{\yng(1)} }_0, \, \overline{ \tiny{\yng(1)} }_1 \right) \; \nonumber \\[5pt]
    &Y_{13} = \left( \tiny{\yng(1)_1, \, \yng(1)}_3 \right) \; , \quad \widetilde{Y}_{23} = \left( \overline{ \tiny{\yng(1)} }_2, \, \overline{ \tiny{\yng(1)} }_3 \right) \; \nonumber \\[5pt]
    &T_{00} = \left( \tiny{\yng(1)_0, \, \yng(1)}_0 \right) \; , \quad \widetilde{T}_{00} = \left( \overline{\tiny{\yng(1)} }_0, \, \overline{\tiny{\yng(1)}}_0 \right) \; \nonumber \\[5pt]
    &T_{33} = \left( \tiny{\yng(1)_3, \, \yng(1)}_3 \right) \; , \quad \widetilde{T}_{33} = \left( \overline{\tiny{\yng(1)}}_3, \, \overline{\tiny{\yng(1)}}_3 \right) \; 
\end{align}
and the superpotential reads 
\begin{align}
    W_{_{0,4,0}}^{\Omega} &= T_{00}\widetilde{Y}_{01}X_{10} - X_{10}X_{02}X_{21} + \widetilde{T}_{00}X_{02}Y_{02} - Y_{02}\widetilde{Y}_{01}X_{12} \nonumber \\[5pt]
    &+ X_{12}X_{23}X_{31} + Y_{13}\widetilde{Y}_{23}X_{21} - T_{33}\widetilde{Y}_{23}X_{23} - \widetilde{T}_{33}X_{31}Y_{13} \; .
\end{align}

Cancellation of gauge-anomalies requires that 
\begin{align}\label{eq:L040Z2OmegaGaugeAnom}
    N_1 - N_2 = 2 \left( \tau_{00} - \widetilde{\tau}_{00} \right) \; .
\end{align}
Proceeding as before, we find the condition for $\beta$-functions to vanish and $R(W)=2$, together with demanding that the $R$-charges of conjugate pairs are equal. Imposing that all ranks are equal, $N_a = N$ $\forall a$, we find 
\begin{align}
      &r_{00} = \widetilde{r}_{00} = r_{33} = \widetilde{r}_{33} = r_{12} = r_{21} = 0 \; , \nonumber \\[5pt] 
      & r_{Y_{02}} = r_{10} \; , \quad r_{Y_{13}} = r_{23} \; , \nonumber \\[5pt] 
      & r_{02} = \widetilde{r}_{Y_{01}} = - 1 - r_{10} \; , \quad r_{31} = \widetilde{r}_{Y_{23}} = - 1 - r_{23} \; ,
\end{align}
in terms of two $R$-charges $r_{10}$ and $r_{23}$. This solution gives at large $N$ 
\begin{align}\label{eq:L040Z2GlobalR}
    &\mathrm{Tr}R=0 \; , \nonumber \\[5pt]
    &\mathrm{Tr}R^3 = 2N^2 \left[ r_{10}^3 + \left( -1 - r_{10} \right)^3 + r_{23}^3 + \left( -1 - r_{23} \right)^3 + 2 \right] \; ,
\end{align}
which is symmetric under the exchange $r_{10} \leftrightarrow r_{23}$. The resulting global anomaly is twice the one of $\left(L^{0,2,0}/\mathbb{Z}_2\right)^{\Omega}$ and the local maximum stays at
\begin{align}\label{eq:L040Z2amax}
    &r_{10} = r_{23} = - \frac{1}{2} \; , \nonumber \\[5pt]
    &a_{_{0,4,0}}^{\Omega} = \frac{27}{32} N^2 \; , 
\end{align}
so that the $R$-charges of tensor fields and vector-like fields is $R=1$ and they do not contribute to 't Hooft anomalies and the superconformal index, while the remaining ones have $R=1/2$.

\subsubsection*{Glide orientifold of $L^{2,2,2}/\mathbb{Z}_2$}

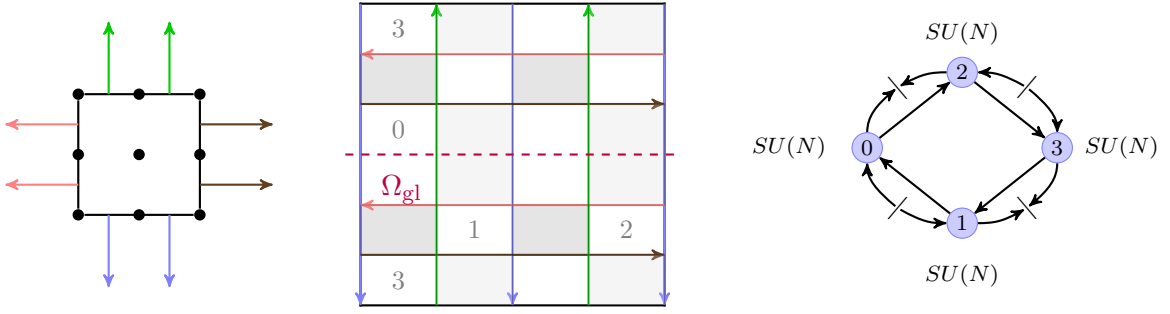
\begin{figure}
\begin{subfigure}{0.25\textwidth}
\centering{
    \begin{tikzpicture}[auto, scale=0.8]
		\node [circle, fill=black, inner sep=0pt, minimum size=1.5mm] (0) at (0,0) {}; 
		\node [circle, fill=black, inner sep=0pt, minimum size=1.5mm] (1h) at (1,0) {};
		\node [circle, fill=black, inner sep=0pt, minimum size=1.5mm] (2h) at (2,0) {};
		\node [circle, fill=black, inner sep=0pt, minimum size=1.5mm] (1l) at (0,1) {}; 
		\node [circle, fill=black, inner sep=0pt, minimum size=1.5mm] (2l) at (0,2) {}; 
		\node [circle, fill=black, inner sep=0pt, minimum size=1.5mm] (1c) at (1,1) {};
		\node [circle, fill=black, inner sep=0pt, minimum size=1.5mm] (2c) at (1,2) {}; 
		\node [circle, fill=black, inner sep=0pt, minimum size=1.5mm] (1r) at (2,1) {};
		\node [circle, fill=black, inner sep=0pt, minimum size=1.5mm] (2r) at (2,2) {}; 
        \draw (0) to (2h) [thick];
        \draw (0) to (2l) [thick];
        \draw (2l) to (2r) [thick];
        \draw (2r) to (2h) [thick];
        \draw (0.5,0) to (0.5,-1.2) [->, thick, blue!50];
        \draw (1.5,0) to (1.5,-1.2) [->, thick, blue!50];
        \draw (0,0.5) to (-1.2,0.5) [->, thick, red!50];
        \draw (0,1.5) to (-1.2,1.5) [->, thick, red!50];
        \draw (2,0.5) to (3.2,0.5) [->, thick, brown!50!black];
        \draw (2,1.5) to (3.2,1.5) [->, thick, brown!50!black];
        \draw (0.5,2) to (0.5,3.2) [->, thick, green!80!black];
        \draw (1.5,2) to (1.5,3.2) [->, thick, green!80!black];
\end{tikzpicture}}
\end{subfigure}
\hfill
\begin{subfigure}{0.35\textwidth}
\centering{
    \begin{tikzpicture}
		\node (0) at (0,0) {}; 
		\node (1) at (4,0) {};
		\node (2) at (4,4) {}; 
		\node (3) at (0,4) {}; 
		\node (c) at (2,2) {};
        \draw (0) to (1) [shorten >=-0.15cm, shorten <=-0.15cm, thick];
        \draw (1) to (2) [shorten >=-0.15cm, shorten <=-0.15cm, thick];
        \draw (2) to (3) [shorten >=-0.15cm, shorten <=-0.15cm, thick];
        \draw (3) to (0) [shorten >=-0.15cm, shorten <=-0.15cm, thick];
        \draw (0,4) to (0,0) [->, thick, blue!50];
        \draw (4,4) to (4,0) [->, thick, blue!50];
        \draw (2,4) to (2,0) [->, thick, blue!50];
        \draw (4,0.67) to (0,0.67) [<-, thick, brown!50!black];
        \draw (4,1.33) to (0,1.33) [->, thick, red!50];
        \draw (4,2.67) to (0,2.67) [<-, thick, brown!50!black];
        \draw (4,3.33) to (0,3.33) [->, thick, red!50];
        \draw (1,0) to (1,4) [->, thick, green!80!black];
        \draw (3,0) to (3,4) [->, thick, green!80!black];
        \draw[fill=gray!80!white, nearly transparent]  (0,0.67) -- (1,0.67) -- (1,1.33) -- (0,1.33) -- cycle;
        \draw[fill=gray!80!white, nearly transparent]  (0,2.67) -- (1,2.67) -- (1,3.33) -- (0,3.33) -- cycle;
        \draw[fill=gray!80!white, nearly transparent]  (2,0.67) -- (3,0.67) -- (3,1.33) -- (2,1.33) -- cycle;
        \draw[fill=gray!80!white, nearly transparent]  (2,2.67) -- (3,2.67) -- (3,3.33) -- (2,3.33) -- cycle;
        \draw[fill=gray!30!white, nearly transparent]  (1,0) -- (1,0.67) -- (2,0.67) -- (2,0) -- cycle;
        \draw[fill=gray!30!white, nearly transparent]  (1,1.33) -- (2,1.33) -- (2,2.67) -- (1,2.67) -- cycle;
        \draw[fill=gray!30!white, nearly transparent]  (1,4) -- (1,3.33) -- (2,3.33) -- (2,4) -- cycle;
        \draw[fill=gray!30!white, nearly transparent]  (3,0) -- (3,0.67) -- (4,0.67) -- (4,0) -- cycle;
        \draw[fill=gray!30!white, nearly transparent]  (3,1.33) -- (4,1.33) -- (4,2.67) -- (3,2.67) -- cycle;
        \draw[fill=gray!30!white, nearly transparent]  (3,4) -- (3,3.33) -- (4,3.33) -- (4,4) -- cycle;
        \draw (-0.2,2) to (4.2,2) [thick, dashed, purple];
        \node [below right=0.05pt of c1, purple] {$\Omega_{\mathrm{gl}}$};
        \node[gray] (00) at (0.5,2.33) {\small{$0$}};
        \node[gray] (11) at (1.5,1) {\small{$1$}};
        \node[gray] (22) at (3.5,1) {\small{$2$}};
        \node[gray] (33) at (0.5,0.33) {\small{$3$}};
        \node[gray] (333) at (0.5,3.67) {\small{$3$}};
    \end{tikzpicture}}
\end{subfigure}
\hfill
\begin{subfigure}{0.35\textwidth}
\centering{
    \begin{tikzpicture}[auto]
        \node [circle, draw=blue!50, fill=blue!20, inner sep=0pt, minimum size=4mm] (0) at (0,0) {\scriptsize{$0$}};
		\node [circle, draw=blue!50, fill=blue!20, inner sep=0pt, minimum size=4mm] (1) at (1.25,-1) {\scriptsize{$1$}};
		\node [circle, draw=blue!50, fill=blue!20, inner sep=0pt, minimum size=4mm] (2) at (1.25,1) {\scriptsize{$2$}};
		\node [circle, draw=blue!50, fill=blue!20, inner sep=0pt, minimum size=4mm] (3) at (2.5,0) {\scriptsize{$3$}};
		\node [left=0.2cm of 0] {\scriptsize{$SU(N)$}};
		\node [below=0.2cm of 1] {\scriptsize{$SU(N)$}};
		\node [above=0.02cm of 2] {\scriptsize{$SU(N)$}};
		\node [right=0.02cm of 3] {\scriptsize{$SU(N)$}};
        \draw (0) to (2) [->, thick];
        \draw (2) to (3) [->, thick];
        \draw (1) to (0) [->, thick];
        \draw (3) to (1) [->, thick];
		\node [rotate=40] (02) at (0.4,0.8) {$|$};
		\node [rotate=-40] (01) at (0.4,-0.8) {$|$};
		\node [rotate=-40] (23) at (2.1,0.8) {$|$};
		\node [rotate=40] (13) at (2.1,-0.8) {$|$};
		\draw (0) to [out=90, in=200, looseness=1] (02) [->, shorten >=-0.15cm, thick];
		\draw (2) to [out=180, in=10, looseness=1] (02) [->, shorten >=-0.15cm, thick];
		\draw (0) to [out=270, in=160, looseness=1] (01) [<-, shorten >=-0.15cm, thick];
		\draw (1) to [out=180, in=350, looseness=1] (01) [<-, shorten >=-0.15cm, thick];
		\draw (2) to [out=0, in=170, looseness=1] (23) [<-, shorten >=-0.15cm, thick];
		\draw (3) to [out=90, in=340, looseness=1] (23) [<-, shorten >=-0.15cm, thick];
		\draw (1) to [out=0, in=190, looseness=1] (13) [->, shorten >=-0.15cm, thick];
		\draw (3) to [out=270, in=20, looseness=1] (13) [->, shorten >=-0.15cm, thick];
    \end{tikzpicture}}
\end{subfigure} 
\caption{The model $L^{2,2,2}/\mathbb{Z}_2$. On the left the toric diagram is drawn, at the center the five-brane and its glide orientifold, on the right the quiver resulting from the projection.}\label{fig:L222Z2Omega}
\end{figure}
    
We now study the glide orientifold of $L^{2,2,2}/\mathbb{Z}_2$, which yields a theory with gauge group $G=SU(N_0)\times SU(N_1)\times SU(N_2) \times SU(N_3)$. The toric diagram of the parent theory, the five-brane and the quiver resulting from the projection are drawn in Fig.~\ref{fig:L222Z2Omega}. The field content is
\begin{align}\label{eq:L222Z2Omegafields}
    &X_{10} = \left( \tiny{\yng(1)_1, \, \overline{\yng(1)}_0} \right) \; , \quad X_{02} = \left( \tiny{\yng(1)_0 \, \overline{\yng(1)}_2} \right) \; \nonumber \\[5pt]
    &X_{23} = \left( \tiny{\yng(1)_2, \, \overline{\yng(1)}_3} \right) \; , \quad X_{31} = \left( \tiny{\yng(1)_3 \, \overline{\yng(1)}_1} \right) \; \nonumber \\[5pt]
    &Y_{02} = \left( \tiny{\yng(1)_0, \, \yng(1)}_2 \right) \; , \quad \widetilde{Y}_{23} = \left( \overline{ \tiny{\yng(1)} }_2, \, \overline{ \tiny{\yng(1)} }_3 \right) \; ; \nonumber \\[5pt]
    &Y_{13} = \left( \tiny{\yng(1)_1, \, \yng(1)}_3 \right) \; , \quad \widetilde{Y}_{01} = \left( \overline{ \tiny{\yng(1)} }_0, \, \overline{ \tiny{\yng(1)} }_1 \right) \; ,
\end{align}
whereas the superpotential is
\begin{align}
    W_{_{2,2,2}}^{\Omega_{\mathrm{gl}}} &= X_{10}Y_{02}\left(X_{02}\right)^T\widetilde{Y}_{01} - X_{10}X_{02}X_{23}X_{31} + X_{31}Y_{13}\left(X_{23}\right)^T\widetilde{Y}_{23} \nonumber \\[5pt] 
    & - \widetilde{Y}_{01}Y_{13}\widetilde{Y}_{23}Y_{02} +  \widetilde{Y}_{23}X_{23}Y_{13}\left(X_{31}\right)^T + Y_{02}\left(X_{10}\right)^T\widetilde{Y}_{01}X_{02} \; .
\end{align} 
The field content is similar to Eq.~\eqref{eq:L040Z2Omegafields} except for the tensor and the vector-like fields. We notice that in this case it is not enough to mass deform the pairs of tensor fields from $\left( L^{0,4,0}/\mathbb{Z}_2 \right)^{\Omega}$, but we need to mass deform also $X_{12}$, $X_{21}$. This happens when the number $k-1$ of vector-like fields is odd, hence for $k$ even.

The set of constraints for the superconformal $R$-charges, together with cancellation of gauge-anomaly and the $\mathbb{Z}_2$-symmetry of the quiver yields
\begin{align}
    & N_0 = N_1 = N_2 = N_3 = N \; , \nonumber \\[5pt]
    & r_{23} = r_{10} \; , \quad r_{02} = r_{31} = -1 - r_{10} \; , \nonumber \\[5pt]
    & r_{Y_{13}} = r_{Y_{02}} \; , \quad \widetilde{r}_{Y_{23}} = \widetilde{r}_{Y_{01}} = -1 - r_{Y_{02}} \; , 
\end{align}    
expressed in terms of the two charges $r_{10}$ and $r_{Y_{02}}$. At large $N$, the global anomalies read
\begin{align}\label{eq:L222Z2GlobalR}
    &\mathrm{Tr}R=0 \; , \nonumber \\[5pt]
    &\mathrm{Tr}R^3 = 2N^2 \left[ r_{10}^3 + \left( -1 - r_{10} \right)^3 + r_{Y_{02}}^3 + \left( -1 - r_{Y_{02}} \right)^3 + 2 \right] \; ,
\end{align}
whose share the same form and local maximum of $\left( L^{0,4,0}/\mathbb{Z}_2 \right)^{\Omega}$ in Eqs.~\eqref{eq:L040Z2GlobalR}-\eqref{eq:L040Z2amax}, as
\begin{align}\label{eq:L222Z2amax}
    &r_{10} = r_{Y_{02}} = - \frac{1}{2} \; , \nonumber \\[5pt]
    &a_{_{0,4,0}}^{\Omega} = \frac{27}{32} N^2 \; , 
\end{align}
The two models orientifold of $L^{0,4,0}/\mathbb{Z}_2$ and of $L^{2,2,2}/\mathbb{Z}_2$ are conformally dual, connected by quadratic marginal deformations.

\subsubsection*{$L^{131}/\mathbb{Z}_2$ and general feature of odd $a$}

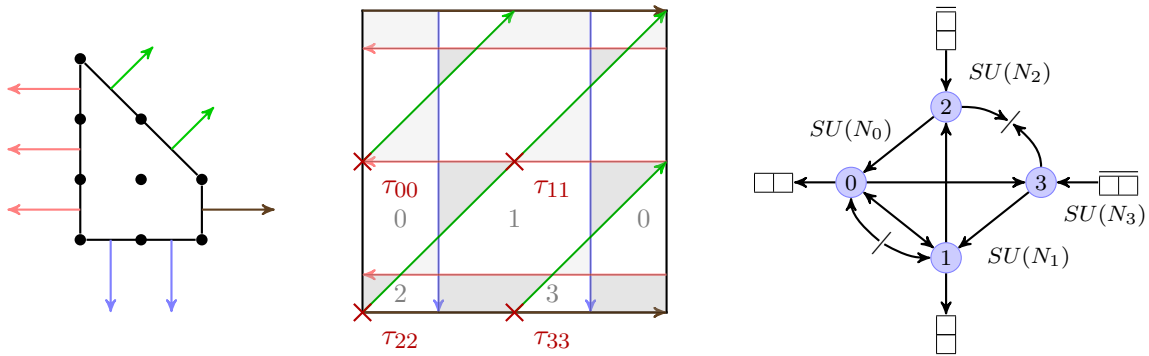
\begin{figure}
\begin{subfigure}{0.25\textwidth}
\centering{
    \begin{tikzpicture}[auto, scale=0.8]
		\node [circle, fill=black, inner sep=0pt, minimum size=1.5mm] (0) at (0,0) {}; 
		\node [circle, fill=black, inner sep=0pt, minimum size=1.5mm] (1h) at (1,0) {};
		\node [circle, fill=black, inner sep=0pt, minimum size=1.5mm] (2h) at (2,0) {};
		\node [circle, fill=black, inner sep=0pt, minimum size=1.5mm] (1l) at (0,1) {}; 
		\node [circle, fill=black, inner sep=0pt, minimum size=1.5mm] (2l) at (0,2) {}; 
		\node [circle, fill=black, inner sep=0pt, minimum size=1.5mm] (3l) at (0,3) {};
		\node [circle, fill=black, inner sep=0pt, minimum size=1.5mm] (1c) at (1,1) {};
		\node [circle, fill=black, inner sep=0pt, minimum size=1.5mm] (2c) at (1,2) {}; 
		\node [circle, fill=black, inner sep=0pt, minimum size=1.5mm] (2r) at (2,1) {}; 
        \draw (0) to (2h) [thick];
        \draw (0) to (3l) [thick];
        \draw (3l) to (2r) [thick];
        \draw (2r) to (2h) [thick];
        \draw (0.5,0) to (0.5,-1.2) [->, thick, blue!50];
        \draw (1.5,0) to (1.5,-1.2) [->, thick, blue!50];
        \draw (0,0.5) to (-1.2,0.5) [->, thick, red!50];
        \draw (0,1.5) to (-1.2,1.5) [->, thick, red!50];
        \draw (0,2.5) to (-1.2,2.5) [->, thick, red!50];
        \draw (2,0.5) to (3.2,0.5) [->, thick, brown!50!black];
        \draw (0.5,2.5) to (1.2,3.2) [->, thick, green!80!black];
        \draw (1.5,1.5) to (2.2,2.2) [->, thick, green!80!black];
\end{tikzpicture}}
\end{subfigure}
\hfill
\begin{subfigure}{0.35\textwidth}
\centering{
    \begin{tikzpicture}
		\node (0) at (0,0) {}; 
		\node (1) at (4,0) {};
		\node (2) at (4,4) {}; 
		\node (3) at (0,4) {}; 
		\node (c) at (2,2) {};
        \draw (0) to (1) [shorten >=-0.15cm, shorten <=-0.15cm, thick];
        \draw (1) to (2) [shorten >=-0.15cm, shorten <=-0.15cm, thick];
        \draw (2) to (3) [shorten >=-0.15cm, shorten <=-0.15cm, thick];
        \draw (3) to (0) [shorten >=-0.15cm, shorten <=-0.15cm, thick];
        \draw (1,4) to (1,0) [->, thick, blue!50];
        \draw (3,4) to (3,0) [->, thick, blue!50];
        \draw (4,0.5) to (0,0.5) [->, thick, red!50];
        \draw (4,2) to (0,2) [->, thick, red!50];
        \draw (4,3.5) to (0,3.5) [->, thick, red!50];
        \draw (0,0) to (4,0) [->, thick, brown!50!black];
        \draw (0,4) to (4,4) [->, thick, brown!50!black];
        \draw (0,0) to (4,4) [->, thick, green!80!black];
        \draw (2,0) to (4,2) [->, thick, green!80!black];
        \draw (0,2) to (2,4) [->, thick, green!80!black];
        \draw[fill=gray!80!white, nearly transparent]  (0,0) -- (0.5,0.5) -- (0,0.5) -- cycle;
        \draw[fill=gray!80!white, nearly transparent]  (1,0) -- (2,0) -- (2.5,0.5) -- (1,0.5) -- cycle;
        \draw[fill=gray!80!white, nearly transparent]  (3,0) -- (4,0) -- (4,0.5) -- (3,0.5) -- cycle;
        \draw[fill=gray!80!white, nearly transparent]  (1,1) -- (1,2) -- (2,2) -- cycle;
        \draw[fill=gray!80!white, nearly transparent]  (3,1) -- (3,2) -- (4,2) -- cycle;
        \draw[fill=gray!80!white, nearly transparent]  (1,3) -- (1,3.5) -- (1.5,3.5) -- cycle;
        \draw[fill=gray!80!white, nearly transparent]  (3,3) -- (3,3.5) -- (3.5,3.5) -- cycle;
        \draw[fill=gray!30!white, nearly transparent]  (0.5,0.5) -- (1,0.5) -- (1,1) -- cycle;
        \draw[fill=gray!30!white, nearly transparent]  (2.5,0.5) -- (3,0.5) -- (3,1) -- cycle;
        \draw[fill=gray!30!white, nearly transparent]  (0,2) -- (1,2) -- (1,3) -- cycle;
        \draw[fill=gray!30!white, nearly transparent]  (2,2) -- (3,2) -- (3,3) -- cycle;
        \draw[fill=gray!30!white, nearly transparent]  (0,3.5) -- (1,3.5) -- (1,4) -- (0,4) -- cycle;
        \draw[fill=gray!30!white, nearly transparent]  (1.5,3.5) -- (3,3.5) -- (3,4) -- (2,4) -- cycle;
        \draw[fill=gray!30!white, nearly transparent]  (3.5,3.5) -- (4,3.5) -- (4,4) -- cycle;
        \node[cross out, minimum size=2mm, draw=red!70!black, inner sep=1mm, thick] (0) at (0,0) {}; 
		\node[cross out, minimum size=2mm, draw=red!70!black, inner sep=1mm, thick] (O2) at (0,2) {};
		\node[cross out, minimum size=2mm, draw=red!70!black, inner sep=1mm, thick] (O3) at (2,2) {};
		\node[cross out, minimum size=2mm, draw=red!70!black, inner sep=1mm, thick] (O4) at (2,0) {};
        \node [below right=0.05pt of O2, red!70!black] {$\tau_{00}$};
        \node [below right=0.05pt of O3, red!70!black] {$\tau_{11}$};
        \node [below right=0.05pt of 0, red!70!black] {$\tau_{22}$};
        \node [below right=0.05pt of O4, red!70!black] {$\tau_{33}$};
        \node[gray] (00) at (0.5,1.25) {\small{$0$}};
        \node[gray] (000) at (3.7,1.25) {\small{$0$}};
        \node[gray] (11) at (2,1.25) {\small{$1$}};
        \node[gray] (22) at (0.5,0.25) {\small{$2$}};
        \node[gray] (33) at (2.5,0.25) {\small{$3$}};
    \end{tikzpicture}}
\end{subfigure}
\hfill
\begin{subfigure}{0.35\textwidth}
\centering{
    \begin{tikzpicture}[auto]
        \node [circle, draw=blue!50, fill=blue!20, inner sep=0pt, minimum size=4mm] (0) at (0,0) {\scriptsize{$0$}};
		\node [circle, draw=blue!50, fill=blue!20, inner sep=0pt, minimum size=4mm] (1) at (1.25,-1) {\scriptsize{$1$}};
		\node [circle, draw=blue!50, fill=blue!20, inner sep=0pt, minimum size=4mm] (2) at (1.25,1) {\scriptsize{$2$}};
		\node [circle, draw=blue!50, fill=blue!20, inner sep=0pt, minimum size=4mm] (3) at (2.5,0) {\scriptsize{$3$}};
		\node [left=12pt of 0] (S) {$\tiny{\yng(2)}$};
		\node [right=12pt of 3] (Sc) {$\overline{\tiny{\yng(2)}}$};
		\node [below=12pt of 1] (A) {$\tiny{\yng(1,1)}$};
		\node [above=12pt of 2] (Ac) {$\overline{\tiny{\yng(1,1)}}$};
		\node [above=0.2cm of 0] {\scriptsize{$SU(N_0)$}};
		\node [right=0.2cm of 1] {\scriptsize{$SU(N_1)$}};
		\node [above right=0.02cm of 2] {\scriptsize{$SU(N_2)$}};
		\node [below right=0.02cm of 3] {\scriptsize{$SU(N_3)$}};
        \draw (2) to (0) [->, thick];
        \draw (1) to (2) [->, thick];
        \draw (3) to (1) [->, thick];
        \draw (1) to (0) [<->, thick];
        \draw (0) to (3) [->, thick];
        \draw (0) to (S) [->, thick, shorten >=-3.5pt];
        \draw (Sc) to (3) [->, thick, shorten <=-3.5pt];
        \draw (1) to (A) [->, thick, shorten >=-3.5pt];
        \draw (Ac) to (2) [->, thick, shorten <=-3.5pt];
		\node [rotate=-40] (01) at (0.4,-0.8) {$|$};
		\node [rotate=-40] (23) at (2.1,0.8) {$|$};
		\draw (0) to [out=270, in=160, looseness=1] (01) [<-, shorten >=-0.15cm, thick];
		\draw (1) to [out=180, in=350, looseness=1] (01) [<-, shorten >=-0.15cm, thick];
		\draw (2) to [out=0, in=170, looseness=1] (23) [->, shorten >=-0.15cm, thick];
		\draw (3) to [out=90, in=340, looseness=1] (23) [->, shorten >=-0.15cm, thick];
    \end{tikzpicture}}
\end{subfigure} 
\caption{The model $L^{1,3,1}/\mathbb{Z}_2$. On the left the toric diagram is drawn, at the center the five-brane and its orientifold projection with fixed points, on the right the quiver resulting from a choice of orientifold projection consistent with gauge anomalies. This does not belong to family $\mathcal{A}$.}\label{fig:L131Z2Omega}
\end{figure}

The set of parent theories $L^{a,b,a}/\mathbb{Z}_2$ with $a+b=4$ involves also the case $a=1$ and $b=3$, i.e. both odd numbers. The orientifold of $L^{1,3,1}/\mathbb{Z}_2$ with four fixed points yields a theory that does not belong to the chain of conformally dual projected theories, i.e. cannot be connected to other models by an exactly marginal deformation that integrates out pairs of conjugate fields, as we instead showed in the previous sections and works~\cite{Antinucci:2021edv, Amariti:2021lhk}. We can see this clearly from Fig.~\ref{fig:L131Z2Omega}. In particular, from the quiver we see that the four tensor fields transform under four different groups, hence they cannot be mass deformed as for the theories in family $\mathcal{A}$. Moreover, cancellation of gauge anomalies gives
\begin{align}
    &N_0 - N_1 - N_2 + N_3 + 4 \tau_{00} = 0 \; , \nonumber \\[5pt]
    -&N_0 + N_1 + N_2 - N_3 + 4 \tau_{11} = 0 \; , \nonumber \\[5pt]
    &N_0 - N_1 - N_2 + N_3 - 4 \tau_{22} = 0 \; , \nonumber \\[5pt]
    -&N_0 + N_1 + N_2 - N_3 - 4 \tau_{33} = 0 \; 
\end{align}
and all ranks equal is not a solution, as it is for family $\mathcal{A}$. 

From the point of view of the five-brane diagram, when $a$ and $b$ are both odd, the $\mathbb{Z}_2$ involution of the orientifold imposes that two horizontal vectors, one oriented to the right and one to the left, pass through two fixed points. See for example Fig.~\ref{fig:L131Z2Omega}, where a red vector oriented to the left lies on $\tau_{00}$ and $\tau_{11}$, while a brown vector oriented to the right lies on $\tau_{22}$ and $\tau_{33}$. Since only two five-branes can meet on a point,\footnote{Or better, in those cases one can describe strongly coupled sectors following \cite{Garcia-Etxebarria:2015hua}.} we need to move either the skew green vectors or the vertical blue vectors. The consequence is that the four fixed points project fields transforming in four different groups. This is general, for all odd $a$ and $b$, and therefore they do not belong to the family $\mathcal{A}$.   

Finally, one can easily see that from the five-brane in Fig.~\ref{fig:L131Z2Omega} the models does not admit a glide orientifold.

\subsection{Generalization}\label{sec:FamAGen}

In this section we want to show the previous results are general and hold for the whole chain 
\begin{align}\label{eq:ChainZ2FamA}
L^{0,2k,0}/\mathbb{Z}_2 \; \to \; L^{2,2(k-1),2}/\mathbb{Z}_2 \ldots \to \; L^{2p,2k-2p,2p}/\mathbb{Z}_2 \; \to \ldots \; \to \;  L^{k,k,k}/\mathbb{Z}_2 \; , 
\end{align}
where $p=1,\, \ldots ,\,  \lfloor \frac{k}{2} \rfloor$. Let us begin with the orientifold of $L^{0,2k,0}/\mathbb{Z}_2$, the generic quiver is drawn in Fig.~\ref{fig:GeneralQuiverFamA}, with all vector-like fields. From the five-brane, we can write down the superpotential as
\begin{align}\label{eq:L02k0SuperPot}
    W^{\Omega}_{0,2k,0} =&\;\; T_{00}\widetilde{Y}_{01}X_{10} + \widetilde{T}_{00}X_{02}Y_{02} - X_{02}X_{21}X_{10} - Y_{02}\widetilde{Y}_{01}X_{12} \nonumber \\[5pt]
    & - T_{2k-1,2k-1}\widetilde{Y}_{2k-2,2k-1}X_{2k-2,2k-1} - \widetilde{T}_{2k-1,2k-1}X_{2k-1,2k-3}Y_{2k-3,2k-1} \nonumber \\[5pt]
    &+ X_{2k-2,2k-1}X_{2k-1,2k-3}X_{2k-3,2k-2} + Y_{2k-3,2k-1}\widetilde{Y}_{2k-2,2k-1}X_{2k-2,2k-3} \nonumber \\[5pt]
    & + \sum_{i=1}^{k-2} \, \left( X_{2i-1,2i}X_{2i,2i+1}X_{2i+1,2i-1} + X_{2i,2i-1}X_{2i-1,2i+2}X_{2i+2,2i} \right) \nonumber \\[5pt]
    & - \sum_{i=1}^{k-2} \, \left( X_{2i+1,2i+2}X_{2i+2,2i}X_{2i,2i+1} + X_{2i+2,2i+1}X_{2i+1,2i-1}X_{2i-1,2i+2} \right) \; .
\end{align}
We need to impose the conditions $R(W)=2$ and that all $\beta$-functions vanish, with all ranks equal, $N_a = N$ $\forall a$.
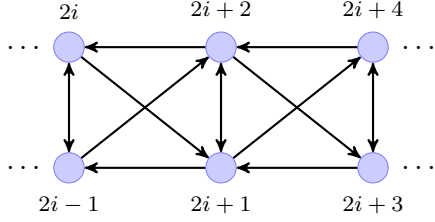
\begin{figure}
\begin{center}
\begin{tikzpicture}[auto, scale=0.4]
		\node [circle, draw=blue!50, fill=blue!20, inner sep=0pt, minimum size=4mm] (a) at (9,-2) {};
		\node [circle, draw=blue!50, fill=blue!20, inner sep=0pt, minimum size=4mm] (b) at (9,2) {};
		\node [circle, draw=blue!50, fill=blue!20, inner sep=0pt, minimum size=4mm] (c) at (14,2) {};
		\node [circle, draw=blue!50, fill=blue!20, inner sep=0pt, minimum size=4mm] (d) at (14,-2) {};
		\node [circle, draw=blue!50, fill=blue!20, inner sep=0pt, minimum size=4mm] (e) at (19,2) {};
		\node [circle, draw=blue!50, fill=blue!20, inner sep=0pt, minimum size=4mm] (f) at (19,-2) {};
		\node [below=1pt of a] (k1) {\scriptsize{$2i-1$}};
		\node [above=1pt of b] (k3) {\scriptsize{$2i$}};
		\node [above=1pt of c] (k2) {\scriptsize{$2i+2$}};
		\node [below=1pt of d] (k4) {\scriptsize{$2i+1$}};
		\node [above=1pt of e] (k5) {\scriptsize{$2i+4$}};
		\node [below=1pt of f] (k6) {\scriptsize{$2i+3$}};
        \draw (c) to (b) [->, thick];
        \draw (d) to (a) [->, thick];
        \draw (b) to (d) [->, thick];
        \draw (a) to (c) [->, thick];
        \draw (e) to (c) [->, thick];
        \draw (f) to (d) [->, thick];
        \draw (c) to (f) [->, thick];
        \draw (d) to (e) [->, thick];
        \draw (a) to (b) [<->, thick];
        \draw (c) to (d) [<->, thick];
        \draw (e) to (f) [<->, thick];
        \node [] (aa) at (7.5,-2) {$\ldots$};
        \node [] (bb) at (7.5,2) {$\ldots$};
        \node [] (cc) at (20.5,2) {$\ldots$};
        \node [] (dd) at (20.5,-2) {$\ldots$};
\end{tikzpicture}
\end{center}
\caption{A portion of generic quiver of the orientifold of $L^{0,2k,0}/\mathbb{Z}_2$ from node $2i$.}\label{fig:PartQuiverFamA}
\end{figure}
Consider a section of the quiver from node $2i$, as in Fig.~\ref{fig:PartQuiverFamA} and the related superpotential terms, whose constraints imply
\begin{align}
    r_{2i,2i-1} =& -1-(r_{2i-1,2i+2}+r_{2i+2,2i}) \; , \nonumber
    \\[5pt]
    r_{2i-1,2i} =& -1-(r_{2i,2i+1}+r_{2i+1,2i-1}) \; ,
\end{align}
as well as the same equations with $i\to i+2$. The sum of this four equations gives:
\begin{equation}
    m_{2i,2i-1} + m_{2i+2,2i+1} =
    -4 - \left(r_{2i-1,2i+2}+r_{2i+2,2i}+r_{2i,2i+1}+r_{2i+1,2i-1} + \lbrace i \to i+2\rbrace \right) \; .
    \label{eq:general_A_W}
\end{equation}
where $m_{2i,2i-1}=r_{2i,2i-1} +  r_{2i-1,2i}$. Vanishing of the beta functions on the nodes $2i+2$ and $2i+1$ imply:
\begin{align}
    m_{2i+2,2i+1} =& - 2 - (r_{2i+2,2i} + r_{2i-1,2i+2} + r_{2i+4,2i+2} + r_{2i+2,2i+3}) \; ,
    \label{eq:general_A_beta1}
    \\[5pt]
    m_{2i+2,2i+1} =& - 2 - (r_{2i+1,2i-1}+r_{2i,2i+1}+r_{2i+1,2i+4}+r_{2i+3,2i+1}) \; .
    \label{eq:general_A_beta2}
\end{align}
The combination of Eqs.~$\eqref{eq:general_A_beta1}+\eqref{eq:general_A_beta2}-\eqref{eq:general_A_W}$ gives
\begin{equation}
    m_{2i,2i-1} = m_{2i+2,2i+1} \; ,
    \quad 
    i=1,\dots,(k-2) \; .
\end{equation}
Similarly one can show that
\begin{equation}
    r_{00} + \tilde{r}_{00}= r_{2k-1,2k-1} + \tilde{r}_{2k-1,2k-1}=m_{2i,2i-1} \; , 
    \quad 
    i=1,\dots,(k-2) \; .
    \label{eq:general_A_masses}
\end{equation}
Finally the beta equation for the node $i=0$ and the superpotential terms that include the tensor fields $T_{00}$ and $\widetilde{T}_{00}$ yield
\begin{equation}
\label{eq:general_A_r0}
    (r_{00} + \tilde{r}_{00}) 2 \tau_{00}=0
\end{equation}
Together with~\eqref{eq:general_A_masses} this implies that the combinations $T_{00} \widetilde{T}_{00}$, $T_{2k-1,2k-1}\widetilde{T}_{2k-1,2k-1}$ and $X_{2i,2i-1}X_{2i-1,2i}$ have fermionic $R$-charge $r=0$ (bosonic $R$-charge $R=2$) and they are marginal deformations. If we impose that conjugate fields have the same $R$-charge, this also means that tensor fields and vector-like have $r$-charge $r=0$. The quadratic marginal operators written above give mass to the fields and we can integrate them out. The resulting effective theory is the orientifold of $L^{2p,2k-2p,2p}/\mathbb{Z}_2$, where $p$ is the number of pairs of conjugate fields that have been integrated out. The conformal mass terms for the pairs of conjugate fields is marginal and does not trigger an RG flow. Another way to see this is that the superpotential of the orientifold of $L^{2p,2k-2p,2p}/\mathbb{Z}_2$ imposes the same constraints in the $r$-charges as the theory $\left(L^{0,2k,0}/\mathbb{Z}_2\right)^{\Omega}$. Indeed from the last two lines of the superpotential in Eq~\eqref{eq:L02k0SuperPot}, using $i \to (i+2)$ for the first term in each line, we can write
\begin{align}
    &r_{2i+2,2i+3} + r_{2i+3,2i+1} + r_{2i+1,2i-1} + r_{2i-1,2i+2} = -2 \; , \nonumber \\[5pt]
    &r_{2i+2,2i} + r_{2i,2i+1} + r_{2i+1,2i+4} + r_{2i+4,2i+2} = -2 \; ,
\end{align}
where we also used \eqref{eq:general_A_r0}.
These are exactly the constraints from the quartic terms after the quadratic deformation. This is due to the fact that one integrates the pairs of conjugate fields out plugging their $F$-terms into the superpotential. Therefore, all we need to study is the first model of the chain in Eq.~\eqref{eq:ChainZ2FamA}. From the superpotential and the $\beta$-functions we have now
\begin{align}
    &r_{00}=\widetilde{r}_{00}=r_{2k-1,2k-1}=\widetilde{r}_{2k-1,2k-1} = r_{2i,2i-1} = r_{2i-1,2i} = 0 \; , \nonumber \\[5pt]
    &r_{2i-1,2i+2} + r_{2i+2,2i} = - 1 \; , \nonumber \\[5pt]
    &r_{2i,2i+1}+r_{2i+1,2i-1} = - 1  \; .
\end{align}
All superpotential terms are generated sequentially shifting $i \to (i+2)$ from the quiver combining a vertical arrow, which does not contribute now, an horizontal one and a diagonal one. Compare Fig.~\ref{fig:ExampleQuiverFamA} and Eq.~\eqref{eq:L02k0SuperPot} in order to see that. Moreover, the generic quiver has a $\mathbb{Z}_2$ symmetry. Hence, we can impose
\begin{align}
    r_{10} &= r_{02} = r_{2k-2,2k-1} = r_{2k-1,2k-3} = r_{2i,2i+1} = r_{2i-1,2i+2} = \ldots \{ i \to (i+2) \} \; , \nonumber \\[5pt]
    r_{Y_{02}} &= \widetilde{r}_{Y_{01}} = r_{Y_{2k-2,2k-1}} = \widetilde{r}_{Y_{2k-3,2k-1}} = r_{2i+2,2i} = r_{2i+1,2i-1} = \ldots \{ i \to (i+2) \} \; ,
\end{align}
so that we can express 't Hooft anomalies only in terms of four $R$-charges, $r_{10}$, $(-1 - r_{10})$, $r_{Y_{02}}$ and $(-1 - r_{Y_{02}})$, and we have a number $k$ of each of them. The central charge reads
\begin{align}
    a^{\Omega}_{_{0,2k,0}} = \frac{9}{32}N^2 k \left[ r_{10}^3 + \left( -1 - r_{10} \right)^3 + r_{Y_{02}}^3 + \left( -1 - r_{Y_{02}} \right)^3 + 2 \right] \; ,
\end{align}
whose local maximum is
\begin{align}
    &r_{10}=r_{Y_{02}}=-\frac{1}{2} \; , \nonumber \\[5pt]
    &a^{\Omega}_{_{0,2k,0}} =  \frac{27}{64}N^2 k \; ,
\end{align}
and the same holds for the orientifold of $L^{2p,2k-2p,2p}/\mathbb{Z}_2$ and $L^{k,k,k}/\mathbb{Z}_2$, as they only differ by fields with $r=0$. Since the quiver is the same and these fields enter in conjugate pairs, 't Hooft anomalies and superconformal index match along the chain of quadratic marginal deformation.

\section{Family $\mathcal{B}$}
\label{sec:famB}

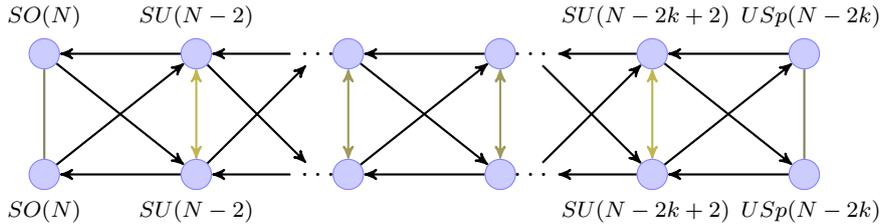
\begin{figure}
\centering{
\begin{tikzpicture}[auto, scale=0.4]
		\node [circle, draw=blue!50, fill=blue!20, inner sep=0pt, minimum size=4mm] (0) at (0,2) {};
		\node [circle, draw=blue!50, fill=blue!20, inner sep=0pt, minimum size=4mm] (1) at (0,-2) {};
		\node [circle, draw=blue!50, fill=blue!20, inner sep=0pt, minimum size=4mm] (2) at (5,2) {};
		\node [circle, draw=blue!50, fill=blue!20, inner sep=0pt, minimum size=4mm] (3) at (5,-2) {};
		\node [circle, draw=blue!50, fill=blue!20, inner sep=0pt, minimum size=4mm] (4) at (10,2) {};
		\node [circle, draw=blue!50, fill=blue!20, inner sep=0pt, minimum size=4mm] (5) at (10,-2) {};
		\node [circle, draw=blue!50, fill=blue!20, inner sep=0pt, minimum size=4mm] (6) at (15,2) {};
		\node [circle, draw=blue!50, fill=blue!20, inner sep=0pt, minimum size=4mm] (7) at (15,-2) {};
		\node [circle, draw=blue!50, fill=blue!20, inner sep=0pt, minimum size=4mm] (8) at (20,2) {};
		\node [circle, draw=blue!50, fill=blue!20, inner sep=0pt, minimum size=4mm] (9) at (20,-2) {};
	    \node [circle, draw=blue!50, fill=blue!20, inner sep=0pt, minimum size=4mm] (10) at (25,2) {};
		\node [circle, draw=blue!50, fill=blue!20, inner sep=0pt, minimum size=4mm] (11) at (25,-2) {};
		\node [above=0.02pt of 0] {\scriptsize{$SO(N)$}};
		\node [above=0.02pt of 2] {\scriptsize{$SU(N-2)$}};
		\node [above=0.02pt of 8] {\scriptsize{$SU(N-2k+2)\;\;$}};
		\node [above=0.02pt of 10] {\scriptsize{$\;\;USp(N-2k)$}};
		\node [below=0.02pt of 1] {\scriptsize{$SO(N)$}};
		\node [below=0.02pt of 3] {\scriptsize{$SU(N-2)$}};
		\node [below=0.02pt of 9] {\scriptsize{$SU(N-2k+2)\;\;$}};
		\node [below=0.02pt of 11] {\scriptsize{$\;\;USp(N-2k)$}};
        \draw (0) to (2) [<-, thick];
        \draw (4) to (6) [<-, thick];
        \draw (8) to (10) [<-, thick];
        \draw (1) to (3) [<-, thick];
        \draw (5) to (7) [<-, thick];
        \draw (9) to (11) [<-, thick];
        \draw (0) to (1) [-, thick, yellow!40!black];
        \draw (2) to (3) [<->, thick, yellow!70!black];
        \draw (4) to (5) [<->, thick, yellow!55!black];
        \draw (6) to (7) [<->, thick, yellow!55!black];
        \draw (8) to (9) [<->, thick, yellow!70!black];
        \draw (10) to (11) [-, thick, yellow!40!black];
        \draw (3) to (0) [<-, thick];
        \draw (7) to (4) [<-, thick];
        \draw (11) to (8) [<-, thick];
        \draw (2) to (1) [<-, thick];
        \draw (6) to (5) [<-, thick];
        \draw (10) to (9) [<-, thick];
        \node [] (aa) at (9,-2) {$\ldots$};
        \node [] (bb) at (9,2) {$\ldots$};
        \draw (2) to (bb) [<-, thick];
        \draw (3) to (aa) [<-, thick];
        \draw (bb) to (3) [<-, thick];
        \draw (aa) to (2) [<-, thick];
        \node [] (cc) at (16,2) {$\ldots$};
        \node [] (dd) at (16,-2) {$\ldots$};
        \draw (cc) to (8) [<-, thick];
        \draw (dd) to (9) [<-, thick];
        \draw (8) to (dd) [<-, thick];
        \draw (9) to (cc) [<-, thick];
\end{tikzpicture}}
\caption{The general quiver of family $\mathcal{B}$ models. Colored fields are the mass deformed pairs.}\label{fig:GeneralQuiverFamB}
\end{figure}

As already anticipated in section \ref{sec:glide}, the second family of conformally dual models that we consider are orientifolds of the chiral  $\mathbb{Z}_2$ orbifolds of $L^{a,b,a}/\mathbb{Z}_2$ (with fixed $a+b=2k$) that give rise to four real gauge groups. As shown in Fig. \ref{fig:five-braneFamB}, for $a\neq b$ (and again both $a$ and $b$ even) the orientifold projection is realized on the five-brane diagram by means of fixed points that are shifted horizontally by a quarter of a period with respect to the case of family $\mathcal{A}$, while for $a=b$ the projection is realized by means of fixed lines. The resulting quiver in drawn in Fig.  \ref{fig:GeneralQuiverFamB}, where as in the previous case the colored fields have unit $R$-charge and they are progressively integrated out along the chain.

Again, the results are based on the computation and comparison  of the central charges, that we denote as $a^{\Omega}_{a,b,a}$ as in the previous section.\footnote{Given that these models are not compared to the models in the previous sections, we assume that this will not cause any confusion to the reader.}
We will discuss in more detail the $k=1$  and $k=2$ cases, showing that 
the central charges of the orientifolds of $L^{0,2,0}/\mathbb{Z}_2$ and $L^{1,1,1}/\mathbb{Z}_2$ coincide. The analysis reveals a direct analogy with the models in family $\mathcal{A}$, and as a consequence the generalization to any $k$ will be given with fewer details.

\subsection{Orbifold with $k=1$}

\subsubsection*{Orientifold projection of $L^{0,2,0}/\mathbb{Z}_2$ with fixed points}

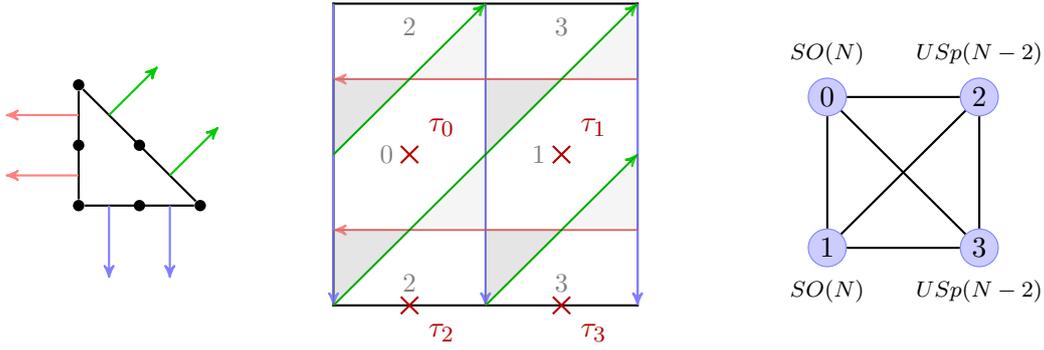
\begin{figure}
\begin{subfigure}{0.25\textwidth}
\centering{
    \begin{tikzpicture}[auto, scale=0.8]
		\node [circle, fill=black, inner sep=0pt, minimum size=1.5mm] (0) at (0,0) {}; 
		\node [circle, fill=black, inner sep=0pt, minimum size=1.5mm] (1h) at (1,0) {};
		\node [circle, fill=black, inner sep=0pt, minimum size=1.5mm] (2h) at (2,0) {};
		\node [circle, fill=black, inner sep=0pt, minimum size=1.5mm] (1l) at (0,1) {}; 
		\node [circle, fill=black, inner sep=0pt, minimum size=1.5mm] (2l) at (0,2) {}; 
		\node [circle, fill=black, inner sep=0pt, minimum size=1.5mm] (1c) at (1,1) {}; 
        \draw (0) to (2h) [thick];
        \draw (0) to (2l) [thick];
        \draw (2l) to (2h) [thick];
        \draw (0.5,0) to (0.5,-1.2) [->, thick, blue!50];
        \draw (1.5,0) to (1.5,-1.2) [->, thick, blue!50];
        \draw (0,0.5) to (-1.2,0.5) [->, thick, red!50];
        \draw (0,1.5) to (-1.2,1.5) [->, thick, red!50];
        \draw (0.5,1.5) to (1.3,2.3) [->, thick, green!80!black];
        \draw (1.5,0.5) to (2.3,1.3) [->, thick, green!80!black];
\end{tikzpicture}}
\end{subfigure}
\hfill
\begin{subfigure}{0.35\textwidth}
\centering{
    \begin{tikzpicture}
		\node (0) at (0,0) {}; 
		\node (1) at (4,0) {};
		\node (2) at (4,4) {}; 
		\node (3) at (0,4) {}; 
		\node (c) at (2,2) {};
        \draw (0) to (1) [shorten >=-0.15cm, shorten <=-0.15cm, thick];
        \draw (1) to (2) [shorten >=-0.15cm, shorten <=-0.15cm, thick];
        \draw (2) to (3) [shorten >=-0.15cm, shorten <=-0.15cm, thick];
        \draw (3) to (0) [shorten >=-0.15cm, shorten <=-0.15cm, thick];
        \draw (0,4) to (0,0) [->, thick, blue!50];
        \draw (4,4) to (4,0) [->, thick, blue!50];
        \draw (2,4) to (2,0) [->, thick, blue!50];
        \draw (4,1) to (0,1) [->, thick, red!50];
        \draw (4,3) to (0,3) [->, thick, red!50];
        \draw (0,0) to (4,4) [->, thick, green!80!black];
        \draw (2,0) to (4,2) [->, thick, green!80!black];
        \draw (0,2) to (2,4) [->, thick, green!80!black];
        \draw[fill=gray!80!white, nearly transparent]  (0,0) -- (1,1) -- (0,1) -- cycle;
        \draw[fill=gray!80!white, nearly transparent]  (0,2) -- (1,3) -- (0,3) -- cycle;
        \draw[fill=gray!80!white, nearly transparent]  (2,0) -- (3,1) -- (2,1) -- cycle;
        \draw[fill=gray!80!white, nearly transparent]  (2,2) -- (3,3) -- (2,3) -- cycle;
        \draw[fill=gray!30!white, nearly transparent]  (1,1) -- (2,1) -- (2,2) -- cycle;
        \draw[fill=gray!30!white, nearly transparent]  (1,3) -- (2,3) -- (2,4) -- cycle;
        \draw[fill=gray!30!white, nearly transparent]  (3,1) -- (4,1) -- (4,2) -- cycle;
        \draw[fill=gray!30!white, nearly transparent]  (3,3) -- (4,3) -- (4,4) -- cycle;
		\node[cross out, minimum size=2mm, draw=red!70!black, inner sep=1mm, thick] (0) at (1,0) {}; 
		\node[cross out, minimum size=2mm, draw=red!70!black, inner sep=1mm, thick] (O2) at (1,2) {};
		\node[cross out, minimum size=2mm, draw=red!70!black, inner sep=1mm, thick] (O3) at (3,2) {};
		\node[cross out, minimum size=2mm, draw=red!70!black, inner sep=1mm, thick] (O4) at (3,0) {};
        \node [above right=0.05pt of O2, red!70!black] {$\tau_{0}$};
        \node [above right=0.05pt of O3, red!70!black] {$\tau_{1}$};
        \node [below right=0.05pt of 0, red!70!black] {$\tau_{2}$};
        \node [below right=0.05pt of O4, red!70!black] {$\tau_{3}$};
        \node[gray] (00) at (0.7,2) {\small{$0$}};
        \node[gray] (11) at (2.7,2) {\small{$1$}};
        \node[gray] (22) at (1,0.3) {\small{$2$}};
        \node[gray] (22) at (1,3.7) {\small{$2$}};
        \node[gray] (33) at (3,0.3) {\small{$3$}};
        \node[gray] (22) at (3,3.7) {\small{$3$}};
    \end{tikzpicture}}
\end{subfigure}
\hfill
\begin{subfigure}{0.35\textwidth}
\centering{
    \begin{tikzpicture}[auto]
        \node[circle, draw=blue!50, fill=blue!20, inner sep=0pt, minimum size=5mm] (0) at (0,0) {$0$};
        \node[circle, draw=blue!50, fill=blue!20, inner sep=0pt, minimum size=5mm] (1) at (0,-2) {$1$};
        \node[circle, draw=blue!50, fill=blue!20, inner sep=0pt, minimum size=5mm] (2) at (2,0) {$2$};
        \node[circle, draw=blue!50, fill=blue!20, inner sep=0pt, minimum size=5mm] (3) at (2,-2) {$3$};
    	\node [above=1pt of 0] (k0) {\scriptsize{$SO(N)$}};
		\node [below=1pt of 1] (k1) {\scriptsize{$SO(N)$}};
		\node [above=1pt of 2] (k0) {\scriptsize{$USp(N-2)$}};
		\node [below=1pt of 3] (k1) {\scriptsize{$USp(N-2)$}};
        \draw (0) to (1) [-, thick];
        \draw (0) to (2) [-, thick];
        \draw (2) to (3) [-, thick];
        \draw (3) to (1) [-, thick];
        \draw (0) to (3) [-, thick];
        \draw (2) to (1) [-, thick];
    \end{tikzpicture}}
\end{subfigure} 
\caption{The model $L^{0,2,0}/\mathbb{Z}_2$. On the left the toric diagram is drawn, at the center the five-brane and its orientifold projection with fixed points, on the right the quiver resulting from the orientifold projection.}\label{fig:L020Z2OmegaB}
\end{figure}

We draw in Fig. \ref{fig:L020Z2OmegaB} the toric diagram of the parent $L^{0,2,0}/\mathbb{Z}_2$ theory, the five-brane diagram with the location of the fixed points and the resulting quiver. Imposing that the superpotential has $R$-charge 2 gives the constraints
\begin{align}
     & r_{03} + r_{01} + r_{13} =-1  \nonumber \\[5pt]
     & r_{03}+ r_{02} + r_{23} =-1 \nonumber \\[5pt]
     & r_{12} + r_{01  } + r_{02} =-1 \nonumber \\[5pt]
     & r _{12}+ r_{13} + r_{23} =-1 \ . \label{superpotconstraintL020Bmodel}
\end{align}
We assign the same $\tau$'s and the same ranks on the groups 0 and 1 and the groups 2 and 3 respectively, so that the quiver possesses a $\mathbb{Z}_2$ symmetry under flip with respect to a horizontal axis, which implies that $r_{02}= r_{13}$ and $r_{03}= r_{12}$. As a consequence, Eqs. \eqref{superpotconstraintL020Bmodel} are reduced to two independent equations implying $r_{01} = r_{23}$, and plugging this into the condition that the $\beta$-function vanishes for each gauge group gives
\begin{equation}
r_{01 } =\frac{N_2 - N_0 + 2 \tau_0}{N_0 - N_2 } = \frac{N_2 - N_0 -2 \tau_2}{N_0 - N_2} \ .    \label{r01r23Z020Bmodel}
\end{equation}
This condition clearly imposes $\tau_0 =-\tau_2$, and we can set $\tau_0=1$ to get the groups in Fig. \ref{fig:L020Z2OmegaB}. It we also assign the ranks as in the figure,\footnote{This implies that $N$ must be even.} Eq. \eqref{r01r23Z020Bmodel} gives $r_{01}=0$. Therefore the fields $X_{01}$ and $X_{23}$ have $R$-charge equal to 1, and maximizing the $a$ central charge one can show that the other $R$-charges are all equal to $\frac{1}{2}$, which can also be more directly deduced observing that the quiver has an additional symmetry under flip of the nodes 2 and 3. 
The value of the central charge at large $N$ is
\begin{equation}
    a^{\Omega}_{0,2,0} = \frac{27}{64} N^2 \ . \label{centralcharge020B}
\end{equation}

As an aside, we observe that once again there is a different assignment of the ranks of the gauge groups, namely $N_2 = N_0 -3 $, which results in a conformal field theory with all $R$-charges equal to $\frac{2}{3}$, and central charge equal to $\frac{1}{2}N^2$. This corresponds to two $SO(N)$ and two $USp(N-3)$ gauge groups.\footnote{Obviously $N$ must be odd in this case.} The occurrence of these orientifolds is a feature of all $L^{0,2k,0}/\mathbb{Z}_2$ models.

\subsubsection*{Orientifold projection of $L^{1,1,1}/\mathbb{Z}_2$ with fixed lines}

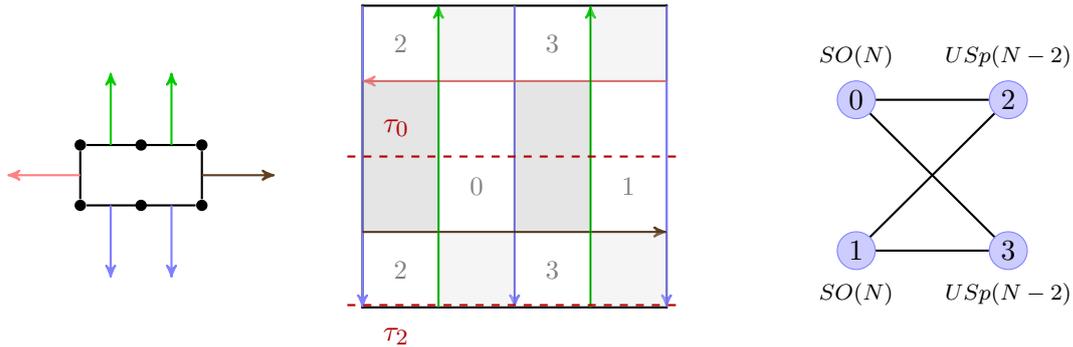
\begin{figure}
\begin{subfigure}{0.25\textwidth}
\centering{
    \begin{tikzpicture}[auto, scale=0.8]
		\node [circle, fill=black, inner sep=0pt, minimum size=1.5mm] (0) at (0,0) {}; 
		\node [circle, fill=black, inner sep=0pt, minimum size=1.5mm] (1h) at (1,0) {};
		\node [circle, fill=black, inner sep=0pt, minimum size=1.5mm] (2h) at (2,0) {};
		\node [circle, fill=black, inner sep=0pt, minimum size=1.5mm] (0v) at (0,1) {}; 
		\node [circle, fill=black, inner sep=0pt, minimum size=1.5mm] (1v) at (1,1) {}; 
		\node [circle, fill=black, inner sep=0pt, minimum size=1.5mm] (2v) at (2,1) {}; 
        \draw (0) to (2h) [thick];
        \draw (2h) to (2v) [thick];
        \draw (0v) to (2v) [thick];
        \draw (0) to (0v) [thick];
        \draw (0.5,0) to (0.5,-1.2) [->, thick, blue!50];
        \draw (1.5,0) to (1.5,-1.2) [->, thick, blue!50];
        \draw (0,0.5) to (-1.2,0.5) [->, thick, red!50];
        \draw (2,0.5) to (3.2,0.5) [->, thick, brown!50!black];
        \draw (0.5,1) to (0.5,2.2) [->, thick, green!80!black];
        \draw (1.5,1) to (1.5,2.2) [->, thick, green!80!black];
\end{tikzpicture}}
\end{subfigure}
\hfill
\begin{subfigure}{0.35\textwidth}
\centering{
    \begin{tikzpicture}
		\node (0) at (0,0) {}; 
		\node (1) at (4,0) {};
		\node (2) at (4,4) {}; 
		\node (3) at (0,4) {}; 
		\node (c) at (2,2) {};
        \draw (0) to (1) [shorten >=-0.15cm, shorten <=-0.15cm, thick];
        \draw (1) to (2) [shorten >=-0.15cm, shorten <=-0.15cm, thick];
        \draw (2) to (3) [shorten >=-0.15cm, shorten <=-0.15cm, thick];
        \draw (3) to (0) [shorten >=-0.15cm, shorten <=-0.15cm, thick];
        \draw (0,4) to (0,0) [->, thick, blue!50];
        \draw (4,4) to (4,0) [->, thick, blue!50];
        \draw (2,4) to (2,0) [->, thick, blue!50];
        \draw (0,1) to (4,1) [->, thick, brown!50!black];
        \draw (4,3) to (0,3) [->, thick, red!50!];
        \draw (1,0) to (1,4) [->, thick, green!80!black];
        \draw (3,0) to (3,4) [->, thick, green!80!black];
        \draw[fill=gray!80!white, nearly transparent]  (0,1) -- (1,1) -- (1,3) -- (0,3) -- cycle;
        \draw[fill=gray!80!white, nearly transparent]  (2,1) -- (3,1) -- (3,3) -- (2,3) -- cycle;
        \draw[fill=gray!30!white, nearly transparent]  (1,0) -- (2,0) -- (2,1) -- (1,1) -- cycle;
        \draw[fill=gray!30!white, nearly transparent]  (1,3) -- (2,3) -- (2,4) -- (1,4) -- cycle;
        \draw[fill=gray!30!white, nearly transparent]  (3,0) -- (4,0) -- (4,1) -- (3,1) -- cycle;
        \draw[fill=gray!30!white, nearly transparent]  (3,3) -- (4,3) -- (4,4) -- (3,4) -- cycle;
        \draw (-0.2,2) to (4.2,2) [thick, dashed, red!70!black];
        \draw (-0.2,0.03) to (4.2,0.03) [thick, dashed, red!70!black];
        \node[above right=0.05 pt of c1, red!70!black] {$\tau_0$};
        \node[below right=0.04 pt of 0, red!70!black] {$\tau_2$};
        \node[gray] (00) at (1.5,1.6) {\small{$0$}};
        \node[gray] (11) at (0.5,0.5) {\small{$2$}};
        \node[gray] (22) at (0.5,3.5) {\small{$2$}};
        \node[gray] (00) at (3.5,1.6) {\small{$1$}};
        \node[gray] (11) at (2.5,0.5) {\small{$3$}};
        \node[gray] (22) at (2.5,3.5) {\small{$3$}};
    \end{tikzpicture}}
\end{subfigure}
\hfill
\begin{subfigure}{0.35\textwidth}
\centering{
    \begin{tikzpicture}[auto]
        \node[circle, draw=blue!50, fill=blue!20, inner sep=0pt, minimum size=5mm] (0) at (0,0) {$0$};
        \node[circle, draw=blue!50, fill=blue!20, inner sep=0pt, minimum size=5mm] (1) at (0,-2) {$1$};
        \node[circle, draw=blue!50, fill=blue!20, inner sep=0pt, minimum size=5mm] (2) at (2,0) {$2$};
        \node[circle, draw=blue!50, fill=blue!20, inner sep=0pt, minimum size=5mm] (3) at (2,-2) {$3$};
    	\node [above=1pt of 0] (k0) {\scriptsize{$SO(N)$}};
		\node [below=1pt of 1] (k1) {\scriptsize{$SO(N)$}};
		\node [above=1pt of 2] (k0) {\scriptsize{$USp(N-2)$}};
		\node [below=1pt of 3] (k1) {\scriptsize{$USp(N-2)$}};
        \draw (0) to (2) [-, thick];
        \draw (3) to (1) [-, thick];
        \draw (0) to (3) [-, thick];
        \draw (2) to (1) [-, thick];
    \end{tikzpicture}}
\end{subfigure} 
\caption{The model $L^{1,1,1}/\mathbb{Z}_2$. On the left the toric diagram is drawn, at the center the five-brane and its orientifold with fixed lines, on the right the quiver resulting from the orientifold projection.}\label{fig:L111Z2OmegaB}
\end{figure}

The second and last orientifold in the $k =1$ chain is the $L^{1,1,1}/\mathbb{Z}_2$ orientifold described in Fig. \ref{fig:L111Z2OmegaB}. This theory has a quartic superpotential which leads to the constraints
\begin{align}
& r_{02} + r_{12} =-1 \nonumber \\[5pt]
& r_{03}+ r_{13} =-1 \nonumber \\[5pt]
& r_{03 } + r_{02} =-1 \nonumber \\[5pt]
& r_{13 } + r_{12} =-1 \ ,
\end{align}
and substituting them in  the  $\beta$-function conditions implies  the $\tau$'s and the rank assignment that  can  be read from the quiver.  
Imposing $a$-maximization one can show that the $R$ charges must all be equal to $\frac{1}{2}$, as one could also easily deduce  by symmetry arguments. This  results in a central charge identical to the one in Eq. \eqref{centralcharge020B}. 

\subsection{Orbifold with $k=2$}

\subsubsection*{Orientifold projection of $L^{0,4,0}/\mathbb{Z}_2$ with fixed points}

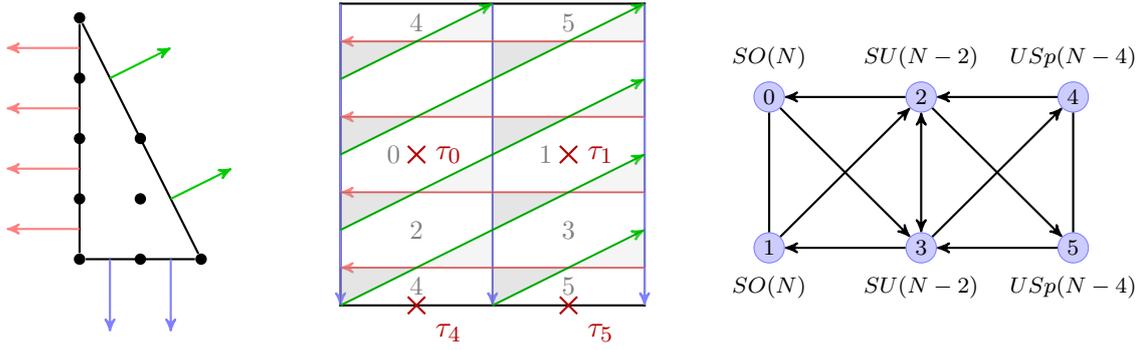
\begin{figure}
\begin{subfigure}{0.25\textwidth}
\centering{
    \begin{tikzpicture}[auto, scale=0.8]
		\node [circle, fill=black, inner sep=0pt, minimum size=1.5mm] (0) at (0,0) {}; 
		\node [circle, fill=black, inner sep=0pt, minimum size=1.5mm] (1h) at (1,0) {};
		\node [circle, fill=black, inner sep=0pt, minimum size=1.5mm] (2h) at (2,0) {};
		\node [circle, fill=black, inner sep=0pt, minimum size=1.5mm] (1l) at (0,1) {}; 
		\node [circle, fill=black, inner sep=0pt, minimum size=1.5mm] (2l) at (0,2) {}; 
		\node [circle, fill=black, inner sep=0pt, minimum size=1.5mm] (3l) at (0,3) {};
		\node [circle, fill=black, inner sep=0pt, minimum size=1.5mm] (4l) at (0,4) {}; 
		\node [circle, fill=black, inner sep=0pt, minimum size=1.5mm] (1c) at (1,1) {};
		\node [circle, fill=black, inner sep=0pt, minimum size=1.5mm] (2c) at (1,2) {}; 
        \draw (0) to (2h) [thick];
        \draw (0) to (4l) [thick];
        \draw (4l) to (2h) [thick];
        \draw (0.5,0) to (0.5,-1.2) [->, thick, blue!50];
        \draw (1.5,0) to (1.5,-1.2) [->, thick, blue!50];
        \draw (0,0.5) to (-1.2,0.5) [->, thick, red!50];
        \draw (0,1.5) to (-1.2,1.5) [->, thick, red!50];
        \draw (0,2.5) to (-1.2,2.5) [->, thick, red!50];
        \draw (0,3.5) to (-1.2,3.5) [->, thick, red!50];
        \draw (0.5,3) to (1.5,3.5) [->, thick, green!80!black];
        \draw (1.5,1) to (2.5,1.5) [->, thick, green!80!black];
\end{tikzpicture}}
\end{subfigure}
\hfill
\begin{subfigure}{0.35\textwidth}
\centering{
    \begin{tikzpicture}
		\node (0) at (0,0) {}; 
		\node (1) at (4,0) {};
		\node (2) at (4,4) {}; 
		\node (3) at (0,4) {}; 
		\node (c) at (2,2) {};
        \draw (0) to (1) [shorten >=-0.15cm, shorten <=-0.15cm, thick];
        \draw (1) to (2) [shorten >=-0.15cm, shorten <=-0.15cm, thick];
        \draw (2) to (3) [shorten >=-0.15cm, shorten <=-0.15cm, thick];
        \draw (3) to (0) [shorten >=-0.15cm, shorten <=-0.15cm, thick];
        \draw (0,4) to (0,0) [->, thick, blue!50];
        \draw (4,4) to (4,0) [->, thick, blue!50];
        \draw (2,4) to (2,0) [->, thick, blue!50];
        \draw (4,0.5) to (0,0.5) [->, thick, red!50];
        \draw (4,1.5) to (0,1.5) [->, thick, red!50];
        \draw (4,2.5) to (0,2.5) [->, thick, red!50];
        \draw (4,3.5) to (0,3.5) [->, thick, red!50];
        \draw (0,0) to (4,2) [->, thick, green!80!black];
        \draw (0,2) to (4,4) [->, thick, green!80!black];
        \draw (2,0) to (4,1) [->, thick, green!80!black];
        \draw (0,1) to (4,3) [->, thick, green!80!black];
        \draw (0,3) to (2,4) [->, thick, green!80!black];
        \draw[fill=gray!80!white, nearly transparent]  (0,0) -- (1,0.5) -- (0,0.5) -- cycle;
        \draw[fill=gray!80!white, nearly transparent]  (0,1) -- (1,1.5) -- (0,1.5) -- cycle;
        \draw[fill=gray!80!white, nearly transparent]  (0,2) -- (1,2.5) -- (0,2.5) -- cycle;
        \draw[fill=gray!80!white, nearly transparent]  (0,3) -- (1,3.5) -- (0,3.5) -- cycle;
        \draw[fill=gray!80!white, nearly transparent]  (2,0) -- (3,0.5) -- (2,0.5) -- cycle;
        \draw[fill=gray!80!white, nearly transparent]  (2,1) -- (3,1.5) -- (2,1.5) -- cycle;
        \draw[fill=gray!80!white, nearly transparent]  (2,2) -- (3,2.5) -- (2,2.5) -- cycle;
        \draw[fill=gray!80!white, nearly transparent]  (2,3) -- (3,3.5) -- (2,3.5) -- cycle;
        \draw[fill=gray!30!white, nearly transparent]  (1,0.5) -- (2,0.5) -- (2,1) -- cycle;
        \draw[fill=gray!30!white, nearly transparent]  (1,1.5) -- (2,1.5) -- (2,2) -- cycle;
        \draw[fill=gray!30!white, nearly transparent]  (1,2.5) -- (2,2.5) -- (2,3) -- cycle;
        \draw[fill=gray!30!white, nearly transparent]  (1,3.5) -- (2,3.5) -- (2,4) -- cycle;
        \draw[fill=gray!30!white, nearly transparent]  (3,0.5) -- (4,0.5) -- (4,1) -- cycle;
        \draw[fill=gray!30!white, nearly transparent]  (3,1.5) -- (4,1.5) -- (4,2) -- cycle;
        \draw[fill=gray!30!white, nearly transparent]  (3,2.5) -- (4,2.5) -- (4,3) -- cycle;
        \draw[fill=gray!30!white, nearly transparent]  (3,3.5) -- (4,3.5) -- (4,4) -- cycle;
        \node[cross out, minimum size=2mm, draw=red!70!black, inner sep=1mm, thick] (0) at (1,0) {}; 
		\node[cross out, minimum size=2mm, draw=red!70!black, inner sep=1mm, thick] (O2) at (1,2) {};
		\node[cross out, minimum size=2mm, draw=red!70!black, inner sep=1mm, thick] (O3) at (3,2) {};
		\node[cross out, minimum size=2mm, draw=red!70!black, inner sep=1mm, thick] (O4) at (3,0) {};
        \node [right=0.05pt of O2, red!70!black] {$\tau_{0}$};
        \node [right=0.05pt of O3, red!70!black] {$\tau_{1}$};
        \node [below right=0.05pt of 0, red!70!black] {$\tau_{4}$};
        \node [below right=0.05pt of O4, red!70!black] {$\tau_{5}$};
        \node[gray] (00) at (0.7,2) {\small{$0$}};
        \node[gray] (11) at (2.7,2) {\small{$1$}};
        \node[gray] (22) at (1,1) {\small{$2$}};
        \node[gray] (33) at (3,1) {\small{$3$}};
        \node[gray] (33) at (1,0.25) {\small{$4$}};
        \node[gray] (333) at (1,3.75) {\small{$4$}};
        \node[gray] (33) at (3,0.25) {\small{$5$}};
        \node[gray] (333) at (3,3.75) {\small{$5$}};
    \end{tikzpicture}}
\end{subfigure}
\hfill
\begin{subfigure}{0.35\textwidth}
\centering{
    \begin{tikzpicture}[auto]
        \node[circle, draw=blue!50, fill=blue!20, inner sep=0pt, minimum size=4mm] (0) at (0,0) {\scriptsize{$0$}};
        \node[circle, draw=blue!50, fill=blue!20, inner sep=0pt, minimum size=4mm] (1) at (0,-2) {\scriptsize{$1$}};
        \node[circle, draw=blue!50, fill=blue!20, inner sep=0pt, minimum size=4mm] (2) at (2,0) {\scriptsize{$2$}};
        \node[circle, draw=blue!50, fill=blue!20, inner sep=0pt, minimum size=4mm] (3) at (2,-2) {\scriptsize{$3$}};
        \node[circle, draw=blue!50, fill=blue!20, inner sep=0pt, minimum size=4mm] (4) at (4,0) {\scriptsize{$4$}};
        \node[circle, draw=blue!50, fill=blue!20, inner sep=0pt, minimum size=4mm] (5) at (4,-2) {\scriptsize{$5$}};
    	\node [above=1pt of 0] (k0) {\scriptsize{$SO(N)$}};
		\node [below=1pt of 1] (k1) {\scriptsize{$SO(N)$}};
		\node [above=1pt of 2] (k2) {\scriptsize{$SU(N-2)$}};
		\node [below=1pt of 3] (k3) {\scriptsize{$SU(N-2)$}};
		\node [above=1pt of 4] (k4) {\scriptsize{$USp(N-4)$}};
		\node [below=1pt of 5] (k5) {\scriptsize{$USp(N-4)$}};
        \draw (0) to (1) [-, thick];
        \draw (2) to (0) [->, thick];
        \draw (4) to (2) [->, thick];
        \draw (3) to (1) [->, thick];
        \draw (5) to (3) [->, thick];
        \draw (0) to (3) [->, thick];
        \draw (1) to (2) [->, thick];
        \draw (2) to (5) [->, thick];
        \draw (3) to (4) [->, thick];
        \draw (2) to (3) [<->, thick];
        \draw (4) to (5) [-, thick];
    \end{tikzpicture}}
\end{subfigure} 
\caption{The model $L^{0,4,0}/\mathbb{Z}_2$. On the left the toric diagram is drawn, at the center the five-brane and its orientifold projection with fixed points, on the right the quiver resulting from the orientifold projection.}\label{fig:L040Z2OmegaB}
\end{figure}

The next model we consider is the $L^{0,4,0}/\mathbb{Z}_2$ orientifold, whose five-brane diagram and quiver, together with the toric diagram of the parent theory, are drawn in Fig.~\ref{fig:L040Z2OmegaB}. We assign the same rank to the groups 0 and 1, 2 and 3 and 4 and 5, and we also require $\tau_0 = \tau_1$ and $\tau_4 = \tau_5$. Again, this implies a symmetry under flip with respect to an horizontal axis. Imposing that the superpotential has $R$-charge 2 then implies $r_{01} = r_{23} = r_{45}$, and requiring that the $\beta$-functions vanish gives
\begin{equation}
r_{01 } =\frac{N_2 - N_0 + 2 \tau_0}{N_0 - N_2 } = \frac{N_4 - N_2 -2 \tau_4}{N_2 - N_4}     \label{r01r23r45Z040Bmodel}
    \end{equation}
together with the further condition on the ranks 
\begin{equation}
    N_2 = \frac{N_0 + N_4}{2} \ . \label{ranks204L040Borientifold}
  \end{equation}
One can then immediately notice that the rank and $\tau$
assignment in the quiver in Fig.~\ref{fig:L040Z2OmegaB} implies that $X_{01}$, $X_{23}$ and $X_{45}$ have $R$-charge 1, while all the remaining fields have $R$-charge $\frac{1}{2}$. The value of the central charge at large $N$ is
\begin{equation}
    a^{\Omega}_{0,4,0} = \frac{27}{32} N^2 \ . \label{centralcharge040B}
\end{equation}
From Eqs.~\eqref{r01r23r45Z040Bmodel} and~\eqref{ranks204L040Borientifold} we deduce that imposing instead that there is a shift of 3  in the ranks, i.e. $N_2 =N_0-3$ and $N_4 = N_2 -3$, results in a conformal model in which all the $R$-charges are equal to $\frac{2}{3}$ and the central charge is equal to $N^2$ at large $N$.

\subsubsection*{Orientifold projection of $L^{2,2,2}/\mathbb{Z}_2$ with fixed lines}

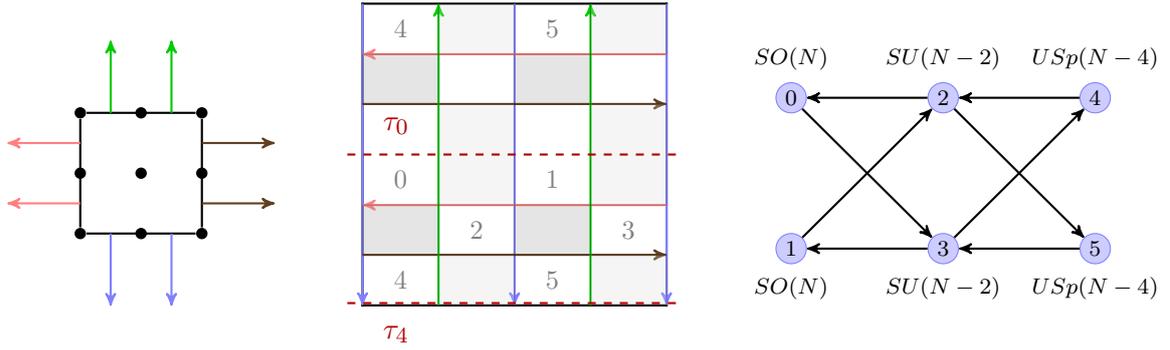
\begin{figure}
\begin{subfigure}{0.25\textwidth}
\centering{
    \begin{tikzpicture}[auto, scale=0.8]
		\node [circle, fill=black, inner sep=0pt, minimum size=1.5mm] (0) at (0,0) {}; 
		\node [circle, fill=black, inner sep=0pt, minimum size=1.5mm] (1h) at (1,0) {};
		\node [circle, fill=black, inner sep=0pt, minimum size=1.5mm] (2h) at (2,0) {};
		\node [circle, fill=black, inner sep=0pt, minimum size=1.5mm] (1l) at (0,1) {}; 
		\node [circle, fill=black, inner sep=0pt, minimum size=1.5mm] (2l) at (0,2) {}; 
		\node [circle, fill=black, inner sep=0pt, minimum size=1.5mm] (1c) at (1,1) {};
		\node [circle, fill=black, inner sep=0pt, minimum size=1.5mm] (2c) at (1,2) {}; 
		\node [circle, fill=black, inner sep=0pt, minimum size=1.5mm] (1r) at (2,1) {};
		\node [circle, fill=black, inner sep=0pt, minimum size=1.5mm] (2r) at (2,2) {}; 
        \draw (0) to (2h) [thick];
        \draw (0) to (2l) [thick];
        \draw (2l) to (2r) [thick];
        \draw (2r) to (2h) [thick];
        \draw (0.5,0) to (0.5,-1.2) [->, thick, blue!50];
        \draw (1.5,0) to (1.5,-1.2) [->, thick, blue!50];
        \draw (0,0.5) to (-1.2,0.5) [->, thick, red!50];
        \draw (0,1.5) to (-1.2,1.5) [->, thick, red!50];
        \draw (2,0.5) to (3.2,0.5) [->, thick, brown!50!black];
        \draw (2,1.5) to (3.2,1.5) [->, thick, brown!50!black];
        \draw (0.5,2) to (0.5,3.2) [->, thick, green!80!black];
        \draw (1.5,2) to (1.5,3.2) [->, thick, green!80!black];
\end{tikzpicture}}
\end{subfigure}
\hfill
\begin{subfigure}{0.35\textwidth}
\centering{
    \begin{tikzpicture}
		\node (0) at (0,0) {}; 
		\node (1) at (4,0) {};
		\node (2) at (4,4) {}; 
		\node (3) at (0,4) {}; 
		\node (c) at (2,2) {};
        \draw (0) to (1) [shorten >=-0.15cm, shorten <=-0.15cm, thick];
        \draw (1) to (2) [shorten >=-0.15cm, shorten <=-0.15cm, thick];
        \draw (2) to (3) [shorten >=-0.15cm, shorten <=-0.15cm, thick];
        \draw (3) to (0) [shorten >=-0.15cm, shorten <=-0.15cm, thick];
        \draw (0,4) to (0,0) [->, thick, blue!50];
        \draw (4,4) to (4,0) [->, thick, blue!50];
        \draw (2,4) to (2,0) [->, thick, blue!50];
        \draw (4,0.67) to (0,0.67) [<-, thick, brown!50!black];
        \draw (4,1.33) to (0,1.33) [->, thick, red!50];
        \draw (4,2.67) to (0,2.67) [<-, thick, brown!50!black];
        \draw (4,3.33) to (0,3.33) [->, thick, red!50];
        \draw (1,0) to (1,4) [->, thick, green!80!black];
        \draw (3,0) to (3,4) [->, thick, green!80!black];
        \draw[fill=gray!80!white, nearly transparent]  (0,0.67) -- (1,0.67) -- (1,1.33) -- (0,1.33) -- cycle;
        \draw[fill=gray!80!white, nearly transparent]  (0,2.67) -- (1,2.67) -- (1,3.33) -- (0,3.33) -- cycle;
        \draw[fill=gray!80!white, nearly transparent]  (2,0.67) -- (3,0.67) -- (3,1.33) -- (2,1.33) -- cycle;
        \draw[fill=gray!80!white, nearly transparent]  (2,2.67) -- (3,2.67) -- (3,3.33) -- (2,3.33) -- cycle;
        \draw[fill=gray!30!white, nearly transparent]  (1,0) -- (1,0.67) -- (2,0.67) -- (2,0) -- cycle;
        \draw[fill=gray!30!white, nearly transparent]  (1,1.33) -- (2,1.33) -- (2,2.67) -- (1,2.67) -- cycle;
        \draw[fill=gray!30!white, nearly transparent]  (1,4) -- (1,3.33) -- (2,3.33) -- (2,4) -- cycle;
        \draw[fill=gray!30!white, nearly transparent]  (3,0) -- (3,0.67) -- (4,0.67) -- (4,0) -- cycle;
        \draw[fill=gray!30!white, nearly transparent]  (3,1.33) -- (4,1.33) -- (4,2.67) -- (3,2.67) -- cycle;
        \draw[fill=gray!30!white, nearly transparent]  (3,4) -- (3,3.33) -- (4,3.33) -- (4,4) -- cycle;
        \draw (-0.2,2) to (4.2,2) [thick, dashed, red!70!black];
        \draw (-0.2,0.03) to (4.2,0.03) [thick, dashed, red!70!black];
        \node[above right=0.05 pt of c1, red!70!black] {$\tau_0$};
        \node[below right=0.04 pt of 0, red!70!black] {$\tau_4$};
        \node[gray] (00) at (0.5,1.67) {\small{$0$}};
        \node[gray] (11) at (2.5,1.67) {\small{$1$}};
        \node[gray] (22) at (1.5,1) {\small{$2$}};
        \node[gray] (33) at (3.5,1) {\small{$3$}};
        \node[gray] (44) at (0.5,0.33) {\small{$4$}};
        \node[gray] (444) at (0.5,3.67) {\small{$4$}};
        \node[gray] (55) at (2.5,0.33) {\small{$5$}};
        \node[gray] (555) at (2.5,3.67) {\small{$5$}};
    \end{tikzpicture}}
\end{subfigure}
\hfill
\begin{subfigure}{0.35\textwidth}
\centering{
    \begin{tikzpicture}[auto]
        \node[circle, draw=blue!50, fill=blue!20, inner sep=0pt, minimum size=4mm] (0) at (0,0) {\scriptsize{$0$}};
        \node[circle, draw=blue!50, fill=blue!20, inner sep=0pt, minimum size=4mm] (1) at (0,-2) {\scriptsize{$1$}};
        \node[circle, draw=blue!50, fill=blue!20, inner sep=0pt, minimum size=4mm] (2) at (2,0) {\scriptsize{$2$}};
        \node[circle, draw=blue!50, fill=blue!20, inner sep=0pt, minimum size=4mm] (3) at (2,-2) {\scriptsize{$3$}};
        \node[circle, draw=blue!50, fill=blue!20, inner sep=0pt, minimum size=4mm] (4) at (4,0) {\scriptsize{$4$}};
        \node[circle, draw=blue!50, fill=blue!20, inner sep=0pt, minimum size=4mm] (5) at (4,-2) {\scriptsize{$5$}};
    	\node [above=1pt of 0] (k0) {\scriptsize{$SO(N)$}};
		\node [below=1pt of 1] (k1) {\scriptsize{$SO(N)$}};
		\node [above=1pt of 2] (k2) {\scriptsize{$SU(N-2)$}};
		\node [below=1pt of 3] (k3) {\scriptsize{$SU(N-2)$}};
		\node [above=1pt of 4] (k4) {\scriptsize{$USp(N-4)$}};
		\node [below=1pt of 5] (k5) {\scriptsize{$USp(N-4)$}};
        \draw (2) to (0) [->, thick];
        \draw (4) to (2) [->, thick];
        \draw (3) to (1) [->, thick];
        \draw (5) to (3) [->, thick];
        \draw (0) to (3) [->, thick];
        \draw (1) to (2) [->, thick];
        \draw (2) to (5) [->, thick];
        \draw (3) to (4) [->, thick];
    \end{tikzpicture}}
\end{subfigure} 
\caption{The model $L^{2,2,2}/\mathbb{Z}_2$. On the left the toric diagram is drawn, at the center the five-brane and its orientifold with fixed lines, on the right the quiver resulting from the projection.}\label{fig:L222Z2OmegaB}
\end{figure}

For the sake of completeness, we also briefly discuss the other model with $k=2$, namely the $L^{2,2,2}/\mathbb{Z}_2$ orientifold with fixed lines. The toric diagram of the parent theory, together with the projected five-brane diagram and quiver, are given in Fig.~\ref{fig:L222Z2OmegaB}, and as usual we assign the ranks so that the quiver possesses a symmetry with respect to the horizontal axis. This fully constrains the model, because imposing the relations on the $R$-charges coming from the superpotential and requiring that the $\beta$-functions vanish give
\begin{equation}
    N_2 = N_0 -2 \tau_0 \qquad N_2 = \frac{N_0 + N_4}{2} \qquad N_4 = N_2 + 2 \tau_2 \ .
\end{equation}
From this we read that the $\tau$'s must be opposite, and choosing for instance $\tau_0=1 $ one gets $N_2= N_0-2 $
 and $N_4 = N_2 -2$, which gives precisely the groups and their corresponding ranks as in the quiver in Fig. \ref{fig:L222Z2OmegaB}.
 One can finally realize that  the central charge $a^{\Omega}_{2,2,2}$ matches exactly, i.e. at all orders in $N$, the central charge in Eq. \eqref{centralcharge040B}. 
 
\subsection{Generalization}

In the following we generalize the set of constraints for the $R$-charges in order to find a superconformal point. Since the line of reasoning follows closely section~\ref{sec:FamAGen}, we show the generic solution in a more compact way. In this family of models the $\tau$'s of the four fixed points project four of the $(2k+2)$ gauge factor as $(\tau_0, \, \tau_1, \, \tau_{2k}, \, \tau_{2k+1})$, which must be equal in pairs as $(\pm , \, \pm, \, \mp, \, \mp )$, in order to yield the same theory of the last model in the family with fixed lines. Let us use the upper signs without loss of generality and consider $L^{0,2k,0}/\mathbb{Z}_2$. All interactions are cubic and the generic terms read
\begin{align}
  \pm \, X_{2j+2,2j} X_{2j, 2j+3} X_{2j+3,2j+2} \mp \, X_{2j+3,2j+1} X_{2j+1, 2j+2} X_{2j+2,2j+3} \; .   
\end{align}
Using Fig.~\ref{fig:GeneralQuiverFamB}, we choose the ranks such that they are symmetric around the horizontal axis, i.e. $N_{2i} = N_{2i+1}$ with $i=0,\ldots,k$. Moreover, all fields enter iteratively in the superpotential and we call the $R$-charge of the fermions $r_x$ for all horizontal fields, $r_y$ for all diagonal ones and $r_z$ for all vertical one. The latter are precisely the fields that are integrated out in the chain Eq.~\ref{eq:ChainZ2}.\footnote{With the choice of the orientifold of family $\mathcal{B}$.} Note that the choices we made corresponds to require that the non-anomalous baryonic symmetries do not contribute at the fixed point. The condition from the superpotential together with the beta functions impose 
\begin{align}
    &r_x + r_y + r_z = -1 \; , \nonumber \\[5pt]
    & N_{2i} - N_{2i+2} = \frac{2}{r_z+1} \; .
\end{align}
We see that choosing a pattern for the ranks fixes the $R$-charge of the vertical fields in Fig.~\ref{fig:GeneralQuiverFamB}. For instance, with $N_{2i+2} = N_{2i} - 2$,\footnote{Which is the choice in Fig.~\ref{fig:GeneralQuiverFamB}.} $r_z=0$ and the mechanism described in the previous sections arises, as we can integrate out the vertical fields. The crucial point is that their mass term is exactly marginal. The resulting theory is the orientifold of $L^{2p,2k-2p,2p}/\mathbb{Z}_2$ with quartic terms in the superpotential that give $2 r_x + 2 r_y = -2$. Therefore, the solution for $L^{0,2k,0}/\mathbb{Z}_2$ still holds. 

Finally, there are $2k$ fields with fermionic $R$-charge $r_x$ and $2k$ with $r_y = -1 - r_x$, so at large $N$ the local maximum of the central charge reads
\begin{align}
    & r_{x} = - \frac{1}{2} \; , \nonumber \\[5pt]
    & a^{\Omega}_{_{0,2k,0}} = \frac{27}{64} k N^2 \; .
\end{align}

If we use instead $N_{2i+2} = N_{2i} - 3$, we find the solution where $R_X=R_Y=R_Z=2/3$, that holds only for $L^{0,2k,0}/\mathbb{Z}_2$, i.e. the orbifolds of flat space, since the $R$-charges are not compatible with a marginal mass term.

\section{Conclusions}
\label{sec:conc}

In this work we have generalized the mechanism studied in \cite{Antinucci:2021edv,Amariti:2021lhk} to chiral $\mathbb{Z}_2$ orbifolds of $L^{a,b,a}$ models. The $L^{a,b,a}$ family exhausts the class of non-chiral toric models. However there exists one (and only one) non-chiral orbifold which does not belong to this infinite class, corresponding to the $\mathbb{C}^3/(\mathbb{Z}_2 \times \mathbb{Z}_2)$ theory studied in section \ref{sec:famA}. We have observed that the fixed point orientifold of this theory is conformally dual to the glide orientifold of $L^{2,2,2}$, corresponding to the non-chiral $\mathbb{Z}_2$ orbifold of the conifold (i.e. $L^{1,1,1}/\mathbb{Z}_2$). The presence of the glide orientifold is the key ingredient that has allowed us to generalize the above construction to an infinite family of dualities analogous to the case of $L^{a,b,a}$ studied in \cite{Antinucci:2021edv,Amariti:2021lhk}.
We explicitly verified that this is not possible for the $L^{a,b,a}$ non-chiral theories studied in \cite{Antinucci:2021edv,Amariti:2021lhk}, so the additional $\mathbb{Z}_2$ orbifold is a necessary condition for the new infinite family to exist.

With the exception of the ``seed'' duality between the non-chiral $\mathbb{C}^3/(\mathbb{Z}_2 \times \mathbb{Z}_2)$ orbifold and $L^{1,1,1}/\mathbb{Z}_2$, this generalization involves chiral models such as $\mathbb{C}^3/(\mathbb{Z}_2 \times \mathbb{Z}_{2k})=L^{0,2k,0}/\mathbb{Z}_2$ and $L^{k,k,k}/\mathbb{Z}_2$. For $k \geq 2$ we have observed the presence of intermediate models $L^{2p,2k-2p,2p}/\mathbb{Z}_2$ with fixed point projections. They altogether form a family of dual projected models that we named family $\mathcal{A}$ in section \ref{sec:famA}. The field theory interpretation of the duality is the presence of an exactly marginal quadratic deformation. In the $L^{a,b,a}$ case such deformation is realized by a pairing of (conjugate) two-index tensor fields with $R=1$ \cite{Antinucci:2021edv,Amariti:2021lhk}. In the orientifolded $L^{a,b,a}/\mathbb{Z}_2$ chiral orbifolds studied here we have observed the possibility of realizing the quadratic deformation via a pair of conjugate bifundamentals.
Note that a different glide orientifold, i.e. around the other axis (see figure \ref{fig:L222Z2Omega} for example) does not give rise to conformally dual models.
    
By considering a different fixed point projection of $L^{2p,2k-2p,2p}/\mathbb{Z}_2$ with $0\leq p < \lfloor k/2 \rfloor$ we have found the existence of a second family, dubbed family $\mathcal{B}$ in section \ref{sec:famB}, involving the fixed line orientifold of $L^{k,k,k}/\mathbb{Z}_2$. This represents a more conventional family of dualities from the perspective of \cite{Antinucci:2021edv,Amariti:2021lhk} since it does not involve a glide orientifold.
    
Another difference between family $\mathcal{A}$ and $\mathcal{B}$ is that in the former all gauge group ranks are equal, whereas in the latter the ranks are assigned in a way that resembles what done in the non-chiral cases of \cite{Antinucci:2021edv,Amariti:2021lhk}. Moreover, in the orientifold models of \cite{Antinucci:2021edv,Amariti:2021lhk} there is always an $\mathcal{N}=2$ mother theory with the same choice of the ranks that flows, upon adding a relevent deformation, to the $\mathcal{N}=1$ models. This is the reason why those dual models inherit part of the action of $S$-duality. We have not been able to identify the mother counterpart for the theories studied here, neither the origin of their duality. However, in both families $\mathcal{A}$ and $\mathcal{B}$, the extremal case $L^{0,2k,0}/\mathbb{Z}_2$ has always a different rank assignment such that $R=2/3$ for all fields and all $\beta$-functions vanish. This choice has two interesting features. First, it is always a shift by one unit w.r.t the rank choice that yields the conformal duality discussed here. It is unclear if this can be understood in terms of fractional branes present in the system, and if there exists any relevant deformation that connects to the conformally dual models. Second, its central charge is always 27/32 times smaller than the central charge of the conformally dual theories, which usually happens when supersymmetry is broken via mass deformation from $\mathcal{N}=2$ to $\mathcal{N}=1$ \cite{Tachikawa:2009tt}. Clearly, this in not the case here, but an explanation is that the Cartan of $SU(2)_R$ survives the extra orbifold $\mathbb{Z}_2$ and enters in the combination with the $U(1)_R$ as in \cite{Tachikawa:2009tt}.

In all orientifold models studied so far, both here and in \cite{Antinucci:2021edv,Amariti:2021lhk}, there are empirical rules at the level of toric diagrams which are necessary but not sufficient to have a conformal duality, namely that the numbers of internal and external points are separately equal and that all internal points sit on a line. The geometric deformation that allow us to pass from a toric diagram to another with these rules is associated to the quadratic deformation on the field theory side, which integrates out fields with $R=1$.
    
Another pair of toric quiver gauge theories describing different singularities before the orientifold projection but that give rise to a pair of chiral models on the same conformal manifold has been obtained in \cite{Antinucci:2020yki}, relating a fixed line projection of PdP$_{3b}$ to a fixed point projection of PdP$_{3c}$. In this case there is no notion of geometric deformation, i.e. the possibility of deforming the superpotential by an exactly marginal massive chiral operator. Even though PdP$_{3c}$ is actually $L^{1,2,1}/\mathbb{Z}_2$, it does not belong to any of the families studied here, because the sum $a+b$ is odd. Nevertheless the model of \cite{Antinucci:2020yki} behaves as the models studied here from the toric perspective, i.e. the toric diagrams in the two phases have separately the same number of internal and external points. For this reason, the models of \cite{Antinucci:2020yki} represent a seed for another infinite family of dual orientifold models. In a forthcoming publication we are planning to show the generalization of this model, similarly to the cases discussed here. 

Finally, let us discuss some interesting avenues of future investigation. 
A possible generalizations involve orientifolds in presence of extra flavors. In \cite{Giacomelli:2022drw} different projections of the same orbifold result in dual unoriented theories. One may also ask if dualities similar to the ones studied here exist in lower or higher dimensional SCFTs. In lower dimensions it would also be interesting to apply the orientifold projections, denoted as $Spin(7)$ orientifolds in \cite{Forcella:2009jj,Franco:2021ixh}, which break holomorphy while preserving some supersymmetry. 

Another aspect that we would like to stress is that differently from the pure $L^{a,b,a}$ cases, where four families have been identified \cite{Amariti:2021lhk}, here for the chiral $L^{a,b,a}/\mathbb{Z}_2$ orbifolds we only found two families giving rise to a conformal duality. The two missing families correspond to the S-dual quivers studied in \cite{Amariti:2021lhk}. Here, by inspection, we have not found the generalization of models with both real gauge groups and tensor matter. If they do not exist, we would like to understand why.

It would be desirable to have a geometric interpretation of the conformal duality from the perspective of the 10d string setup and/or the holographic dual. For instance for the conformal dualities of non-chiral $L^{a,b,a}$ models one can understand them as being inherited from S-duality of the $\mathcal{N}=2$ parent theories. (See \cite[Sec. 4.2]{Amariti:2021lhk}.) The latter are engineered as type IIA elliptic models of D4's, NS5's and O6-planes, where by tilting some of the NS5-branes (or the O6-planes) one can halve supersymmetry. In turn, by lifting the $\mathcal{N}=2$ models to M-theory one can understand S-duality relating two type IIA configurations as two different classical degenerations of the M-theory torus \cite{Uranga:1998uj}. Having at hand a similar picture for the $\mathcal{N}=1$ dualities studied here would clarify their string theory origin. A possible starting point is the conformal duality between one of the fixed point orientifolds of $\mathbb{C}^3/\mathbb{Z}_2$ and the fixed line orientifold of $L^{1,1,1}$, i.e. the conifold. For the latter, both type IIB and IIA (elliptic) configurations have been constructed in \cite{Ahn:2001hy} without relying on the brane tiling technology. One would then need to construct the type IIA engineering of the former, and lift the two IIA setups to M-theory to try to understand the addition of the exactly marginal quadratic deformation in field theory (responsible for the $\mathcal{N}=1$ conformal duality) via a chain of string dualities.

Lastly, another promising piece of geometric technology is K-stability \cite{Collins:2012dh,Collins:2015qsb} of the SCFT \cite{Collins:2016icw,Benvenuti:2017lle,Amariti:2019pky,Fazzi:2019gvt,Alday:2021twg,Bao:2020ugf,Collins:2022nux}, which can be understood as a criterion to check whether the SCFT is stable in the IR against certain deformations of its superpotential, captured by the chiral ring. In favorable situations (such as for toric theories, but also for classes of non-toric ones) these deformations can be classified, and are related to complex deformations of the hypersurface singularity probed by the $N$ D3-branes. It would be interesting to study whether the conformal dualities of this paper admit an interpretation in terms of deformations of the chiral ring of the SCFTs.

\section*{Acknowledgments}

We would like to thank R.~Argurio, P.~Benetti Genolini, M.~Bertolini, F.~Fucito, E.~Garc\'ia-Valdecasas, S.~Giacomelli, J.~F.~Morales, N.~Mekareeya, R.~Savelli, and L.~Tizzano for interesting discussions and comments, as well as the organizers of the ``Theories of Fundamental Interactions 2022'' workshop in Venice where part of this work was performed. MF would like to thank SISSA, Trieste for hospitality during the completion of this work. The work of AA and SR is supported in part by MIUR-PRIN contract 2017CC72MK-003. The work of MB, SM and FR is partially supported by the MIUR PRIN Grant 2020KR4KN2 ``String Theory as a bridge between Gauge Theories and Quantum Gravity''. The work of MF is supported in part by the European Union's Horizon 2020 research and innovation programme under the Marie Skłodowska-Curie grant agreement No. 754496 - FELLINI. 

\bibliographystyle{JHEP}
\bibliography{biblio}

\end{document}